\documentclass{ws-ijmpd}
\usepackage[super,compress]{cite}
\pdfoutput=1
\usepackage{graphicx,epsf,amsmath}
\usepackage[]{latexsym}
\usepackage{color}
\usepackage{appendix}
\DeclareMathOperator\erf{erf}
\DeclareMathOperator\erfc{erfc}
\newcommand{\be}{\begin{eqnarray}}
\newcommand{\ee}{\end{eqnarray}}

\newcommand{\ket}[1]{\mbox{$\mid #1\,\rangle$}}

\newcommand{\pro}[2]{\mbox{$\langle\, #1 \mid #2\,\rangle$}}
\newcommand{\expec}[1]{\mbox{$\langle\, #1\,\rangle$}}

\renewcommand{\d}{\mbox{${\rm d}$}}
\newcommand{\lp}{\ell_{\rm p}}
\newcommand{\mpl}{m_{\rm p}}

\newcommand{\rh}{r_{\rm H}}
\newcommand{\Rh}{R_{\rm H}}

\newcommand{\gd}{G_{\rm D}}

\newcommand{\E}{\mathrm{E}}
\newcommand{\psis}{{\psi}_{\rm S}}

\newcommand{\psih}{{\psi}_{\rm H}}

\begin{document}

\markboth{R.~Casadio, A.~Giugno and O.~Micu}
{Horizon Quantum Mechanics: A hitchhiker's guide to Quantum Black Holes}

%
\catchline{}{}{}{}{}
%

\title{HORIZON QUANTUM MECHANICS \\
\small A hitchhiker's guide to Quantum Black Holes}

\author{Roberto~Casadio}

\address{Dipartimento di Fisica e Astronomia,\\
Alma Mater Universit\`a di Bologna,\\
via~Irnerio~46, 40126~Bologna, Italy\\
casadio@bo.infn.it}

\author{Andrea~Giugno}

\address{Dipartimento di Fisica e Astronomia,\\
Alma Mater Universit\`a di Bologna,\\
via~Irnerio~46, 40126~Bologna, Italy\\
andrea.giugno2@unibo.it}

\author{Octavian Micu}

\address{Institute of Space Science, Bucharest,\\
P.O.~Box MG-23, RO-077125 Bucharest-Magurele, Romania\\
octavian.micu@spacescience.ro}

\maketitle

\begin{history}
\received{Day Month Year}
\revised{Day Month Year}
\end{history}

\begin{abstract}
It is congruous with the quantum nature of the world to view the space-time geometry 
as an emergent structure that shows classical features only at some observational level. 
One can thus conceive the space-time manifold as a purely theoretical arena, 
where quantum states are defined, with the additional freedom of changing coordinates
like any other symmetry.
Observables, including positions and distances, should then be described by suitable
operators acting on such quantum states. In principle, the top-down (canonical) quantisation 
of Einstein-Hilbert gravity falls right into this picture, but is notoriously very involved. 
The complication stems from allowing all the classical canonical variables that appear 
in the (presumably) fundamental action to become quantum observables acting 
on the ``superspace'' of all metrics, regardless of whether they play any role 
in the description of a specific physical system.
On can instead revisit the more humble ``minisuperspace'' approach and choose
the gravitational observables not simply by imposing some symmetry,
but motivated by their proven relevance in the (classical) description of a given system. 
In particular, this review focuses on compact, spherically symmetric, 
quantum mechanical sources, in order to determine the probability they
are black holes rather than regular particles.
The gravitational radius is therefore lifted to the status of a
quantum mechanical operator acting on the ``horizon wave-function'', the latter
being determined by the quantum state of the source.
This formalism is then applied to several sources with a mass around the
fundamental scale, which are viewed as natural candidates of quantum black holes.
\end{abstract}
\keywords{Horizon Wave-Function; Quantum Mechanics; Black Holes.}
\ccode{PACS numbers:}
%
%
\section{Introduction}
After Einstein introduced the theory of Special Relativity~\cite{einstein1}, we have grown
accustomed to thinking of the space-time as the geometrical space where things happen. 
In this respect, Special Relativity just adds one dimension to the three-dimensional space 
of Newtonian physics, which is the natural arena for describing mathematically 
our intuitive notion of motion, or object displacements. 
However, we should not forget Einstein's first great achievement came 
from a rethinking of the concept of time and length as being related to actual measurements, 
which in turn require synchronised clocks. Quantum physics emerged around the same time 
from the very same perspective: a proper description of atoms and elementary particles, 
and other phenomena mostly occurring at microscopic scales, required a more refined analysis 
of how variables involved in such phenomena are actually measured. 
Since measuring means interacting with the system under scrutiny, 
the uncertainty principle due to a finite Planck constant then came out as a fact of life, 
like the Lorentz transformations come out from the finite speed of light. 
This gave rise to the mathematical structure of the complex Hilbert space of states, 
on which observables are given by operators with suitable properties, 
and the outcome of any measurements could then be predicted with at best a certain probability. 
In Special Relativity one can nonetheless think of the space-time coordinates 
as being labels of actual space-time points, observables in principle, 
as they implicitly define an inertial observer.
\par
Then came General Relativity~\cite{einstein2}, which allows for the use of any coordinates 
to identify space-time points, in a way that let us describe physics 
again much closer to what experimentalists do. The price to pay is that space-time 
correspondingly becomes a manifold endowed with a general Lorentzian metric, 
which acts as the ``potential'' for the universal gravitational force. 
This metric, in practice, determines the causal structure that was before given 
by the fixed Minkowski metric, and black holes (BHs) were found in this theory. 
The quantisation of matter fields on these metric manifolds led to the discovery 
of paradoxes and other difficulties, which are often pinpointed as the smoking gun 
that these two theories, of Quantum matter (fields) and General Relativity, are hard to unify. 
But if one looks back at how these two pillars of modern physics precisely emerged 
from the rethinking of the interplay between a physical system and the observer, 
the path to follow should become clear, at least ideally: one should give up 
as many assumptions as possible, and set up the stage for describing the most
fundamental processes  that involve both. In so doing, one preliminary question
we can try to address is  what are the best variables to use (for each specific system),
regardless of  what we have come to accept as ``fundamental'' or ``elementary''.
The very concept of space-time, as a ``real'' entity, should be put through this
rethinking process. 
If the aim of our quantum theory is to describe the motion of objects, 
the space-time geometry is just an effective picture that we can conveniently
employ in classical General Relativity, but which might be too difficult to describe
fully in the quantum theory~\cite{sakharov}. 
In fact, the first step in this construction should be to give a clear modelling 
of the detection process by which we observe something somewhere: 
which observables should we employ then, and what are the physical restrictions
we expect on them?
All we wrote above is in fact nothing new.
Any attempt at quantising canonically the Einstein-Hilbert action~\cite{dirac,Bergmann,DeWitt,rovelli} 
falls into this scheme, in which the space-time is just a mathematical arena,
and the metric becomes the basic observable, along with matter variables.
Unfortunately, a mathematical treatment of  the so called ``superspace'' of
wave-functions describing all the possible states of the metric is extremely complicated.
In fact, DeWitt himself, in his famous 1967 paper~\cite{DeWitt}, immediately reverted
to a simplified formulation in order to apply it to cosmology. 
His choice was based on preserving isotropy and homogeneity of the universe
at the quantum level, which leads to the Friedman-Robertson-Walker family of metrics,
with one degree of freedom, the scale factor. 
The corresponding space of quantum states is greatly simplified and referred to as
the FRW ``minisuperspace''.
\par
On the other hand, one of the most relevant scenarios where we expect
a quantum theory of gravitation could lead to strong predictions is the collapse
of compact objects and the possible formation of BHs. 
This physical process cannot be realistically modelled as isotropic or homogeneous
in all of its aspects, both because of the high non-linearity of the underlying 
relativistic dynamics and for the presence of many mechanisms, e.g.~generating outgoing
radiation~\cite{Vaidya,Lemaitre,Linquist,Misner}.
After the seminal papers of Oppenheimer and co-workers~\cite{OS,OV}, the literature
on the subject has grown immensely, but many issues are still open in General Relativity
(see, e.g.~Ref.~\cite{Penrose:1969pc,poisson,joshi,Bekenstein:2004eh},
and references therein).
This is not to mention the conceptual and technical difficulties one faces 
when the quantum nature of the collapsing matter is taken into account.
Assuming quantum gravitational fluctuations are small, one can describe matter
by means of Quantum Field Theory on the curved background space-time~\cite{birrell},
an approach which has produced remarkable results, like the discovery of the
Hawking evaporation~\cite{hawking,hawking2}.
However, the use of a fixed background is directly incompatible with the description
of a self-gravitating system representing a collapsing object, for which the evolution
of the background and possible emergence of non-trivial causal structures cannot
be reliably addressed perturbatively.
\par
A general property of the Einstein theory is that the gravitational interaction is always
attractive and we are thus not allowed to neglect its effect on the causal structure
of space-time if we pack enough energy in a sufficiently small volume.
This can occur, for example, if two particles (for simplicity, 
of negligible spatial extension and
total angular momentum) collide with an impact parameter $b$ shorter than the Schwarzschild
radius corresponding to the total center-of-mass energy $E$ of the system,
that is~\footnote{We shall use units with $c=k_B=1$,
and always display the Newton constant $G=\lp/\mpl$, where $\lp$ and $\mpl$
are the Planck length and mass, respectively, so that $\hbar=\lp\,\mpl$.}
\be
b \leq 2\,\lp\,\frac{E}{\mpl}
\equiv
\rh
\ .
\label{hoop}
\ee
This hoop conjecture~\cite{Thorne:1972ji} has been checked and verified
theoretically in a variety of situations, but it was initially formulated for BHs
of (at least) astrophysical sizes~\cite{payne,D'Eath:1992hd,D'Eath:1992qu}, 
for which the very concept of a classical background metric 
and related horizon structure should be reasonably safe
(for a review of some problems, see the bibliography in Ref.~\cite{Senovilla:2007dw}).
Whether the concepts involved in the above conjecture can also be trusted for
masses approaching the Planck size, however, is definitely more challenging.
In fact, for masses in that range, quantum effects may hardly be neglected
(for a recent discussion, see, e.g., Ref.~\cite{acmo})
and it is reasonable that the picture arising from General Relativistic BHs must
be replaced in order to include the possible existence of ``quantum BHs''.
Although a clear definition of such objects is still missing, 
most would probably agree that their production cross-section should (approximately)
comply with the hoop conjecture, and that they do not decay thermally
(see, e.g., Refs.~\cite{hsu,carr,calmet,calmet2,calmet3,casX}).
\par
The main complication in studying the Planck regime is that we do not have any
experimental insight thereof, which makes it very difficult to tell whether any theory
we could come up with is physically relevant.
We might instead start from our established concepts and knowledge of nature,
and push them beyond the present experimental limits.
If we set out to do so, we immediately meet with a conceptual challenge:
how can we describe a system containing both Quantum Mechanical objects
(such as the elementary particles of the Standard Model) and classically defined
horizons?
The aim of this review is precisely to show how one can introduce an operator
(observable) for the gravitational radius, and define a corresponding horizon
wave-function (HWF)~\cite{Casadio:2013tma}, which can be associated with any
localised Quantum Mechanical particle or source~\cite{Maggiore:1993kv,kempf}.
This horizon quantum mechanics (HQM) then provides a quantitative (albeit probabilistic)
condition that distinguishes a BH from a regular particle.
Since this ``transition'' occurs around the Planck scale, the HQM represents a simple
tool to investigate properties of (any models of) quantum BHs in great generality.  
We shall also review how the HQM naturally leads to an effective Generalised Uncertainty Principle
(GUP)~\cite{Casadio:2013aua,scardigli,scardigli2,Scardigli:2007bw,BNSminimallength}
for the particle position, a decay rate for microscopic BHs~\cite{Casadio:2013aua},
and a variety of other results for BHs with mass around the fundamental Planck scale~\cite{casX}
(for a review of the results obtained from the HWF for Bose-Einstein condensate
models of astrophysical size BHs, see Ref.~\cite{Thermal}).
\par
The paper is organised as follows:
in the next Section, we first recall a few relevant notions about horizons in General Relativity
and then illustrate the main ideas that define the HQM~\cite{Casadio:2013tma,Casadio:2014twa}
and how it differs from other attempts at quantising horizon degrees of freedom;
in Section~\ref{Gparticle}, we apply the general HQM to the particularly simple
cases of a particle described by a Gaussian wave-function at rest, electrically
neutral in four~\cite{Casadio:2013aua} and in $(1+D)$ dimensions
(with $D=1$ and $D> 3$)~\cite{Casadio:2015jha}, and with electric charge in four
dimensions~\cite{RN,Casadio:2015sda}; we also consider collisions of two
such particles in one spatial dimension and extend the hoop conjecture
into the quantum realm~\cite{Ctest};
in Section~\ref{Tevo}, we recall a proposal for including the time evolution
in the HQM~\cite{Casadio:2014twa} and, finally, in Section~\ref{Conclusions},
we comment on such findings and outline future applications.
\section{Horizon Quantum Mechanics}
The very first attempt at solving Einstein's field equations resulted in the discovery
of the Schwarzschild metric~\cite{schwarzschild1,schwarzschild2}
\be
\d s^2
=
-f\,\d t^2
+f^{-1}\,\d r^2
+r^2
\left(d\theta^2 + \sin^2 \theta\,d\phi^2\right)
\ ,
\label{gf}
\ee
with
\be
f=
1- \frac{2\, M}{r}
\ ,
\label{schwf}
\ee
and the appearance of the characteristic length $\Rh=2\,M$ associated to the source.
In fact, given a spherically symmetric matter source, the Schwarzschild radius $\Rh$
measures the area of the event horizon, which makes the interior of the sphere
causally disconnected from the outer portion of space-time.
At the same time, Quantum Mechanics (QM) naturally associates a Compton-de~Broglie wavelength
to a particle.
This is the minimum resolvable length scale, according to the Heisenberg uncertainty 
principle, and it can be roughly understood as the threshold below which quantum effects
cannot be neglected.
It is clear that any attempt at quantising gravity should regard those two lengths 
on somewhat equal grounds.
We therefore start with a brief review of these concepts before discussing how to
deal with them consistently in the quantum theory.
\subsection{Gravitational radius and trapping surfaces}
In order to introduce the relevant properties of a classical horizon,
we start by writing down the most general metric for a spherically symmetric 
space-time as~\cite{stephani}
\be
\d s^2
=
g_{ij}(x^k)\,\d x^i\,\d x^j
+
r^2(x^k)\left(\d\theta^2+\sin^2\theta\,\d\phi^2\right)
\ ,
\label{metric}
\ee
where $r$ is the areal coordinate and $x^i=(x^0,x^1)$ are coordinates
on surfaces where the angles $\theta$ and $\phi$ are constant.
It is clear that all the relevant physics takes place on the radial-temporal plane 
and we can safely set $x^0=t$ and $x^1=r$ from now on.
Heuristically, we can think of a (local) ``apparent horizon'' as the place
where the escape velocity equals the speed of light, and we expect its location 
be connected to the energy in its interior by simple Newtonian reasoning.
More technically, in General Relativity, an apparent horizon occurs where
the divergence of outgoing null congruences vanishes~\cite{stephani},
and the radius of this trapping surface in a spherically symmetric space-time
is thus determined by
\be
g^{ij}\,\nabla_i r\,\nabla_j r
=
0
\ ,
\label{th}
\ee
where $\nabla_i r$ is the covector perpendicular to surfaces of constant area
$\mathcal{A}=4\,\pi\,r^2$.
But then General Relativity makes it very hard to come up with a
sensible definition of the amount of energy inside a generic closed surface.
Moreover, even if several proposals of mass functions are available~\cite{faraoni15},
there is then no simple relation between these mass functions and the location of
trapping surfaces. 
Accidentally, spherical symmetry is powerful enough to overcome all of these 
difficulties, in that it allows to uniquely define the total Misner-Sharp mass as
the integral of the classical matter density $\rho=\rho(x^i)$ weighted by the flat
metric volume measure,
\be
m(t,r)
=
\frac{4\,\pi}{3}\int_0^r \rho(t, \bar r)\,\bar r^2\,\d \bar r
\ ,
\label{M}
\ee
as if the space inside the sphere were flat.
This Misner-Sharp function represents the active gravitational mass~\footnote{Roughly
speaking, it is the sum of both matter energy and its gravitational potential energy.}
inside each sphere of radius $r$ and also determines the location of trapping surfaces,
since Einstein equations imply that 
\be
g^{ij}\,\nabla_i r\,\nabla_j r
=
1-\frac{2\,M}{r}
\ ,
\ee
where $M=\lp\,m/\mpl$.
Due to the high non-linearity of gravitational dynamics, it is still very difficult 
to determine how a matter distribution evolves in time and forms surfaces obeying
Eq.~\eqref{th}, but we can claim that a classical trapping surface is found where
the gravitational radius $R=2\,M$ equals the areal radius $r$, that is 
\be
\Rh
\equiv
2\,M(t,r)
=
r
\ ,
\label{CondTheta}
\ee
which is nothing but a generalisation of the hoop conjecture~\eqref{hoop} 
to continuous energy densities.
Of course, if the system is static, the above radius will not change in time and
the rapping surface becomes a permanent proper horizon (which is the case we
shall mostly consider in the following).
\par
It stands out that the above picture lacks of any mass threshold, since the classical
theory does not yield a lower limit for the function $M$. 
Therefore, it seems that one can set the area of the trapping surface to be arbitrarily
small and eventually have BHs of vanishingly small mass.
\subsection{Compton length and BH mass threshold}
As we mentioned above, quantum mechanics provides a length cut-off through
the uncertainty in the spatial localisation of a particle. 
It is roughly given by the Compton length
\be
\lambda_m
\simeq
\lp\,\frac{\mpl}{m}
=
\frac{\lp^2}{M}
\ 
\label{lambdaM}
\ee
if, for the sake of simplicity, we consider a spin-less point-like source of mass $m$.
It is a well-established fact that quantum physics is a more fundamental description 
of the laws of nature than classical physics.
This means that $\Rh$ only makes sense when it is not ``screened'' by $\lambda_m$,
that is
\be
\Rh
\geq
\lambda_m
\ ,
\ee
and, equivalently, the BH mass must satisfy
\be
m
\geq
\mpl
\ ,
\label{clM}
\ee
or $M\geq\lp$.
We want to remark that the Compton length~\eqref{lambdaM} can also be thought
of as a quantity which rules the quantum interaction of $m$ with the local geometry.
Although it is likely that the particle's self-gravity will affect it,
we still safely assume the flat space condition~\eqref{clM} as a reasonable
order of magnitude estimate.
\par
In light of recent developments, the common argument that quantum gravity effects
should become relevant only at scales of order $\mpl$ or higher appears to be 
somewhat questionable, since the condition~\eqref{clM} implies that a classical
description of a gravitational system with $m\gg\mpl$ should be fairly accurate
(whereas for $m\sim \mpl$ the judge remains out).
This is indeed the idea of ``classicalization'' in a nutshell, as it was presented in
Refs.~\cite{dvali,dvali2} and, before that, of models with a minimum length 
and gravitationally inspired GUPs~\cite{Hossenfelder:2012jw}.
The latter are usually presented as fundamental principles for the reformulation 
of quantum mechanics in the presence of gravity, following the canonical steps 
that allow to bring a theory to the quantum level.
In this picture, gravity would then reduce to a ``kinematic effect'' encoded by the modified
commutators for the canonical variables.
In this review, we shall instead follow a different line of reasoning: 
we will start from the introduction of an auxiliary wave-function
that describes the horizon associated with a given localised particle,
and retrieve a modified uncertainty relation as a consistent result~\cite{Casadio:2013aua}.
\subsection{Horizon Wave-Function}
\label{HWF}
We are now ready to formulate the quantum mechanical description of the gravitational
radius in three spatial dimensions in a general fashion~\cite{Casadio:2013tma}.
For the reasons listed above, we shall only consider quantum mechanical
states representing {\em spherically symmetric\/} objects, which are
{\em localised in space\/}.
Since we want to put aside a possible time evolution for the moment (see Section~\ref{Tevo}),
we also choose states {\em at rest\/} in the given reference frame or,
equivalently, we suppose that every function is only taken at a fixed instant of time.
According to the standard procedure, the particle is consequently described by a
wave-function $\psi_{\rm S}\in L^2(R^3)$,
which we assume can be decomposed into energy eigenstates,
\be
\ket{\psi_{\rm S}}
=
\sum_E\,C(E)\,\ket{\psi_E}
\ .
\label{spectral}
\ee
As usual, the sum over the variable $E$ represents the decomposition
on the spectrum of the Hamiltonian,
\be
\hat H\,\ket{\psi_E}=E\,\ket{\psi_E}
\ ,
\ee
regardless of the specific form of the actual Hamiltonian operator $\hat H$.
Note though that the relevant Hamiltonian here should be the analogue of the
flat space energy that defines the Misner-Sharp mass~\eqref{M}.
Once the energy spectrum is known, we can invert the expression of the Schwarzschild radius 
in Eq.~\eqref{hoop} in order to get
\be
E
=
\mpl\,\frac{\rh}{2\,\lp}
\ .
\label{Erh}
\ee
We then define the (unnormalised) HWF as
\be
\psi_{\rm H}(\rh)
=
C\left(\mpl\,{\rh}/{2\,\lp}\right)
\ ,
\ee
whose normalisation is fixed by means of the Schr\"odinger scalar product in spherical symmetry,
\be
\pro{\psi_{\rm H}}{\phi_{\rm H}}
=
4\,\pi\,\int_0^\infty
\psi_{\rm H}^*(\rh)\,\phi_{\rm H}(\rh)\,\rh^2\,\d \rh
\ .
\label{normH}
\ee
In this conceptual framework, we could naively say that the normalised wave-function 
$\psi_{\rm H}$ yields the probability for an observer to detect a gravitational radius
of areal radius $r=\rh$ associated with the particle in the quantum state 
$\psi_{\rm S}$.
The sharply defined classical radius $\Rh$ is thus replaced by the expectation value of
the operator $\hat r_{\rm H}$.
Since the related uncertainty is in general not zero, this gravitational quantity will
necessarily be ``fuzzy'', like the position of the source itself. 
In any case, we stress that the observational meaning of the HQM will appear
only after we introduce a few derived quantities.
\par
In fact, we recall that we aimed at introducing a quantitative way of telling whether the
source is a BH or a regular particle.
Given the wave-function $\psi_{\rm H}$ associated with the quantum state
$\psi_{\rm S}$ of the source, the probability density for the source to lie inside 
its own horizon of radius $r=\rh$ will be the product of two factors, namely
\be
\mathcal{P}_<(r<\rh)
=
P_{\rm S}(r<\rh)\,\mathcal{P}_{\rm H}(\rh)
\ .
\label{PrlessH}
\ee
The first term,
\be
P_{\rm S}(r<\rh)
=
\int_0^{\rh} \mathcal{P}_{\rm S}(r) \, \d r
=
4\,\pi\,\int_0^{\rh}
|\psi_{\rm S}(r)|^2\,r^2\,\d r
\ ,
\ee
is the probability that the particle resides inside the sphere of radius $r=\rh$,
while the second term,
\be
\mathcal{P}_{\rm H}(\rh)
=
4\,\pi\,\rh^2\,|\psi_{\rm H}(\rh)|^2
\ ,
\label{Ph}
\ee
is the probability density that the value of the gravitational radius is $\rh$.
Finally, it seems natural to consider the source is a BH if it lies inside its
horizon, regardless of the size of the latter.
The probability that the particle described by the wave-function $\psi_{\rm S}$
is a BH will then be given by the integral of~\eqref{PrlessH} over all possible
values of the horizon radius $\rh$, namely
\be
P_{\rm BH}
=
\int_0^\infty \mathcal{P}_<(r<\rh)\,\d \rh
\ ,
\label{PBH}
\ee
which is the main outcome of the HQM.
\par
In the following, we shall review the application of this construction to some simple,
yet intriguing examples, in which the source is represented by Gaussian wave-functions.
We anticipate that such states show very large horizon fluctuations and are not good
candidates for describing astrophysical BHs~\cite{Casadio:2013aua} (for which extended
models instead provide a better semiclassical limit~\cite{Thermal}), but appear well-suited
for investigating BHs around the fundamental Planck scale as unstable bound
states~\cite{casX}.
\subsection{Alternative horizon quantizations}
\begin{figure}[t]
\centering
\includegraphics[width=6cm]{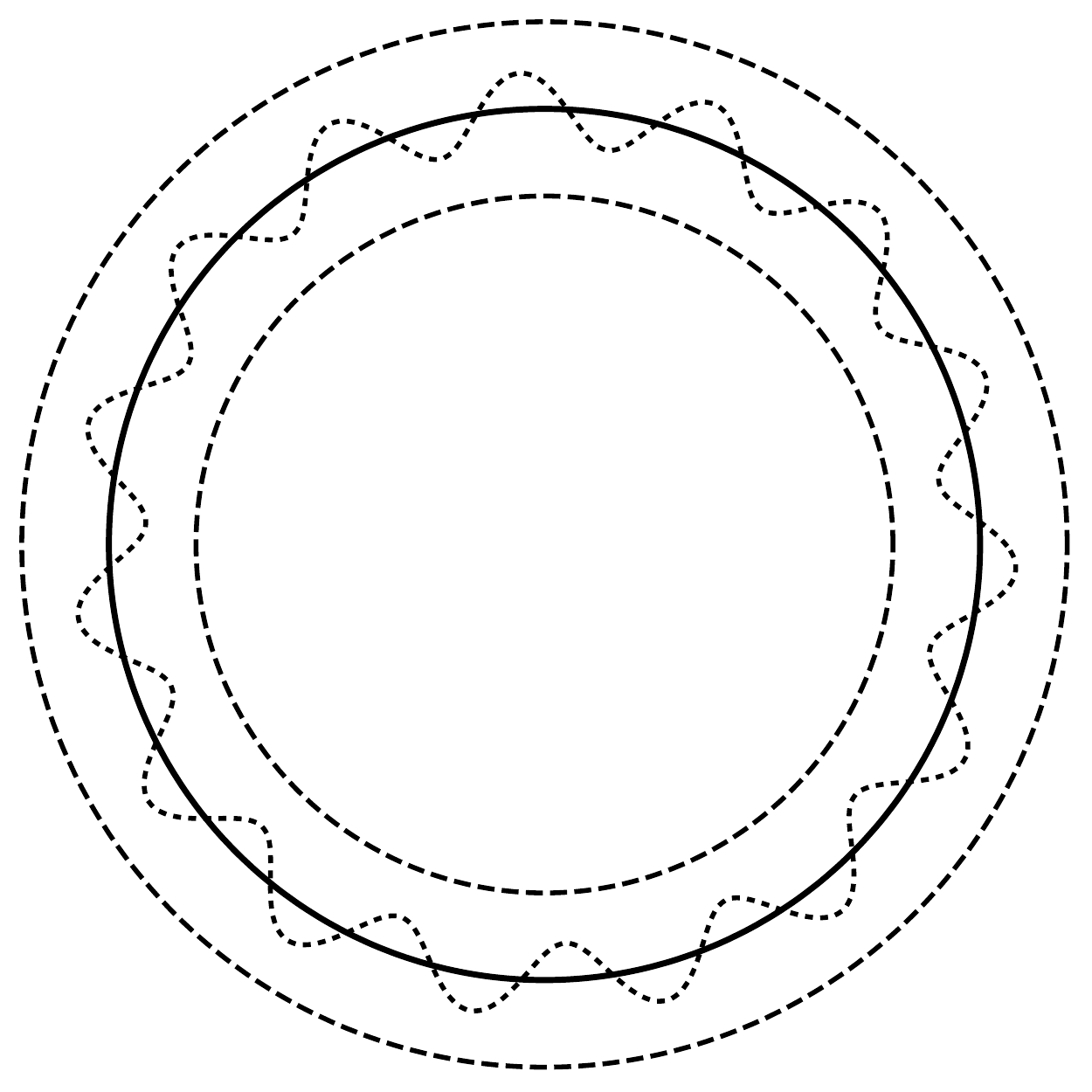}
%
\caption{Pictorial view of the HQM radial fluctuations (dashed lines) and quantum field 
theoretic fluctuations (dotted line) around the classical horizon radius (solid line).
\label{fluct}}
\end{figure}
It is important to remark the differences of the HQM with respect to other approaches
in which the gravitational degrees of freedom of (or on) the horizon are quantised
according to the background field method~\cite{DeWitt:1967ub}
(see, e.g.~Refs.~\cite{tomimatsu,Ashtekar:1997yu,carlip99,Price:1986yy,frozen,
Stephens:1993an,Susskind:1994vu}).
In general, such attempts consider linear perturbations of the metric on this
surface~\cite{frozen}, and apply the standard quantum field construction~\cite{birrell},
which is what one would do with free gravitons propagating on a fixed background.
Of course, the fact that the horizon is a null surface implies that these perturbative
modes enjoy several peculiar properties.
For instance, they can be described by a conformal field theory~\cite{carlip99},
which one can view as the origin of the idea of BHs as
holograms~\cite{Stephens:1993an,Susskind:1994vu}.
\par
In the HQM, one instead only describes those spherical fluctuations of the horizon 
(or, more, precisely, of the gravitational radius) which are determined by the quantum
state of the source.
These fluctuations therefore do not represent independent gravitational degrees of
freedom, although one could suggest that they be viewed as collective perturbations
in the zero point energy of the above-mentioned perturbative modes (see Fig.~\ref{fluct}).
In this respect, the HWF would be analogous to the quantum mechanical state of a hydrogen
atom, whereas the perturbative degrees of freedom would be the quantum field corrections
that lead to the Lamb shift.  
\par
Let us finally point out that the HQM also differs from other quantisations of the canonical
degrees of freedom associated with the Schwarzschild BH
metric~\cite{hajicek,tomimatsu92,casadio01,kuchar,Davidson:2014tda,Brustein:2013uoa},
in that the quantum state for the matter source plays a crucial role in defining
the HWF.
The HQM is therefore complementary to most of the approaches one usually encounters
in the literature.
In fact, it can be combined with perturbative approaches, like it was done in Ref.~\cite{casX},
to show that the poles in the dressed graviton propagator~\cite{calmet3} can indeed be
viewed as (unstable) quantum BHs.
%
%
%
%
\section{Spherically symmetric Gaussian sources}
\label{Gparticle}
We can make the previous formal construction more explicit by describing the massive
particle at rest in the origin of the reference frame with the spherically symmetric
Gaussian wave-function~\cite{Casadio:2013tma,Casadio:2013aua,Casadio:2014twa}
\be
\psi_{\rm S}(r)
=
\frac{e^{-\frac{r^2}{2\,\ell^2}}}{(\ell\,\sqrt{\pi})^{3/2}}
\ .
\label{Gauss}
\ee
We shall often consider the particular case when the width $\ell$ (related to the uncertainty in
the spatial size of the particle) is roughly given by the Compton length~\eqref{lambdaM}
of the particle,
\be
\ell
=
\lambda_m
\simeq
\lp\,\frac{\mpl}{m}
\ .
\label{LLc}
\ee
Even though our analysis holds for independent values of $\ell$ and $m$, one expects
that $\ell\ge \lambda_m$ and Eq.~\eqref{LLc} is therefore a limiting case of maximum localisation
for the source.
It is also useful to recall that the corresponding wave-function in momentum space is 
given by
\be
\tilde\psi_{\rm S}(p)
=
\frac{e^{-\frac{p^2}{2\,\Delta^2}}}{(\Delta\,\sqrt{\pi})^{3/2}}\,
\ ,
\label{psi_p}
\ee
with $p^2=\vec p\cdot\vec p$ being the square modulus of the spatial momentum,
and the width
\be
\Delta
=
\mpl\,\frac{\lp}{\ell}
\simeq
m
\ .
\ee
Note that the mass $m$ is not the total energy of the particle, and $m<\expec{\hat H}$
if the spectrum of $\hat H$ is positive definite.
\subsection{Neutral spherically symmetric BHs}
\label{sec:neutral}
In order to relate the momentum $p$ to the total energy $E$, the latter being the analogue
of the Misner-Sharp mass~\eqref{M}, we simply and consistently assume the relativistic
mass-shell equation in flat space-time,
\be
E^2=p^2+m^2
\label{mass-shell}
\ .
\ee
From Eq.~\eqref{Erh}, and fixing the normalisation in the inner product~\eqref{normH},
we then obtain the HWF~\cite{Casadio:2013tma,Casadio:2013aua,Casadio:2014twa}
\be
\psi_{\rm H}(\rh)
=
\frac{1}{4\,\lp^3}\sqrt{\frac{\ell^3}{\pi\,\Gamma\left(\frac{3}{2},1\right)}}
\, \Theta(\rh-\Rh) \, e^{-\frac{\ell^2\,\rh^2}{8\,\lp^4}}
\ ,
\label{psiHnnorm}
\ee
where we defined $\Rh=2\,\lp\,m/\mpl$ and the Heaviside step function
appears in the above equation because $E\ge m$.
Finally,
\be
\Gamma(s,x)
=
\int_x^\infty t^{s-1} \, e^{-t} \, \d t
\ ,
\ee
is the upper incomplete Gamma function.
In general, one has two parameters, the particle mass $m$ and the Gaussian width
$\ell$.
The HWF will therefore depend on both and so will the probability
$P_{\rm BH}=P_{\rm BH}(\ell,m)$, which can be computed only
numerically~\cite{{Casadio:2014twa}} (see also section~\ref{Tevo}).
\begin{figure}[t]
\centering
\includegraphics[width=8cm]{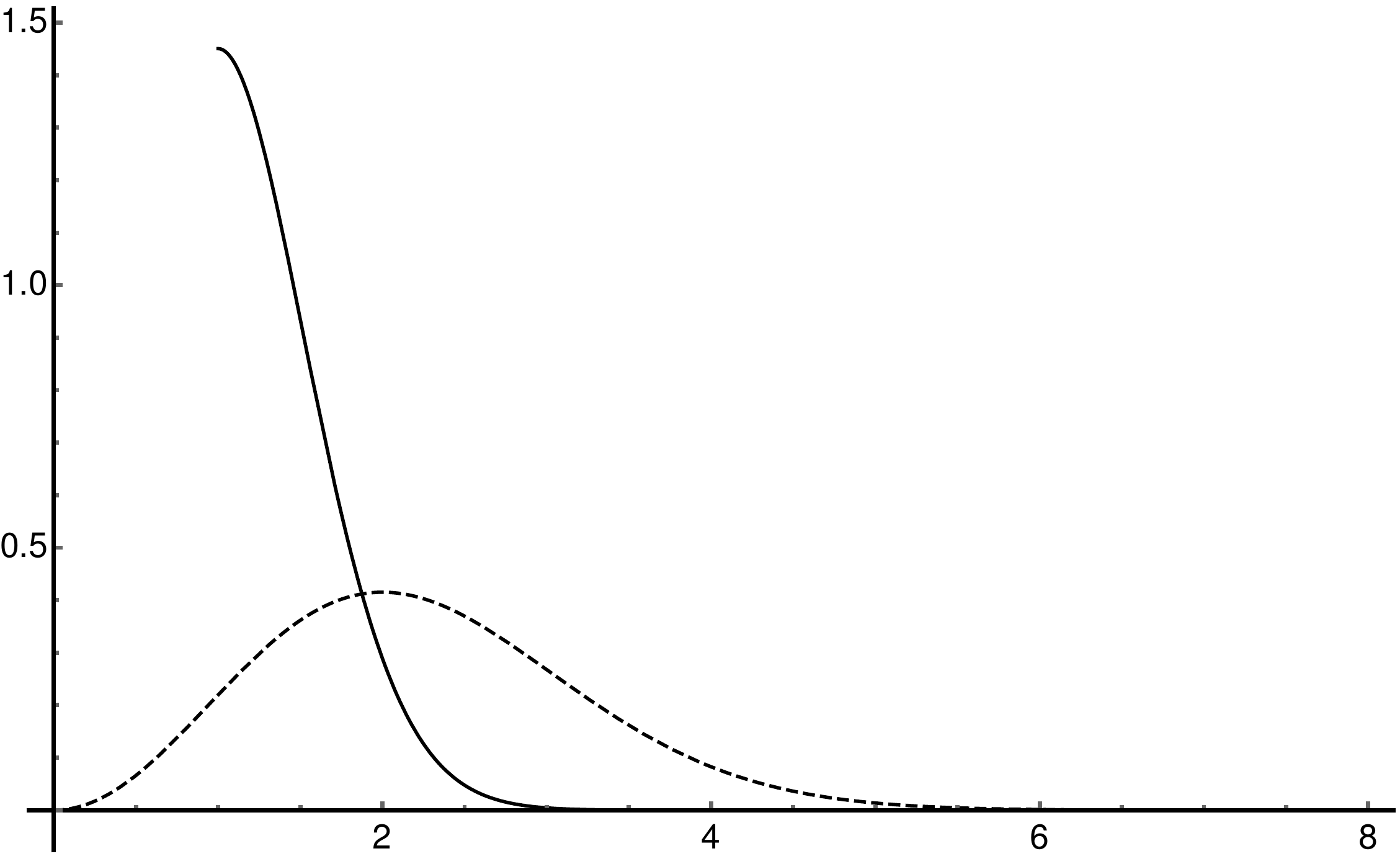}
$r/\lp$
\\
(A)
\\
\includegraphics[width=8cm]{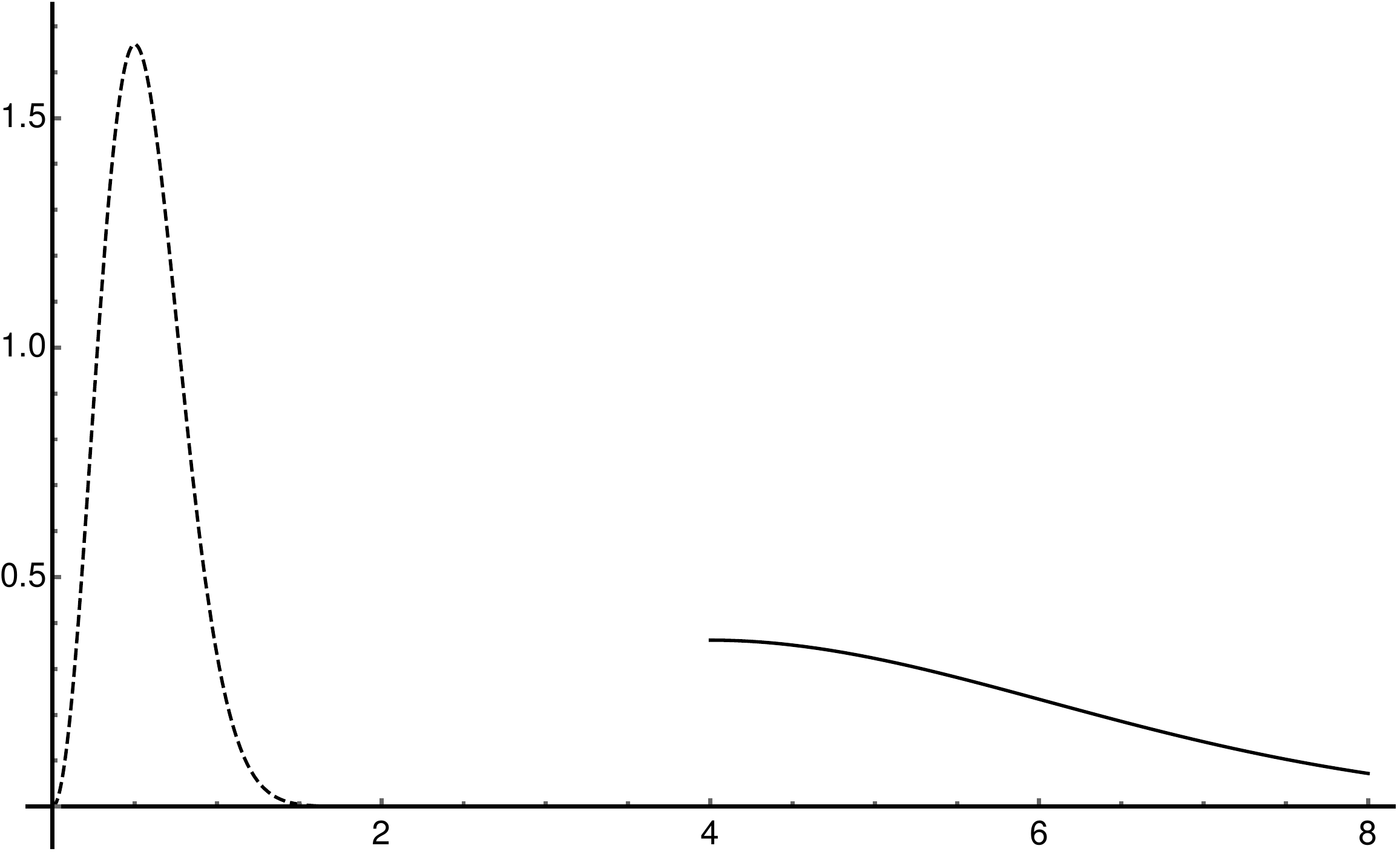}
$r/\lp$
\\
(B)
%
\caption{Probability densities $\mathcal{P}_{\rm H}$ in Eq.~\eqref{Ph} (solid line) 
and $\mathcal{P}_{\rm S}$ (dashed line) 
for $m=\mpl/2$ (upper panel) and $m=2\,\mpl$ (lower panel), assuming $m\sim \ell^{-1}$.
\label{2Psi}}
\end{figure}
\begin{figure}[t]
\centering
\raisebox{3.5cm}{$\mathcal{P}_<$}
\includegraphics[width=8cm]{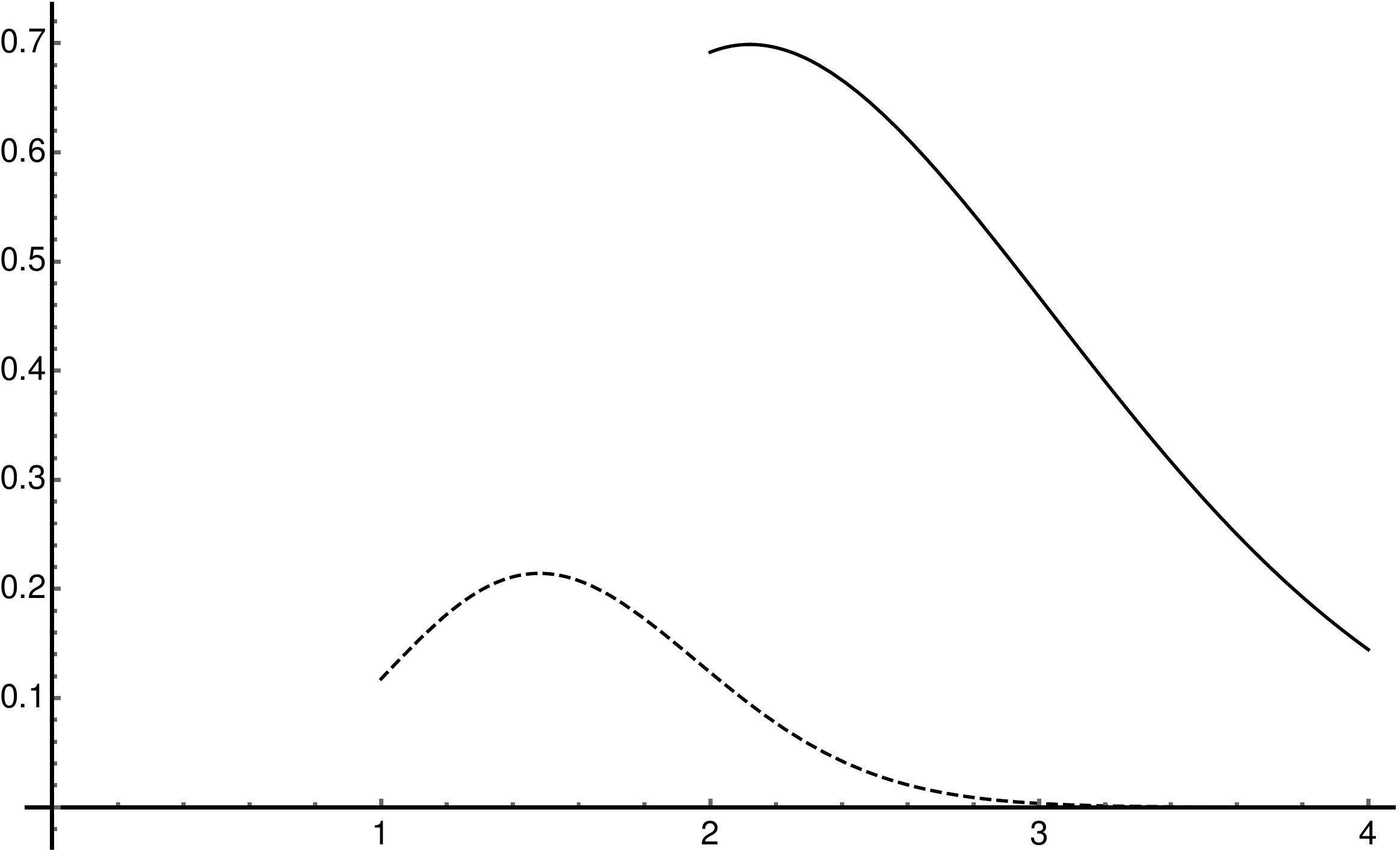}
$\rh/\lp$
%
\caption{Probability density $\mathcal{P}_<$ in Eq.~\eqref{Pin} 
that particle is inside its horizon of radius $\rh\ge \Rh=2\,\lp\,m/\mpl$,
for $\ell=\lp$ (solid line) and for $\ell=2\,\lp$ (dashed line), assuming $m\sim \ell^{-1}$.
\label{probdens<}}
\end{figure}
\begin{figure}[t]
\centering
\raisebox{3.5cm}{$P_{\rm BH}$}
\includegraphics[width=8cm]{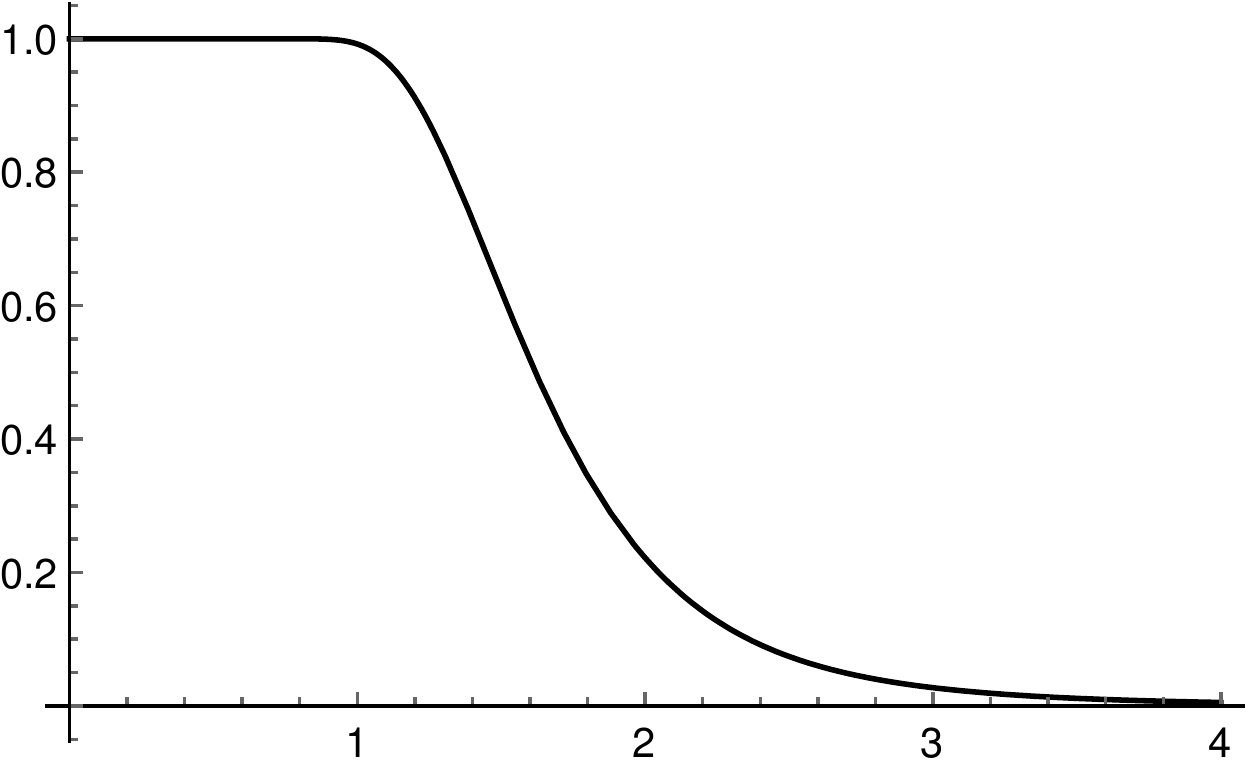}
$\ell/\lp$
%
\caption{Probability $P_{\rm BH}$ in Eq.~\eqref{Pbh} that particle of width $\ell\sim m^{-1}$ 
is a BH.
\label{probL}}
\end{figure}
\begin{figure}[t]
\centering
\raisebox{3.5cm}{$P_{\rm BH}$}
\includegraphics[width=8cm]{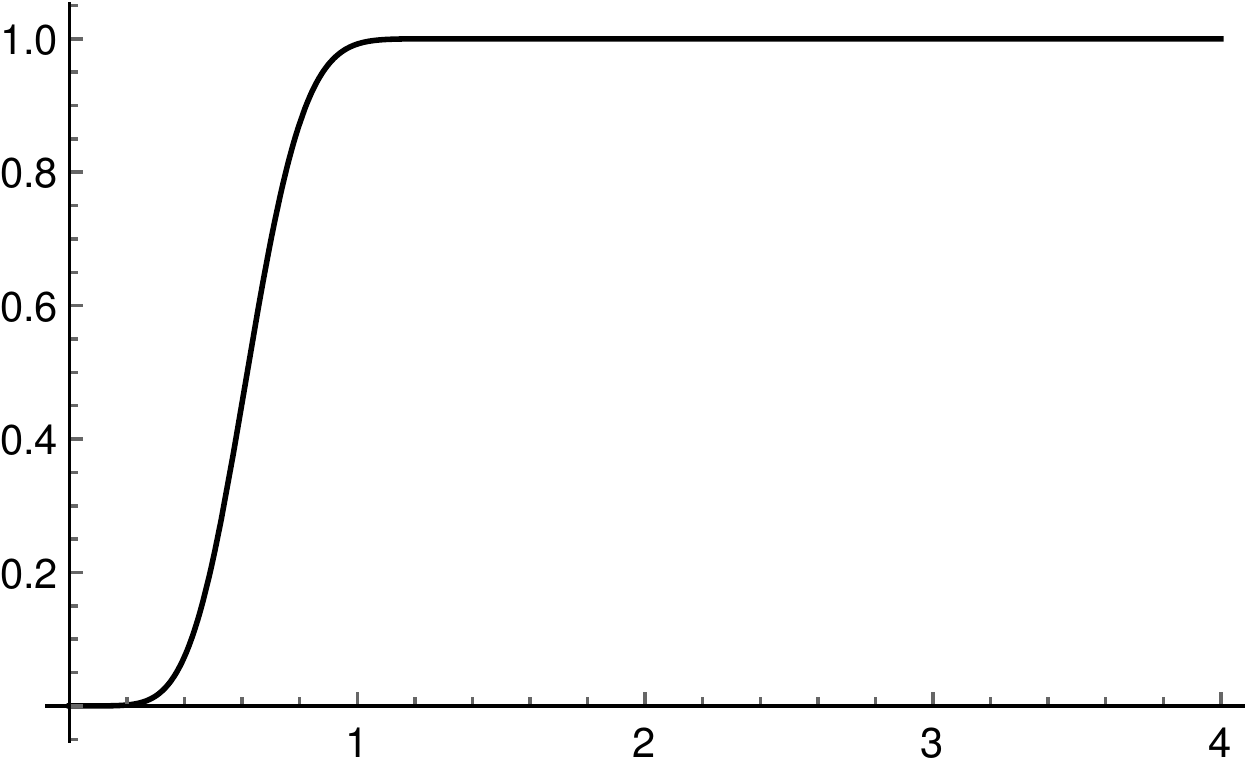}
$m/\mpl$
%
\caption{Probability $P_{\rm BH}$ in Eq.~\eqref{Pbh} that particle of mass $m\sim\ell^{-1}$ 
is a BH.
\label{probM}}
\end{figure}
\par
As we mentioned previously, it seems sensible to assume $\ell\gtrsim\lambda_m$.
In particular, the condition $\ell\sim m^{-1}$ in Eq.~\eqref{LLc} precisely leads to a BH mass
threshold of the form given in Eq.~\eqref{clM}.
We indeed expect that the particle will be inside its own horizon if 
$\expec{\hat r^2}\lesssim \expec{\hat r_{\rm H}^2}$, and Eq.~\eqref{clM} then follows
straightforwardly from $\expec{\hat r^2}\simeq \ell^2$ and
$\expec{\hat r_{\rm H}^2}\simeq \lp^4/\ell^2$.
For example, this conclusion is illustrated in Fig.~\ref{2Psi}, where the density
$\mathcal{P}_{\rm H}$ is plotted along with the probability density
$\mathcal{P}_{\rm S}=4\,\pi\,r^2\,|\psis(r)|^2$
for $m<\mpl$ and $m>\mpl$.
In the former case, the horizon is more likely found within a smaller radius than
the particle's, with the opposite situation occurring in the latter.
As a matter of fact, the probability density~\eqref{PrlessH} can be explicitly computed,
\be
\mathcal{P}_<
=
\frac{\ell^3}{2\,\sqrt{\pi}\,\lp^6}\,
\frac{\gamma\left(\frac{3}{2},\frac{\rh^2}{\ell^2}\right)}{\Gamma\left(\frac{3}{2},1\right)}\,
\Theta(\rh-\Rh) \, e^{-\frac{\ell^2\rh^2}{4\,\lp^4}} \, \rh^2
\ ,
\label{Pin}
\ee
where $\gamma(s,x)=\Gamma(s)-\Gamma(s,x)$ is the lower incomplete Gamma function.
One can integrate the density~\eqref{Pin} for $\rh$ from $\Rh$ to infinity and the
probability~\eqref{PBH} for the particle to be a BH is finally given by
\be
P_{\rm BH}(\ell)
&=&
\erf\left(\frac{2\lp^2}{\ell^2}\right)+
\frac{\sqrt{\pi}}{2}\,\frac{\erfc\left(\frac{2\lp^2}{\ell^2}\right)}
{\Gamma\left(\frac{3}{2},1\right)}
-\frac{2\lp^2/\ell^2}{\sqrt{\pi}\,\Gamma\left(\frac{3}{2},1\right)}
\,\frac{\left(3+\frac{4\lp^4}{\ell^4}\right)}
{\left(1+\frac{4\lp^4}{\ell^4}\right)^2}\, e^{-\left(1+\frac{4\lp^4}{\ell^4}\right)}
\notag
\\
&&
-\frac{2\sqrt{\pi}}{\Gamma\left(\frac{3}{2},1\right)}\,T\left(\frac{2\sqrt{2}\lp^2}{\ell^2},
\frac{\ell^2}{2\lp^2}\right)
\ ,
\label{Pbh}
\ee
where $T$ is the Owen's function~\eqref{Towen}~\footnote{More detailed calculations
of cumbersome integrals are given in~\ref{Integrals}.
In this particular case, the variable $x=\ell\, \rh/2\,\lp^2$, and we made use of
Eq.~\eqref{I3} with $A=2\,\lp^2/\ell^2$.}.
Since we are assuming that ${\ell}/{\lp}={\mpl}/{m}$, this probability can also be written
as a function of the mass $m$ as
\be
P_{\rm BH}(m)
&=&
\erf\left(\frac{2m^2}{\mpl^2}\right)+
\frac{\sqrt{\pi}}{2}\,\frac{\erfc\left(\frac{2m^2}{\mpl^2}\right)}
{\Gamma\left(\frac{3}{2},1\right)}
-\frac{2m^2/\mpl^2}{\sqrt{\pi}\,\Gamma\left(\frac{3}{2},1\right)}
\,\frac{\left(3+\frac{4m^4}{\mpl^4}\right)}
{\left(1+\frac{4m^4}{\mpl^4}\right)^2}\, e^{-\left(1+\frac{4m^4}{\mpl^4}\right)}
\notag
\\
&&
-\frac{2\sqrt{\pi}}{\Gamma\left(\frac{3}{2},1\right)}\,T\left(\frac{2\sqrt{2}m^2}{\mpl^2},
\frac{\mpl^2}{2m^2}\right)
\ .
\label{Pbh2}
\ee
In Fig.~\ref{probdens<}, we plot the probability density~\eqref{Pin}, for different values
of the Gaussian width $\ell\sim m^{-1}$.
It is already clear that such a probability decreases with $m$ (eventually vanishing below
the Planck mass).
In fact, in Fig.~\ref{probL}, we show the probability~\eqref{Pbh} that the particle is a BH
as a function of the width $\ell\sim m^{-1}$, and in Fig.~\ref{probM} the same probability
as a function of the particle mass $m\sim\ell^{-1}$.
From these plots of $P_{\rm BH}$, we can immediately infer that the particle is most likely
a BH, namely $P_{\rm BH}\simeq 1$, for $\ell\lesssim\lp$ or - equivalently - $m\gtrsim\mpl$.
We have therefore derived the condition~\eqref{clM} from a totally Quantum Mechanical
picture.
\par
\begin{figure}[t]
\centering
\raisebox{3.5cm}{$P_{\rm BH}$}
\includegraphics[width=8cm]{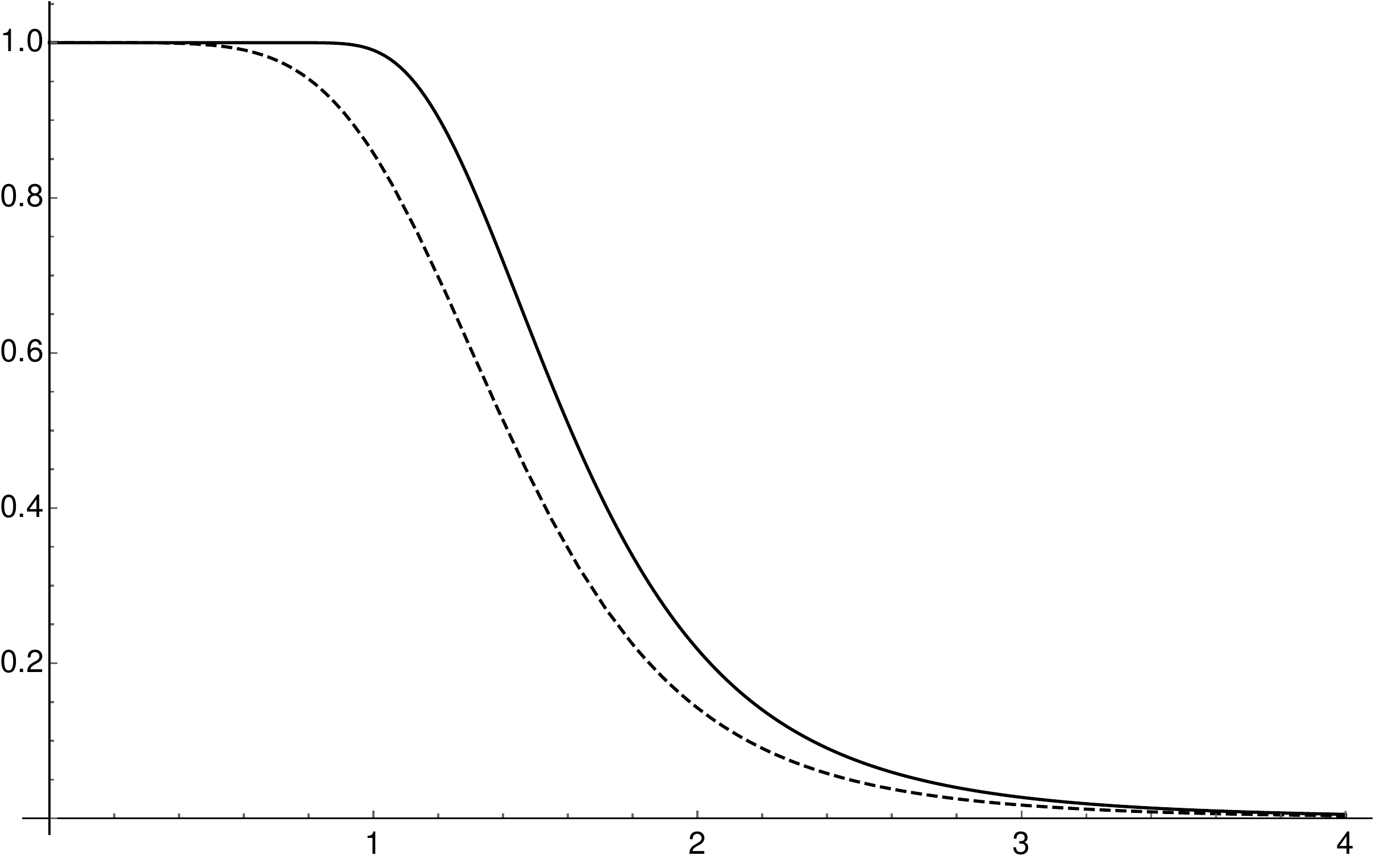}
$\ell/\lp$
%
\caption{Plot of exact $P_{\rm BH}$ in Eq.~\eqref{Pbh} (straight line) and its approximation
in Eq.~\eqref{PBHapprox} (dashed line).
\label{PBHcomp}}
\end{figure}

We conclude by recalling that a simple analytic approximation is obtained by taking the limit 
$\Rh\to0$ in Eq.~\eqref{psiHnnorm}, namely~\cite{Casadio:2013tma,Casadio:2013aua,Casadio:2014twa}
\be
\psi_{\rm H}(\rh)
=
\left(\frac{\ell}{2\sqrt{\pi} \, \lp^2}\right)^{3/2}
e^{-\frac{\ell^2\,\rh^2}{8\lp^4}}
\ ,
\ee
from which follows the approximate probability 
\be
P_{\rm BH}(\ell)
=
\frac{2}{\pi}\left[
\arctan\left(2\,\frac{\lp^2}{\ell^2}\right)
-
\frac{2\,\ell^2\,(\ell^4/\lp^4-4)}{\lp^2\,(4+\ell^4/\lp^4)^2}
\right]
\ .
\label{PBHapprox}
\ee
Fig.~\ref{PBHcomp} shows graphically that this approximation slightly underestimates
the exact probability in Eq.~\eqref{Pbh}.
\subsubsection{Effective GUP and horizon fluctuations}
\label{effCUPqf}
\begin{figure}[t]
\centering
\raisebox{3.5cm}{$\frac{\Delta R}{\lp}$}
\includegraphics[width=8cm]{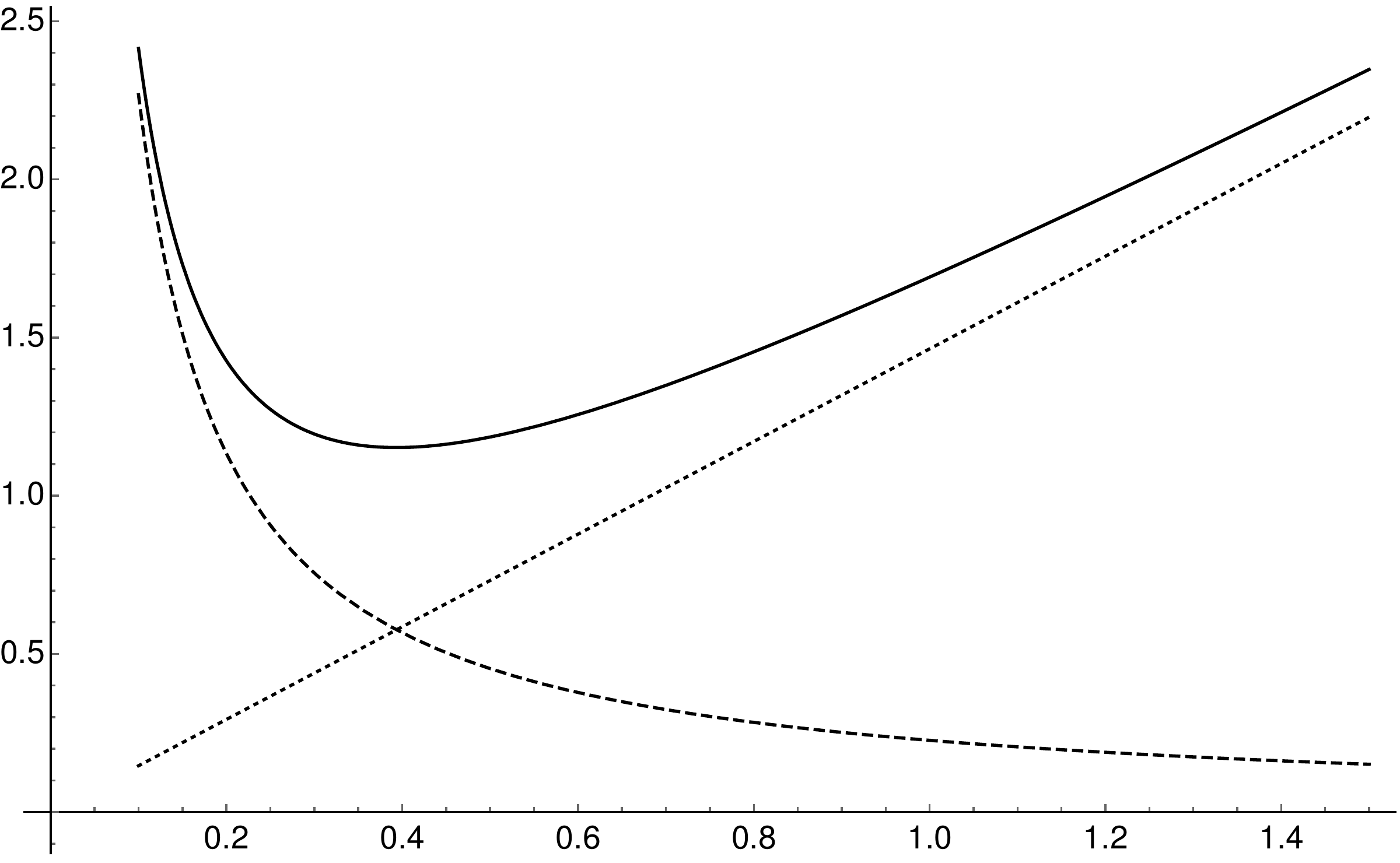}
$\frac{\Delta p}{\mpl}$
%
\caption{Uncertainty relation~\eqref{effGUP} (solid line) as a combination	
of the Quantum Mechanical uncertainty (dashed line) and the uncertainty
in horizon radius (dotted line).
\label{pGUP}}
\end{figure}

From the Gaussian wave-function~\eqref{Gauss}, we easily find that the uncertainty
in the particle's size is given by 
\be
\Delta r^2
&\equiv&
4\,\pi\,\int_0^{\infty}
|\psi_{\rm S}(r)|^2\,r^4\,\d r
-
\left(
4\,\pi\,\int_0^{\infty}
|\psi_{\rm S}(r)|^2\,r^3\,\d r
\right)^2
\nonumber
\\
&=&
\Delta_{\rm QM}\,
\ell^2
\ ,
\label{Dr}
\ee
where
\be
\Delta_{\rm QM}
=
\frac{3\,\pi-8}{2\,\pi}
\ .
\ee
Analogously, the uncertainty in the horizon radius results in
\be
\Delta \rh^2
&\equiv&
4\,\pi\,\int_0^{\infty}
|\psi_{\rm H}(\rh)|^2\,\rh^4\,\d \rh
-
\left(
4\,\pi\,\int_0^{\infty}
|\psi_{\rm H}(\rh)|^2\,\rh^3\,\d \rh
\right)^2
\nonumber
\\
&=&
4\,\lp^4\left[\frac{E_{-\frac{3}{2}}(1)}{E_{-\frac{1}{2}}(1)}
-\left(\frac{E_{-1}(1)}{E_{-\frac{1}{2}}(1)}\right)^2\right]
\frac{1}{\ell^2}
\ ,
\label{DeltaRH}
\ee
where
\be
E_{n}(x)
=
\int_1^\infty \frac{e^{-xt}}{t^n} \, \d t
\label{ExpInt}
\ ,
\ee
is the generalised exponential integral.
Since
\be
\Delta p^2
&\equiv&
4\,\pi\,\int_0^{\infty}
|\psi_{\rm S}(p)|^2\,p^4\,\d p
-
\left(
4\,\pi\,\int_0^{\infty}
|\psi_{\rm S}(p)|^2\,p^3\,\d p
\right)^2
\nonumber
\\
&=&
\Delta_{\rm QM}\,\frac{\lp^2}{\ell^2}\,\mpl^2
\ ,
\ee
we can write the width of the Gaussian as
$
\ell^2
=
\Delta_{\rm QM}\,\lp^2\,{\mpl^2}/{\Delta p^2}
$,
and, finally, assume the total radial uncertainty is a linear combination of
Eqs.~\eqref{Dr} and \eqref{DeltaRH}, thus obtaining~\cite{Casadio:2013aua} 
\be
\frac{\Delta R}{\lp}
&\equiv&
\frac{\Delta r + \xi\, \Delta \rh}{\lp}
\nonumber
\\
&=&
\Delta_{\rm QM}\,\frac{\mpl}{\Delta p}
+
\xi\,\Delta_{\rm H}\,\frac{\Delta p}{\mpl}
\ ,
\label{effGUP}
\ee
where $\xi$ is an arbitrary coefficient (presumably of order one), and
\be
\Delta_{\rm H}^2
=
\frac{4}{\Delta_{\rm QM}}
\left[\frac{E_{-\frac{3}{2}}(1)}{E_{-\frac{1}{2}}(1)}
-\left(\frac{E_{-1}(1)}{E_{-\frac{1}{2}}(1)}\right)^2\right]
\ .
\ee
This GUP is plotted in Fig.~\ref{pGUP} (for $\xi=1$), and is precisely of the kind
considered in Ref.~\cite{scardigli}, leading to a minimum measurable length
\be
\Delta R
=
2\,\sqrt{\xi\,\Delta_{\rm H}\,\Delta_{\rm QM}}\,\lp
\simeq
1.15\,\sqrt{\xi}\,\lp
\ ,
\ee
obtained for
\be
\Delta p
=
\sqrt{\frac{\Delta_{\rm QM}}{\xi\,\Delta_{\rm H}}}\,\mpl
\simeq
0.39 \, \frac{\mpl}{\sqrt{\xi}}
\ .
\ee
Of course, this is not the only possible way to define a combined uncertainty,
but nothing forces us to consider a GUP instead of making direct
use of the HWF.
\par
One of the main conclusions for the HQM of Gaussian
states can now be drawn from Eq.~\eqref{DeltaRH}, that is
\be
\Delta r_{\rm H}
\sim 
\ell^{-1}
\sim
m
\ ,
\ee
which means the size of the corresponding horizon shows fluctuations
of magnitude $\Delta r_{\rm H}\sim \rh\sim \Rh$.
This is clearly not acceptable for BHs with mass $m\gg\mpl$, which we expect
to behave (semi)classically. 
In other words, the classical picture of a BH as the vacuum geometry generated
by a (infinitely) thin matter source does not seem to survive in the quantum description,
and one is led to consider alternative models for astrophysical size
BHs~\cite{dvaligomez,DvaliGomez,Dvali:2012en,DvaliGomez4,kuhnel,Thermal,
Casadio:2013ulk,mueckPT,BEC_BH,Casadio:2015bna}.
\subsubsection{Quantum BH evaporation}
One of the milestones of contemporary theoretical physics is the discovery that BHs
radiate thermally at a characteristic temperature~\cite{hawking,hawking2}
\be
T_{\rm H}
=
\frac{\mpl^2}{8\,\pi\,m}
\ .
\ee
However, if we try to extrapolate this temperature to vanishingly small mass $M$,
we see that $T_{\rm H}$ diverges.
\par
\begin{figure}[t]
\centering
\raisebox{3.5cm}{$\frac{T}{\mpl}$}
\includegraphics[width=8cm]{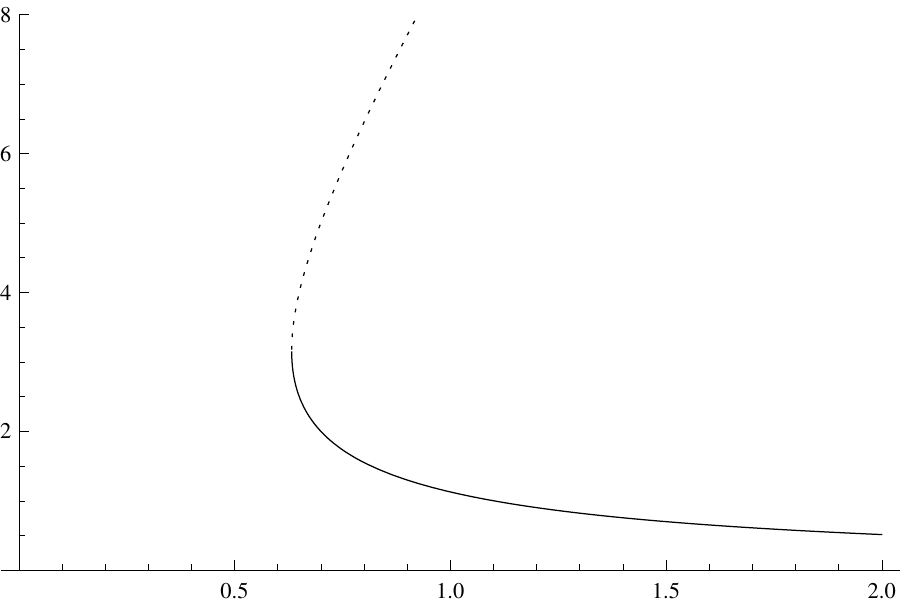}
\\
\hspace{6cm}$m/\mpl$
\caption{Temperature vs.~mass according to Eq.~\eqref{mT}
with $\beta=1/10$:
solid line reproduces the Hawking behaviour for large $m\gg\mpl$;
dotted line is the unphysical branch, and their meeting point
represents the BH with minimum mass.
\label{mTp}}
\end{figure}

One can derive improved BH temperatures for $m\simeq\mpl$ from the GUP 
(see Refs.~\cite{hayward,Scardigli:1995qd,Adler,Cavaglia,Scardigli:06,
Nouicer,Scardigli:2008jn,ChenGrub,Casadio:2013aua} for detailed computations). 
Here, we just recall that one obtains~\footnote{The parameter $\xi$ here is analogue,
but not necessarily equal, to the parameter $\xi$ in Eq.~\eqref{effGUP}.}
\be
m
=
\frac{\mpl^2}{8\,\pi\, T}
+2\,\pi\,\xi\,T
\ ,
\label{mT}
\ee
with the condition $\xi>0$, which is necessary for the existence of a minimum
BH mass (see Fig.~\ref{mTp}).
We remark that this is consistent with our previous analysis, since we stated
repeatedly that a particle with a mass significantly smaller than $\mpl$ 
should not be a BH, i.e.~$P_{\rm BH}\ll 1$ whenever $m\ll\mpl$.
It is straightforward to extremise~\eqref{mT} and get
\be
m_{\rm{min}}
=
\sqrt{\xi}\,\mpl
\ ,
\qquad
T_{\rm{max}}
=
\frac{\mpl}{4\,\pi\,\sqrt{\xi}}
\ .
\ee
Moreover, we can invert~\eqref{mT} in order to obtain $T=T(m)$ and consider
the ``physical'' branch, which reproduces the Hawking behaviour $T=0$ for $m\gg\mpl$.
When $0<\xi<1$ we can expand the result for $m$ around $\mpl$, hence
\be
\frac{T}{\mpl}
&=&
\frac{1}{4\,\pi\, \xi\,\mpl}
\left(
m-\sqrt{m^2-\xi\,\mpl^2}
\right)
\nonumber
\\
&=&
\frac{1-\sqrt{1-\xi}}{4\,\pi\,\xi}
\left(
1-
\frac{m-\mpl}{\sqrt{1-\xi}\,\mpl}
\right)
+
{\mathcal O}\left[(1-m/\mpl)^2\right]
\ .
\ee
We note that such an expansion for $T$ is well-defined even for $\xi<0$,
suggesting that the microscopic structure of the space-time may be
arranged as a lattice~\cite{WC}.
In the same approximation, we can also expand the canonical decay rate
\be
-\frac{\d m}{\d t}
&=&
\frac{8\, \pi^3\,m^2\,T^4}{15\,\mpl^5\,\lp}
\label{SB}
\\
&\simeq&
\beta \,\frac{m^2}{\mpl\,\lp}
+
{\mathcal O}(m-\mpl)
\ ,
\label{GUPdm}
\ee
where $4\cdot10^{-5} < \beta < 7\cdot10^{-4}$ when $0<\xi<1$~\cite{Casadio:2013aua}.
\par
The reader may deem unlikely that an object with a mass of the order
of $\mpl$ can be faithfully described by the same standard thermodynamics
which arises from a (semi-)classical description of BHs.
On the other hand, the HQM is specifically designed to hold in a
quantum regime. 
We can therefore guess that the decay of a Planck size BH will be related
to the probability $P_{\rm T}$ that the particle is found outside
its own horizon~\footnote{The subscript T stands
for tunnelling, which alludes to the understanding of the Hawking emission 
as a tunnelling process through the horizon~\cite{PW}.}~\cite{Casadio:2013aua}.
Of course, if the mass $m\ll\mpl$, the HWF tells us the particle is
most likely not a BH to begin with, so the above interpretation must be restricted
to $m\simeq \mpl$ (see again Fig.~\ref{probM}).
We first define the complementary probability density 
\be
\mathcal{P}_>(r>\rh)
=
P_{\rm S}(r>\rh) \, \mathcal{P}_{\rm H}(\rh)
\ ,
\label{PrmoreH}
\ee
where now
\be
P_{\rm S}(r>\rh)
=
4\,\pi\,\int_{\rh}^\infty
|\psi_{\rm S}(r)|^2\,r^2\,\d r
=
\frac{2}{\sqrt{\pi}} \, \Gamma\left(\frac{3}{2},\frac{\rh^2}{\ell^2}\right)
\ .
\ee
Upon integrating the above probability density over all values of $\rh$,
we then obtain
\be
P_{\rm T}(m)
\simeq
a-b\,\frac{m-\mpl}{\mpl}
\ ,
\ee
where $a\simeq 0.008$ and $b\simeq 0.14$ are positive constants.
We can accordingly estimate the amount of particle's energy 
outside the horizon as
\be
\Delta m
\simeq
m\,P_{\rm T}
\simeq
a\,m
+
{\mathcal O}(m-\mpl)
\ .
\ee
On the other hand, from the time-energy uncertainty relation,
$
\Delta E\,\Delta t
\simeq
\mpl\,\lp
$,
one gets the typical emission time
\be
\Delta t
\simeq
\frac{\lp^2}{\Delta \rh}
\simeq
\ell
\ ,
\ee
employing \eqref{hoop} and \eqref{DeltaRH}.
Putting the two pieces together, we find that the flux emitted by a Planck size black
hole would satisfy~\cite{Casadio:2013aua}
\be
-\frac{\Delta m}{\Delta t}
\simeq
a\,\frac{m}{\ell}
\simeq
a\,\frac{m^2}{\mpl\,\lp}
\ ,
\label{qbh}
\ee
whose functional behaviour agrees with the result~\eqref{GUPdm}
obtained from a GUP.
\par
There is a large discrepancy between the numerical coefficients in Eq.~\eqref{GUPdm}
and those in Eq.~\eqref{qbh}. 
First, we note that Eq.~\eqref{SB} holds in the canonical ensemble of statistical
mechanics, and the disparity may therefore arise because a Planck mass particle
cannot be consistently described by standard thermodynamics,
which in turn requires the BH is in quasi-equilibrium with its own radiation~\cite{PW,Angheben}.
In fact, the canonical picture does not even enforce energy conservation,
which is instead granted in the microcanonical formalism~\cite{CasHarms,Casadio:2011pd}.
However, the HQM is insensitive to thermodynamics and it is therefore remarkable
that the HQM and the GUP yield qualitatively similar results. 
In any case, the above analysis of BH evaporation is very preliminary and significant
changes are to be expected when considering a better description of the microscopic structure of
quantum BHs~\cite{casX,Dvali:2012en,Casadio:2013ulk,BEC_BH,mueckPT}.
\subsection{Electrically charged sources}
An extension of the original HQM regards the case of electrically charged massive
sources~\cite{Greenwood:2008ht,Saini:2014qpa,Vachaspati:2006ki,Vachaspati:2007hr,Wang:2009ay},
and was obtained in Refs.~\cite{RN,Casadio:2015sda} from the Reissner-Nordstr\"om (RN)
metric~\cite{reissner,nordstrom}.
The latter is of the form~\eqref{gf} with
\be
f=
1- \frac{2\, \lp\, m}{\mpl \,r}+\frac{Q^2}{r^2}
\ ,
\label{RNf}
\ee
where $m$ is again the ADM mass and $Q$ is the charge of the source.
In the following, it will be convenient to employ the specific charge
\be
\alpha
=
\frac{|Q|\, \mpl}{\lp\, m}
\ . 
\label{alpha}
\ee
The case $\alpha=0$ reduces to the neutral Schwarzschild metric.
For $0<\alpha<1$, the above function $f$ has two zeroes, namely 
\be
R_{\pm}
&=&
\lp\, \frac{m}{\mpl}\pm\sqrt{\left(\lp\, \frac{m}{\mpl}\right)^2 - Q^2}
\nonumber
\\
&=&
\lp\, \frac{m}{\mpl}\left(1\pm\sqrt{1- \alpha^2}\right)
\ ,
\label{R+-}
\ee
and the RN metric therefore describes a BH.
Moreover, the two horizons coincide for $\alpha=1$ and the BH is 
said to be {\em extremal\/}, while the singularity is naked, i.e.~accessible
to an external observer, for $\alpha>1$.
\subsubsection{Inner Horizon}
The case $0<\alpha \le 1$ was considered in Ref.~\cite{RN}, where the HQM
was extended for the presence of more than one trapping surface.  
A procedure similar to the neutral case was followed for each of the two
horizon radii~\eqref{R+-}:
one initially determines the HWFs and then uses them to compute the probability
for each horizon to exist.
Eqs.~\eqref{R+-} is lifted to the quantum level by introducing the operators
$\hat r_\pm$ and $\hat H$, which replace their classical counterparts $R_\pm$ and $m$.
Moreover, these operators are chosen to act multiplicatively on the
respective wave-functions, whereas the specific charge $\alpha$ remains
a simple parameter (c-number)~\footnote{As usual, going from the classical
to the quantum realm is affected by ambiguities, and this choice is not unique.}.
\par
First we note the total energy $\hat H$ can be expressed in terms of the horizon
radii as
\be
\lp\,\frac{\hat H}{\mpl}
=
\frac{\hat r_{+}+\hat r_{-}}{2}
\ ,
\label{EofRpm}
\ee
and one also has
\be
\hat r_{\pm}
=
\hat r_{\mp}\,
\frac{1\pm\sqrt{1-\alpha^2}}{1\mp\sqrt{1-\alpha^2}}
\ .
\label{R-R+}
\ee
We then obtain the HWFs for $r_{+}$ and $r_{-}$ by expressing $p$ from the mass-shell
relation~\eqref{mass-shell} in terms of the eigenvalue $E$ of $\hat H$ in Eq.~\eqref{EofRpm},
and then replacing one of the relations~\eqref{R-R+} into the wave-function representing
the source in momentum space, as in Eq.~\eqref{psi_p}.
For the usual limiting case~\eqref{LLc}, $\ell\sim m^{-1}$, it is straightforward to obtain 
\be
\psi_{\rm H}(r_{\pm})
&=&
\sqrt{\frac{1}{2\pi\,\Gamma\left(\frac{3}{2},1\right)}\,
\left[\frac{\ell}{\lp^2(1\pm\sqrt{1-\alpha^2})}\right]^3}\,
\Theta\left(r_{\pm}-R_\pm \right)
\nonumber
\\
&&
\times\, 
\exp\left\{-\frac{\ell^2\,r_{\pm}^2}{2\lp^4\,(1\pm\sqrt{1-\alpha^2})^2}\right\}
\ ,
\label{psih}
\ee
where the minimum radii are given by
\be
R_\pm
=
\lp\,\frac{m}{\mpl}\left(1\pm\sqrt{1 - \alpha^2}\right)
=
\frac{\lp^2}{\ell}\left(1\pm\sqrt{1 - \alpha^2}\right)
\ .
\label{Rminalpha}
\ee
The probability densities for the source to be found
inside each of the two horizons turn out to be
\be
\mathcal{P}_{<\pm}
&=&
\frac{4}{\sqrt{\pi}\,\Gamma\left(\frac{3}{2},1\right)} \,
\left[\frac{\ell}{\lp^2(1\pm\sqrt{1-\alpha^2})}\right]^3 \,
\Theta(r_\pm -R_\pm) \,
\notag
\\
&&
\times\, \gamma\left(\frac{3}{2},\frac{r^2_\pm}{\ell^2}\right)
\, \exp\left\{-\frac{\ell^2\, r^2_\pm}{\lp^4\,(1\pm\sqrt{1-\alpha^2})^2}\right\} \, r^2_\pm
\ .
\label{PinRN}
\ee
In the neutral case $\alpha=0$, $\mathcal{P}_{<-}$ is of course ill-defined,
while $\mathcal{P}_{<+}$ equals the probability density~\eqref{Pin},
which means that $r_+$ becomes the Schwarzschild radius $\rh$.
\par
\begin{figure}[t]
\centering
\raisebox{4.5cm}{${\mathcal P}_{<+}$}
\includegraphics[width=8cm]{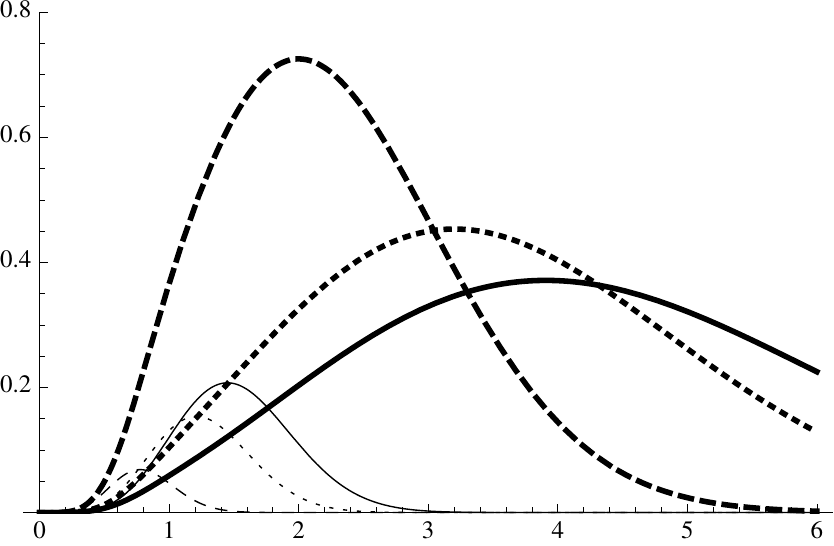}
$\frac{r_+}{\lp}$
%
\caption{Probability density ${\mathcal P}_{<+}$ in Eq.~\eqref{PinRN}
that the particle is inside its outer horizon $r=r_{+}$,
for $\ell=\lp/2$ (thick lines) and $\ell=2\,\lp$ (thin lines)
with $\alpha=0.3$ (continuous lines), $\alpha=0.8$ (dotted lines) and $\alpha=1$
(dashed lines).
For $\alpha=1$, the two horizons coincide and ${\mathcal P}_{<-}={\mathcal P}_{<+}$.
\label{prob<+}}
\end{figure}
\begin{figure}[t]
\centering
\raisebox{4.5cm}{${\mathcal P}_{<-}$}
\includegraphics[width=8cm]{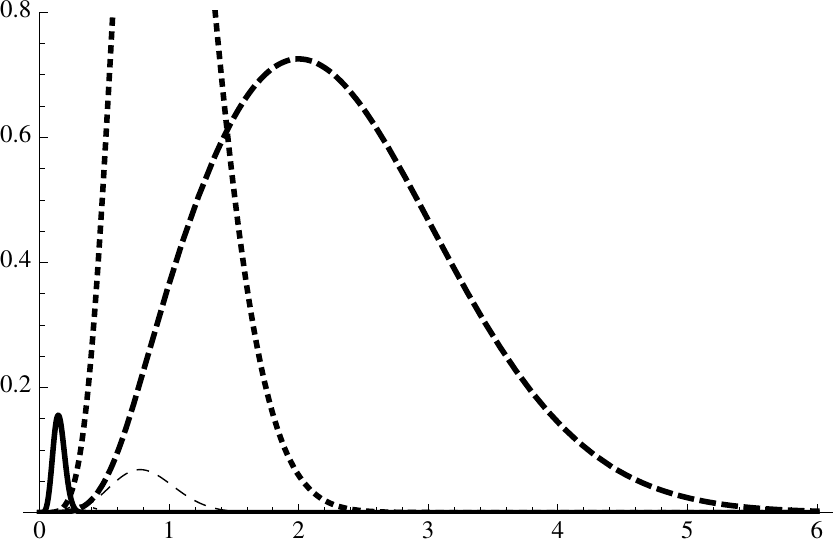}
$\frac{r_-}{\lp}$
%
\caption{Probability density ${\mathcal P}_{<-}$ in Eq.~\eqref{PinRN}
that particle is inside its inner horizon $r=r_{-}$,
for $\ell=\lp/2$ (thick lines) and $\ell=2\,\lp$ (thin lines)
with $\alpha=0.3$ (continuous lines), $\alpha=0.8$ (dotted lines) and $\alpha=1$
(dashed lines). 
For $\alpha=1$, the two horizons coincide and ${\mathcal P}_{<-}={\mathcal P}_{<+}$.
\label{prob<-}}
\end{figure}
Fig.~\ref{prob<+} shows the probability density ${\mathcal P}_{<+}$ for the massive
source to reside inside the external horizon $r=r_+$ for two values
of the width $\ell$ (above and below the Planck scale) and three values of
the specific charge $\alpha$.
The maximum of this function clearly decreases when $\ell$ increases
above $\lp$ or, equivalently, when $m$ gets smaller than the Planck mass.
Fig.~\ref{prob<-} shows the analogous  probability densities ${\mathcal P}_{<-}$
for the inner horizon $r=r_-$.
Obviously, the smaller $\alpha$ the smaller is the probability that a trapping
surface occurs.
Moreover, as we expected from the start, the density profiles coincide in the extremal
case $\alpha=1$ (thick and thin dashed lines), because the two horizons merge.
\begin{figure}[t]
\centering
\raisebox{4.5cm}{$P_{\rm BH}$}
\includegraphics[scale=0.92]{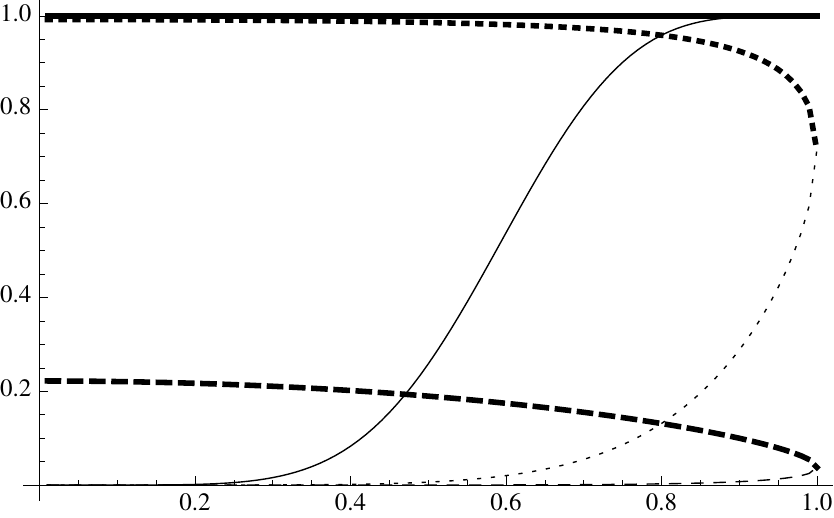}
\\
\hspace{7cm}$\alpha$
\caption{Probability $P_{\rm BH+}$ in Eq.~\eqref{PBHRN} for the particle to be a BH (thick lines)
and $P_{\rm BH-}$ in Eq.~\eqref{PBHRN} for the particle to be inside its inner horizon (thin lines)
as functions of $\alpha$ for $\ell=\lp/2$ (continuous line), $\ell=\lp$ (dotted line)
and $\ell=2\,\lp$ (dashed line).
For $\alpha=1$ the two probabilities merge. 
\label{PBHplot1}}
\end{figure}
\par
Integrating over $r_\pm$, we obtain the probabilities~\footnote{ 
It is convenient to define 
$
x_\pm
=
{\ell\, r_\pm}/{\lp^2(1\pm\sqrt{1-\alpha^2})}
$,
and use again Eq.~\eqref{I3}, with $A=(1\pm\sqrt{1-\alpha^2})\,\lp^2/\ell^2$.}
\be
P_{\rm BH\pm}(\ell,\alpha)
&=&
\erf\left[\frac{\lp^2}{\ell^2}\left(1\pm\sqrt{1-\alpha^2}\right)\right]
+\frac{\sqrt{\pi}}{2}\,\frac{\erfc\left[\frac{\lp^2}
{\ell^2}\left(1\pm\sqrt{1-\alpha^2}\right)\right]}{\Gamma\left(\frac{3}{2},1\right)}
\notag
\\
&&
-\frac{\left(1\pm\sqrt{1-\alpha^2}\right)\lp^2/\ell^2}
{\sqrt{\pi}\,\Gamma\left(\frac{3}{2},1\right)}\,
\frac{3+\frac{\lp^4}{\ell^4}\left(1\pm\sqrt{1-\alpha^2}\right)^2}
{\left[1+\frac{\lp^4}{\ell^4}(1\pm\sqrt{1-\alpha^2})^2\right]^2}
\,e^{-\left[1+\left(1\pm\sqrt{1-\alpha^2}\right)^2\,\frac{\lp^4}{\ell^4}\right]}
\notag
\\
&&
-\frac{2\sqrt{\pi}}{\Gamma\left(\frac{3}{2},1\right)} \, 
T\left[\sqrt{2}\frac{\lp^2}{\ell^2}\left(1\pm\sqrt{1-\alpha^2}\right),
\frac{\ell^2}{\lp^2\left(1\pm\sqrt{1-\alpha^2}\right)}\right]
\ .
\label{PBHRN}
\ee
where $T$ is again the Owen's function~\eqref{Towen}.
\par
Fig.~\ref{PBHplot1} shows how these probabilities vary with the parameter $\alpha$
for values of $\ell$ above or below the Planck scale.
For the outer horizon, it is clear that $P_{\rm BH+}\simeq 1$ for widths $\ell\lesssim \lp$
(mass larger than $\mpl$). 
On the contrary, when $\ell\gtrsim\lp$ (or $m\lesssim\mpl$), the probability sensibly
decreases as the specific charge $\alpha$ approaches $1$ from below.
We see that this probability is does not exactly vanish even when $\ell$ exceeds
the Planck length $\lp$.
As an example, for $\ell=2\,\lp$, corresponding to $m=\mpl/2$, we find
$0.15 \lesssim P_{\rm BH+}(\alpha)\lesssim 0.2$ for a large interval of values
of the specific charge.
$P_{\rm BH+}$ only falls below $0.1$ right before the BH becomes 
maximally charged ($\alpha\simeq 1$).
As far as the inner horizon is concerned, the scenario is profoundly different.
The same plot shows that the probability $P_{\rm BH-}\ll 1$ for small
values of $\alpha$ and increases with this parameter.
However, the role of $\ell$ is prominent because the sharper the Gaussian
packet is localised in space (or the more massive it is), the smaller the value of
$\alpha$ for which this probability becomes significant.
To summarise, there is an appreciable range of values of the specific charge
$\alpha$ for which the inner horizon is not likely to exist ($P_{\rm BH-}\ll 1$),
while the system is a BH ($P_{\rm BH+}\simeq 1$). 
\par
\begin{figure}[t]
\centering
\raisebox{4.5cm}{$P_{\rm BH}$}
\includegraphics[scale=0.72]{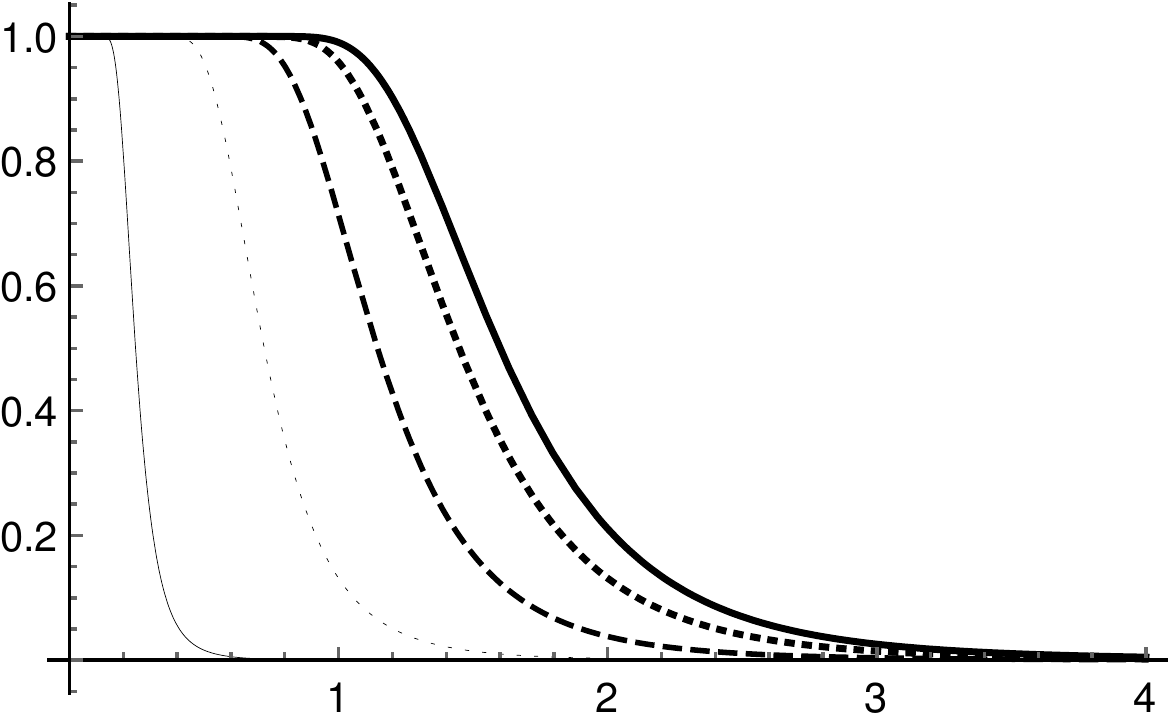}
$\frac{\ell}{\lp}$
\caption{Probability $P_{\rm BH+}$ for the particle to be a BH (thick lines)
and $P_{\rm BH-}$ for the particle to be inside its inner horizon (thin lines), 
in Eq.~\eqref{PBHRN}, as functions of $\ell$, 
for $\alpha=0.3$ (continuous line), $\alpha=0.8$ (dotted line) and
$\alpha=1$ (dashed line).
For $\alpha=1$ thick and thin dashed lines overlap. 
\label{PBHplot2L}}
\end{figure}
\begin{figure}[t]
\centering
\raisebox{4.5cm}{$P_{\rm BH}$}
\includegraphics[scale=0.68]{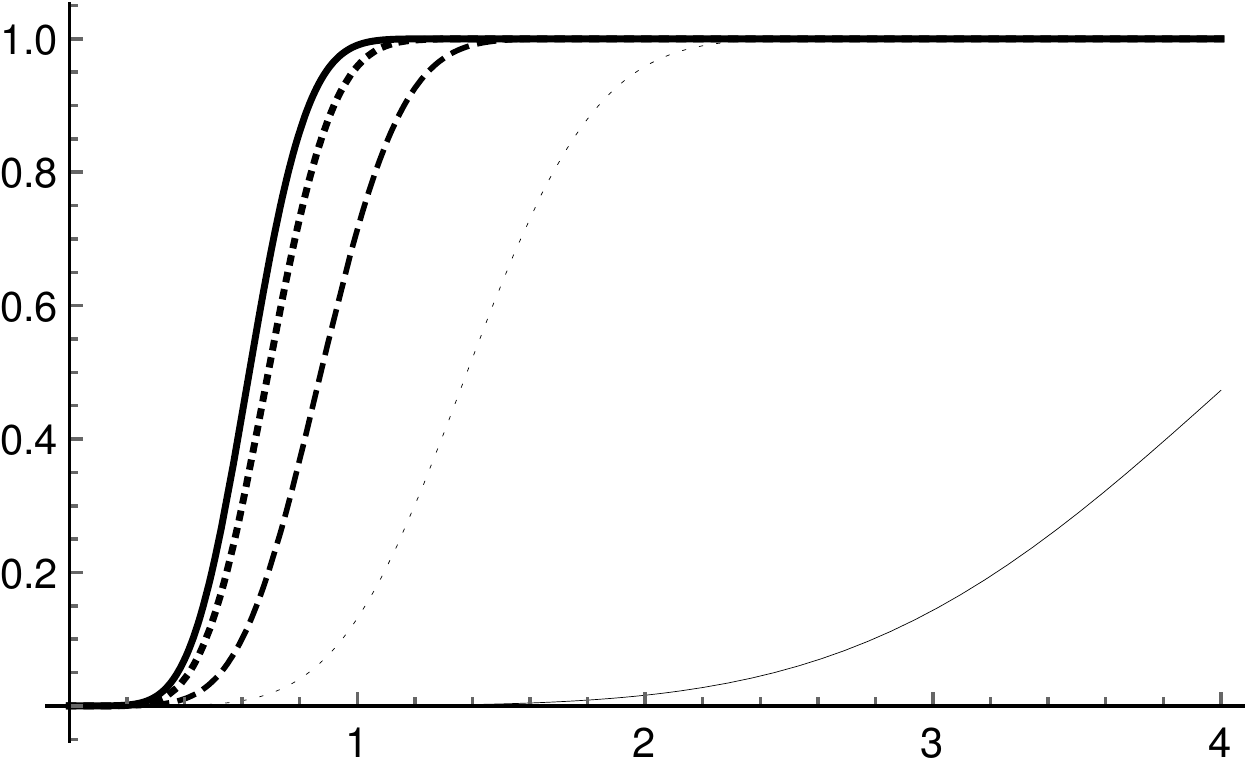}
$\frac{m}{\mpl}$
%
\caption{Probability $P_{\rm BH+}$ for the particle to be a BH (thick lines)
and $P_{\rm BH-}$ for the particle to be inside its inner horizon (thin lines), 
in Eq.~\eqref{PBHRN}, as functions of $m$, 
for $\alpha=0.3$ (continuous line), $\alpha=0.8$ (dotted line) and
$\alpha=1$ (dashed line).
For $\alpha=1$ thick and thin dashed lines overlap. 
\label{PBHplot2M}}
\end{figure}
The probabilities $P_{\rm BH\pm}$ as functions of the width $\ell$ are shown in 
Fig.~\ref{PBHplot2L} and as functions of the mass $m$ in Fig.~\ref{PBHplot2M},
for $\alpha=0.3$, $0.8$ and $1$.
It is evident that smaller values of $\alpha$ allow for $P_{\rm BH+}$ to
approach $1$ for smaller masses $m$.
The specular situation happens when studying the inner probability $P_{\rm BH-}$.
If we focus on the smallest specific charge considered here, $\alpha=0.3$,
we notice that both probabilities are close to $1$ only around $m\simeq 6\,\mpl$,  
and not at the naively expected scale $m\simeq\mpl$.
Hence, there exists a non-negligible interval in the possible values of $m$
(around the Planck scale) for $\alpha< 1$ in which
\be
P_{\rm BH+}
\simeq 
1
\quad
{\rm and}
\quad
P_{\rm BH-}
\ll 
1
\ .
\label{P10}
\ee
In this interval, the system is most likely a BH, because it is the outer horizon which
dictates this property, while the inner horizon is still not very likely to exist. 
Lowering the value of $\alpha$ this range grows larger, while it narrows and
eventually vanishes when approaching the maximally charged limit $\alpha=1$.
\par
We conclude by remarking that we could have guessed this result.
In fact, the smaller $\alpha$, the more the system looks like a neutral (Schwarzschild)
BH, since the mass becomes the dominant parameter and the presence of
charge is (at most) a small perturbation.
However, the existence of an inner horizon at $r=R_{-}$ is phenomenologically
very important, because of the possible instability known as
{\em mass inflation\/}~\cite{Dokuchaev:2013uda,Brown:2010cr,Brown:2011tv}
related to the specific features of such a Cauchy horizon.
Eq.~\eqref{P10} suggests that this instability should not always occur for $0<\alpha\le 1$,
even when the particle is (most likely) a BH.
\subsubsection{Quantum Cosmic Censorship}
Overcharged sources with $\alpha>1$ were analysed in Ref.~\cite{Casadio:2015sda}.
We recall that the cosmic censorship~\cite{Penrose:1969pc}
was conjectured in order to exclude such naked singularities from General Relativity.
It is therefore interesting to investigate whether quantum physics supports
this view or can introduce modifications of any kind.
The analysis is developed by assuming that the overcharged regime $\alpha>1$
is reached by continuing analytically the HWF from the case $0<\alpha\le 1$.
It is clear that this choice is again not unique, but it should be consistent at least when
the specific charge is not much greater than the classical limiting threshold $\alpha=1$.
\par
The first issue that needs to be taken into consideration for $\alpha>1$ is that the operators
$\hat r_\pm$ directly obtained from Eq.~\eqref{R-R+} are not Hermitian.
This could in principle be a reason to give up any observables corresponding to
$\hat r_\pm$ in this classically forbidden region.
Nonetheless, one can follow through and construct a Hermitian radial operator
using only the real parts of the multiplicative operators $\hat r_\pm$.
By continuing analytically Eq.~\eqref{psih} for $\alpha>1$, the square modulus
of the HWF becomes~\cite{Casadio:2015sda}
\be
|\psi_{\rm H}(\rh)|^2
=
\mathcal{N}^2\,
\exp\left\{ - \frac{2-\alpha^2}{\alpha^4}\,\frac{\ell^2\, \rh^2}{\lp^4}\right\}
\ ,
\label{psih1}
\ee
where $\rh$ now replaces both $r_+$ and $r_-$ (which in fact merge at $\alpha=1$)
and $\mathcal{N}$ is a normalisation factor.
It so happens that this HWF is still normalisable in the Schr\"odinger
scalar product~\eqref{normH} if $\rh$ is a real variable and for specific charge
values in the range
\be
1 < \alpha^2 < 2
\ .
\label{a>1}
\ee
This suggests that there must be a quantum obstruction
forbidding the system from crossing $\alpha^2=2$.
We will discuss this issue more completely after determining the full HWF.
\par
One also needs to modify the step function in Eq.~\eqref{psih} when the system
enters the overcharged regime.
First, we note that the real part of the complex Eq.~\eqref{Rminalpha} 
is the same for $r_+$ and $r_-$, and set
\be
\Rh
=
{\rm Re}\left[\frac{\lp^2}{\ell}\left(1\pm\sqrt{1- \alpha^2}\right)\right]
=
\frac{\lp^2}{\ell}
\ .
\label{Rmin>1}
\ee
We can then show that the continuity property which leads to Eq.~\eqref{psih1}
extends to $\hat r_{\rm H}$ when $\rh$ is bounded from below by $\Rh$.
In fact, 
we can compute the expectation value 
\be
\expec{\hat r_{\rm H}}
=
4\,\pi\,\int_{R_{\rm H}}^{\infty}
|\psi_{\rm H}(\rh)|^2\,\rh^3\,\d \rh\
=
\frac{\alpha^2}{\sqrt{2-\alpha^2}}\,
\frac{\Gamma\left(2,\frac{2-\alpha^2}{\alpha^4}\right)}
{\Gamma\left(\frac{3}{2},\frac{2-\alpha^2}{\alpha^4}\right)}\,
\Rh
\ ,
\label{expecRH}
\ee
and observe that this expression matches the analogous expressions
from the regime $0<\alpha\le 1$,
\be
\expec{\hat r_{\pm}}
=
4\,\pi\,\int_{R_\pm}^{\infty}
|\psi_{\rm \pm}(r_{\pm})|^2\,r_{\pm}^3\,\d r_{\pm}
=
\frac{\Gamma\left(2,1\right)}{\Gamma\left(\frac{3}{2},1\right)}\, R_\pm
\ ,
\ee
in the limit $\alpha = 1$, namely
\be
\lim_{\alpha\searrow 1} \expec{ \hat r_{\rm H}}
=
\frac{\Gamma\left(2,1\right)}{\Gamma\left(\frac{3}{2},1\right)}\,\frac{\lp^2}{\ell}
=
\lim_{\alpha\nearrow 1} \expec{ \hat r_{\pm}}
\ .
\label{alphato1}
\ee
Moreover, the same holds for the corresponding uncertainties, that is 
\be
\Delta \rh^2(\ell,\alpha\to 1^+)
=
\Delta r_\pm^2(\ell,\alpha\to 1^-)
\ .
\label{DRHpm}
\ee
We omit the details here~\cite{Casadio:2015sda}, and just remark that,
for $\alpha=1$, the width of the Gaussian $\ell> \expec{\hat r_{\rm H}}$ for 
$
m
<
\sqrt{\Gamma\left(\frac{3}{2},1\right)/\Gamma(2,1)}\, \mpl
\simeq
0.8\,\mpl
$.
The gravitational fluctuations in the size of the source will thus be
subdominant when its mass is sensibly smaller than the
Planck mass $\mpl$, like in the neutral case.
\par
\begin{figure}[t]
\centering
\includegraphics[width=7.5cm]{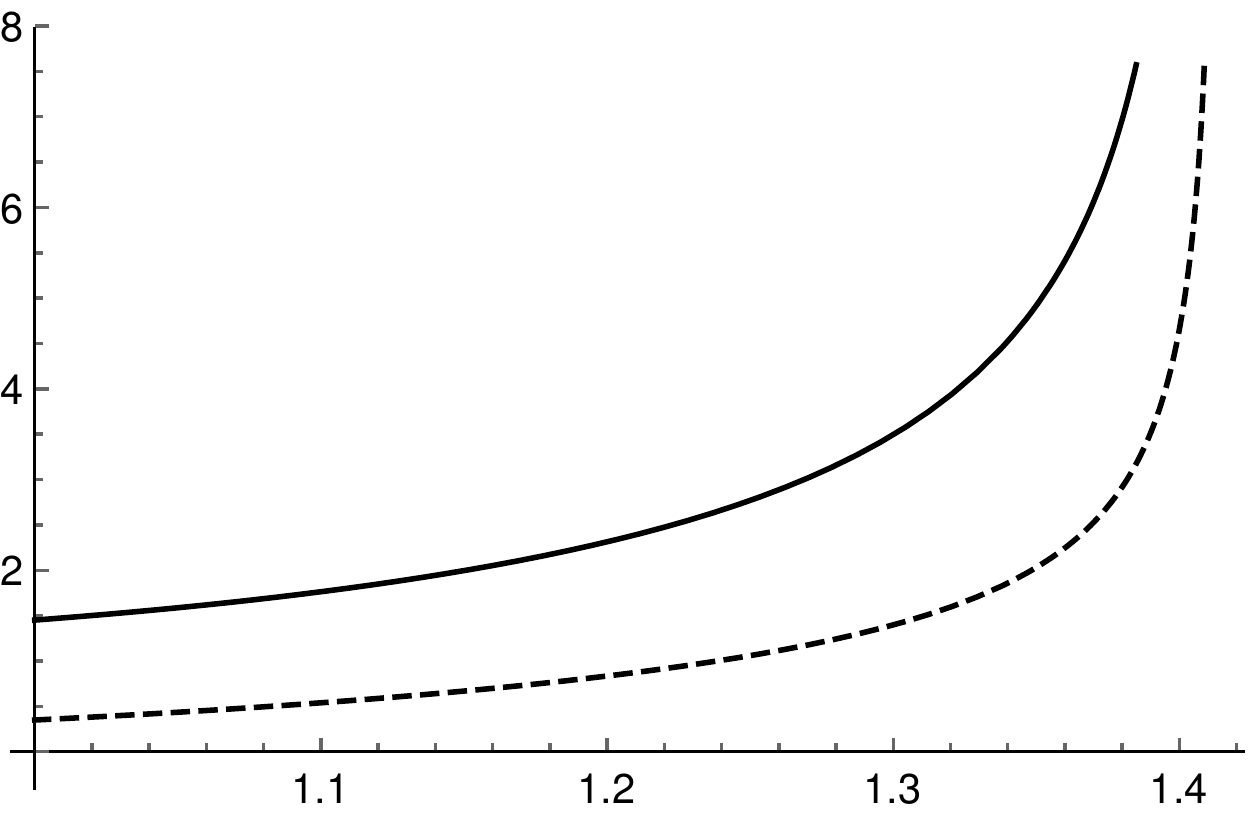}
$\alpha$
%
\caption{Expectation value $\expec{\hat r_{\rm H}}$ (solid line) and its uncertainty
$\Delta \rh$ (dashed line), in units of $\lp$, as functions of the specific charge 
$1<\alpha^2<2$ and $\ell = \lp$ ($m=\mpl$).
\label{ExpR}}
\end{figure}

Let us now consider what happens when approaching the critical specific
charge $\alpha^2= 2$.
One may have already noticed that 
\be
\expec{\hat r_{\rm H}}
\simeq
\frac{8}{\sqrt{\pi\,\left(2-\alpha^2\right)}}\,\frac{\lp^2}{\ell}
\ ,
\ee
so that the ratio $\expec{\hat r_{\rm H}}/\ell$ diverges in the limit $\alpha^2\to 2$,
regardless of the mass $m=\mpl\,\lp/\ell$.
Moreover, since
\be
\Delta \rh
\simeq
\sqrt{{3\,\pi}/{8}-1}\,
\expec{\hat r_{\rm H}}
\simeq
0.4\,
\expec{\hat r_{\rm H}}
\ ,
\ee
the uncertainty $\Delta \rh$ shows the same behaviour for $\alpha^2\to 2$
(see also Fig.~\ref{ExpR}).
\par 
\begin{figure}[h]
\centering
\raisebox{4.5cm}{$P_{\rm BH}$}
\includegraphics[width=8cm]{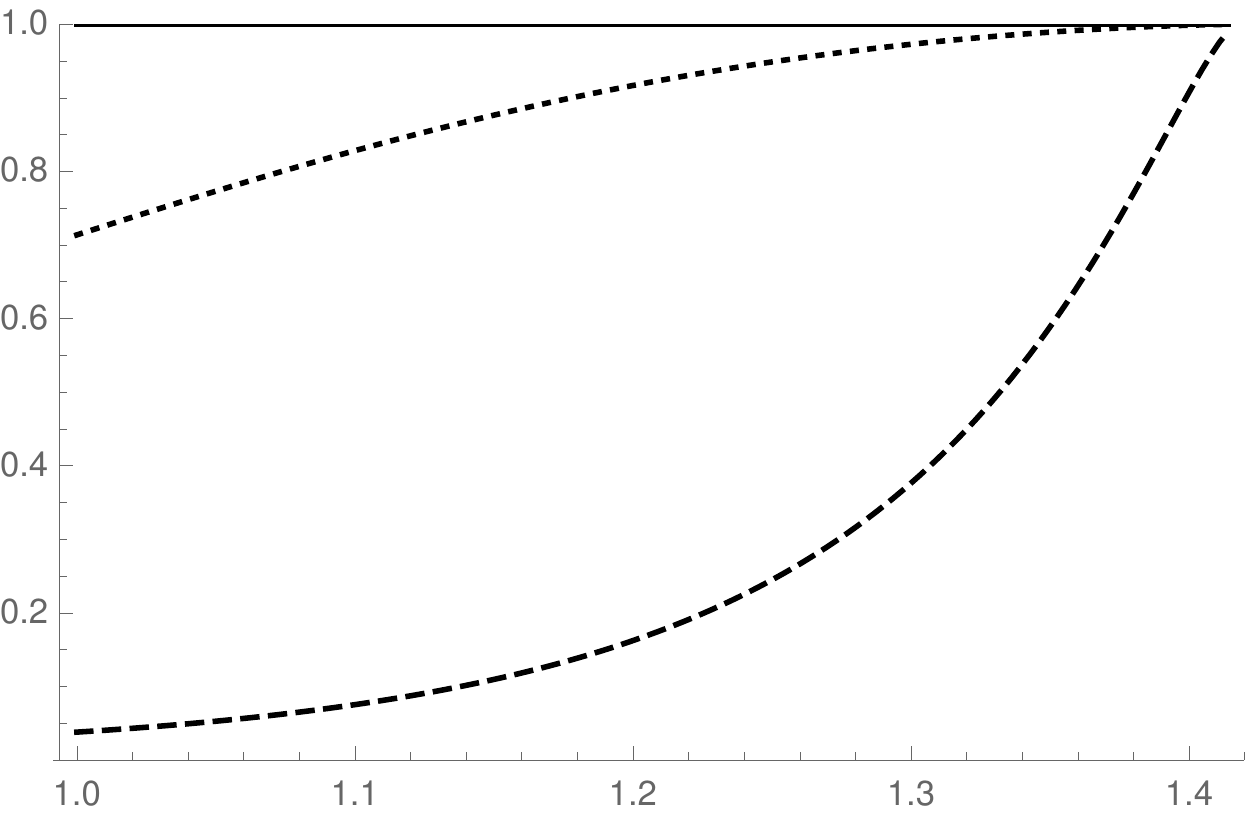}
$\alpha$
%
\caption{$P_{\rm BH}$ as a function of $\alpha$ for $\ell = \lp/2$
(solid line), $\ell= \lp$ (dotted line) and $\ell=2\,\lp$ (dashed line).
Cases with $\ell \ll \lp$ are not plotted since they behave
the same as  $\ell = \lp/2$, i.e.~an object with $1<\alpha^2<2$
must be a BH.
\label{large alpha}}
\end{figure}
In the same way that led to Eq.~\eqref{PBHRN}, we can obtain the probability $P_{\rm BH}$ 
that the particle is a BH for $\alpha$ in the allowed range~\eqref{a>1},
\be
P_{\rm BH}
=
\frac{4}{\sqrt{\pi}\,\Gamma\left(\frac{3}{2},\frac{2-\alpha^2}{\alpha^4}\right)}
\int_{\frac{\sqrt{2-\alpha^2}}{\alpha^2}}^\infty \gamma\left(\frac{3}{2},
\frac{\alpha^4}{2-\alpha^2}\,\frac{\lp^4}{\ell^4}\,x^2 \right)
e^{-x^2}\,x^2\,\d x
\ ,
\ee
where $x\equiv\sqrt{2-\alpha^2}\,\ell\,\rh/\alpha^2\,\lp^2$.
\par
This probability is computed numerically and plotted in Fig.~\ref{large alpha}
as a function of $\alpha$.
One notes that, for a Gaussian  width much smaller than $\lp$, $P_{\rm BH}\simeq 1$ 
throughout the whole range of the specific charge, which extends a similar result
for $0<\alpha\le 1$.
Nevertheless, even when $\ell$ significantly exceeds the Planck length, we see that
the same result is obtained in the limit $\alpha^2\to 2$.
It is important to recall that, when the system is far from the Planck scale, 
$\ell\gg \expec{\hat r_{\rm H}}$, quantum fluctuations in the particle's position
dominate and $P_{\rm BH}\ll1$ accordingly.
However, strong quantum fluctuations in the size of the horizon appear in the 
overcharged regime where the probability $P_{\rm BH}$ is large,
since $\expec{\hat r_{\rm H}}$ and $\Delta \rh$ blow up for $\alpha^2\to 2$.
\par
Bearing all the limitations and ambiguities in the above analysis, the picture that
emerges is that of a {\em quantum\/} version of the cosmic censorship:
first of all it appears that (slightly) overcharged configurations may exist,
but have a large probability of being BHs, rather than naked singularities;
secondly, when the specific charge is larger than a critical value (here
found to be $\alpha\simeq 1.4$), there exist no well-behaved HWF and the
gravitational radius of the system cannot be defined.
Of course, one should not forget that assuming a Gaussian wave-function
for the source already restricts these conclusions to masses of the order
of the Planck scale, and not too much larger, as we recalled in
section~\ref{effCUPqf}.  
\subsection{Particle collisions in $(1+1)$ dimensions}
A straightforward extension of the HQM to a state containing two
free particles colliding head-on in one-dimensional flat space was presented
in Ref.~\cite{Ctest}, where both constituents are represented by Gaussian
wave-functions centred around the positions $X_i$ and having linear
momentum $P_i$ ($i=1$ or $2$),
\be
\pro{x_i;0}{\psi_{\rm S}^{(i)}}
\equiv
\psi_{\rm S}(x_i)
=
e^{-i\,\frac{P_i\,x_i}{\mpl\,\lp}}\,
\frac{e^{-\frac{\left(x_i-X_i\right)^2}{2\,\ell_i}}}{\sqrt{\pi^{1/2}\,\ell_i}}
\ ,
\ee
where dynamical phases are neglected since we will only consider
``snapshots'' of the collision.
Like in the one-particle case, one switches to momentum space
in order to compute the spectral decomposition of the system,
\be
\pro{p_i;0}{\psi_{\rm S}^{(i)}}
\equiv
\psi_{\rm S}(p_i)
=
e^{-i\,\frac{p_i\,X_i}{\mpl\lp}}\,
\frac{e^{-\frac{\left(p_i-P_i\right)^2}{2\,\Delta_i}}}{\sqrt{\pi^{1/2}\,\Delta_i}}
\ ,
\ee
where the width $\Delta_i=\mpl\lp/\ell_i$, 
and we will use the relativistic flat-space dispersion relation
$
E_i^2
=
p_i^2+m_i^2
$,
just like in the single particle case~\eqref{mass-shell}.
It is particularly interesting to consider particles with masses
$m_1\simeq m_2\ll\mpl$, so that the probability that they form
a BH can be significant only in the ultra-relativistic limit
$|P_i|\sim E_i\sim \mpl$, which implies
\be
\ell_i
\simeq
\frac{\lp\,\mpl}{|P_i|}
\ ,
\qquad
\Delta_i
\simeq
|P_i|
\ .
\ee
The two-particle state can be written as
\be
\ket{\psi_{\rm S}^{(1,2)}}
=
\prod_{i=1}^2
\left[
\int\limits_{-\infty}^{+\infty}
\d p_i
\,\psi_{\rm S}(p_i,t)\,\ket{p_i}
\right]
\ ,
\label{PsiPp}
\ee
and the coefficients in the spectral decomposition~\eqref{spectral} are given by
\be
C(E)
=
\!\!
\int\limits_{-\infty}^{+\infty}
\int\limits_{-\infty}^{+\infty}
\psi_{\rm S}(p_1)\,\psi_{\rm S}(p_2)\,
\delta(E-E_1-E_2)\,
\d p_1\,\d p_2
\ .
\label{C(E)}
\ee
The HWF is defined in the rest frame of the possible BH, 
that is in the centre-of-mass coordinate system with
$
P_1=-P_2
\equiv P>0
$.
From $P\sim\mpl\gg m_1\simeq m_2$, we can also set
$
X_1\simeq -X_2
\equiv 
X>0
$.
The unnormalised HWF is then given by~\cite{Ctest}
\be
\psi_{\rm H}
&=&
e^{-\frac{\mpl\rh^2}{16\lp^2\,P}-\frac{X^2P^2}{\lp^2\mpl^2}}
{\rm Erf}\left(1+\frac{\mpl\,\rh}{4\,\lp\,P}+i\,\frac{XP}{\lp\,\mpl}\right)
\nonumber
\\
&&
-e^{-\frac{\mpl\rh^2}{16\lp^2\,P}-\frac{X^2P^2}{\lp^2\mpl^2}}
{\rm Erf}\left(1-\frac{\mpl\,\rh}{4\,\lp\,P}-i\,\frac{XP}{\lp\,\mpl}\right)
\nonumber
\\
&&
+
2\,e^{-1-\frac{2 i X P}{\lp \mpl}-\frac{\mpl \rh^2}{16 \lp^2 P}}
\cosh\left(\frac{\mpl\rh}{2\,\lp P}
+i\,\frac{\rh X}{2\,\lp^2}\right)
\,
{\rm Erf}\left(\frac{\mpl \rh}{4\,\lp P}\right)
\ ,
\label{2PartHWF}
\ee
whose normalisation can be computed numerically for fixed $X$ and $P$.
\begin{figure}[t]
\centering
\raisebox{3cm}{$|\psi_{\rm S}|^2$}
\includegraphics[width=7cm,height=4cm]{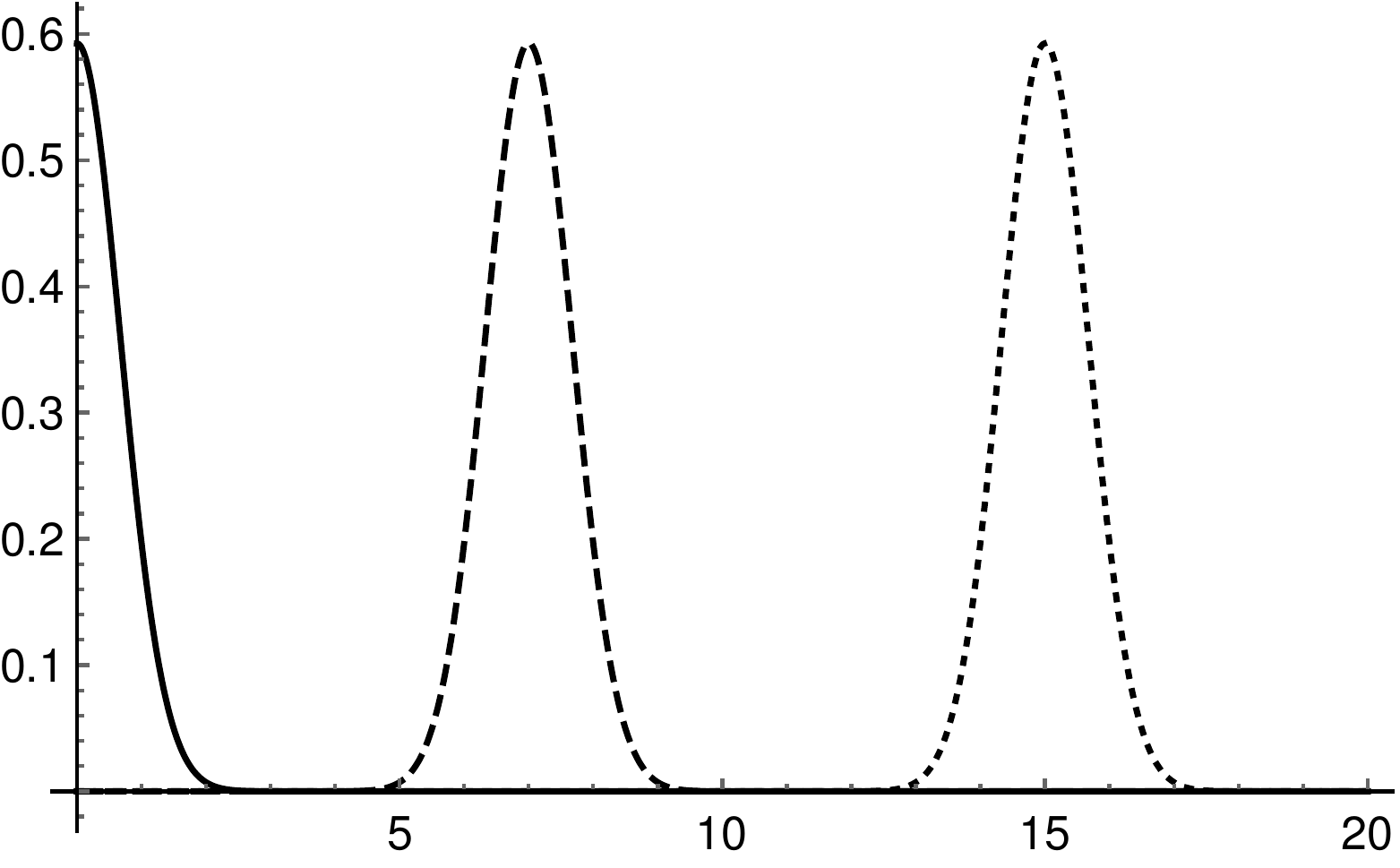}
{$\frac{x}{\lp}$}
\\
\raisebox{3cm}{$|\psi_{\rm H}|^2$}
\includegraphics[width=7cm,height=4cm]{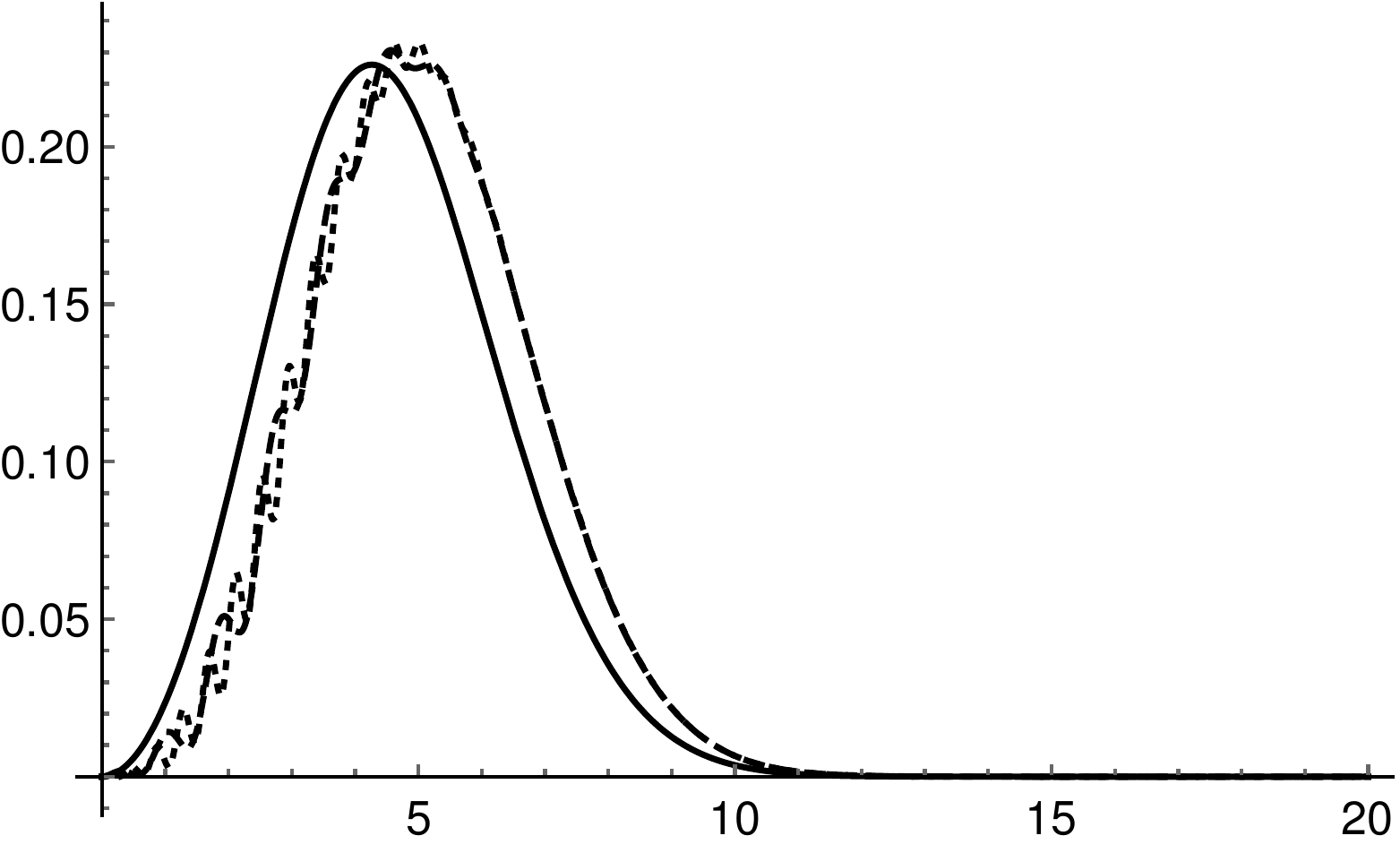}
{$\frac{\rh}{\lp}$}
\caption{Top panel: square modulus of $\psi_{\rm S}$ for $P=\mpl$ and $X=0$ (solid line)
$X=7\,\lp$ (dashed line) and $X=15\,\lp$ (dotted line).
Bottom panel: square modulus of $\psi_{\rm H}$ for $P=\mpl$ and $X=0$ (solid line)
$X=7\,\lp$ (dashed line) and $X=15\,\lp$ (dotted line).
Particles are inside the horizon only for sufficiently small $X$.}
\label{PBHx}
\end{figure}
\begin{figure}[h!]
\centering
\raisebox{3cm}{$|\psi_{\rm S}|^2$}
\includegraphics[width=7cm,height=4cm]{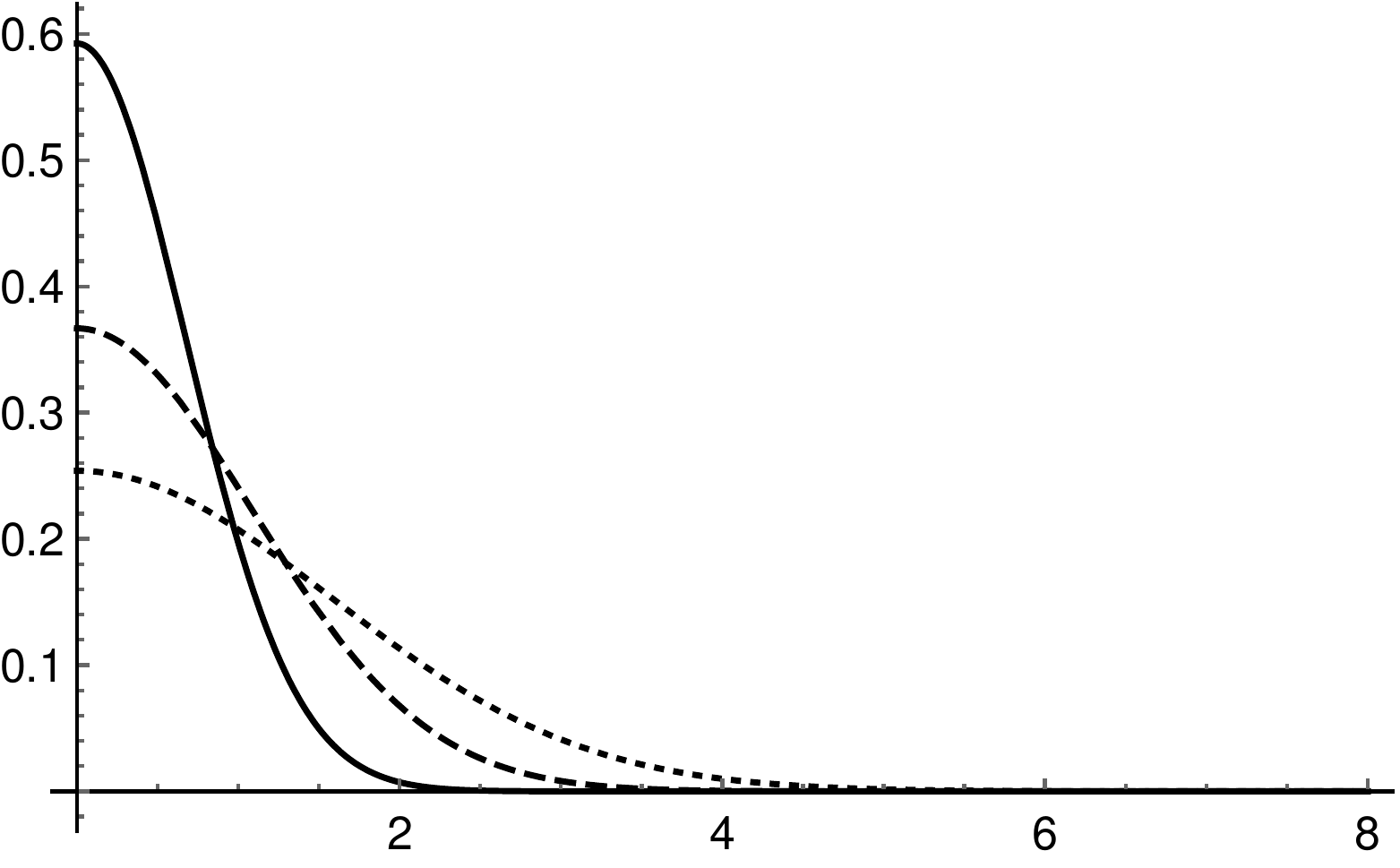}
{$\frac{x}{\lp}$}
\\
\raisebox{3cm}{$|\psi_{\rm H}|^2$}
\includegraphics[width=7cm,height=4cm]{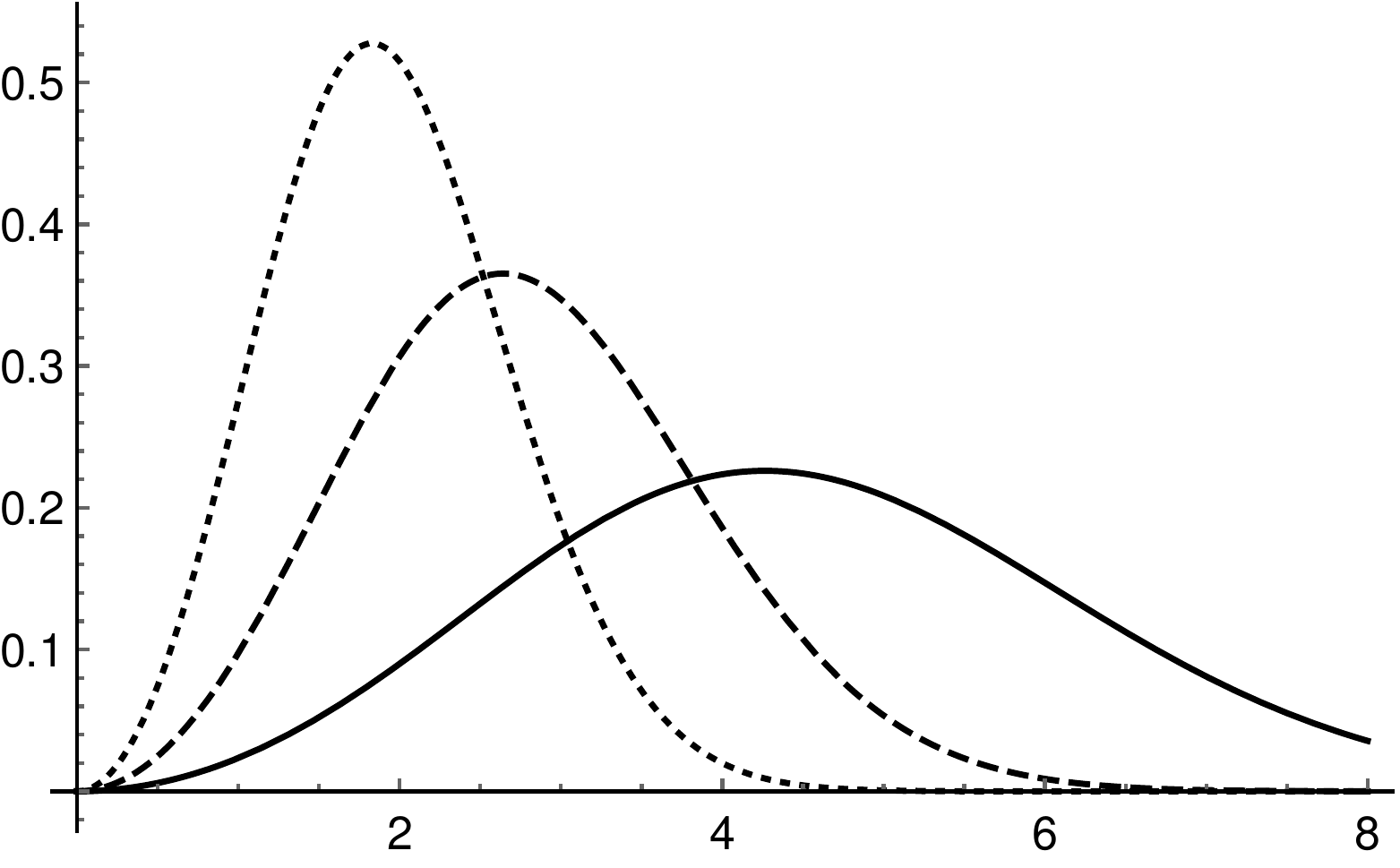}
{$\frac{\rh}{\lp}$}
\caption{Top panel: square modulus of $\psi_{\rm S}$ for $X=0$ and $P=\mpl$ (solid line)
$P=3\,\mpl/5$ (dashed line) and $P=2\,\mpl/5$ (dotted line).
Bottom panel: square modulus of $\psi_{\rm H}$ for $X=0$ and $P=\mpl$ (solid line)
$P=3\,\mpl/5$ (dashed line) and $P=2\,\mpl/5$ (dotted line).
Particles' location is sharper the fuzzier (more spread) the horizon location
and vice versa.}
\label{PBHp}
\end{figure}
\begin{figure*}[t!]
\centering
\includegraphics[width=8cm]{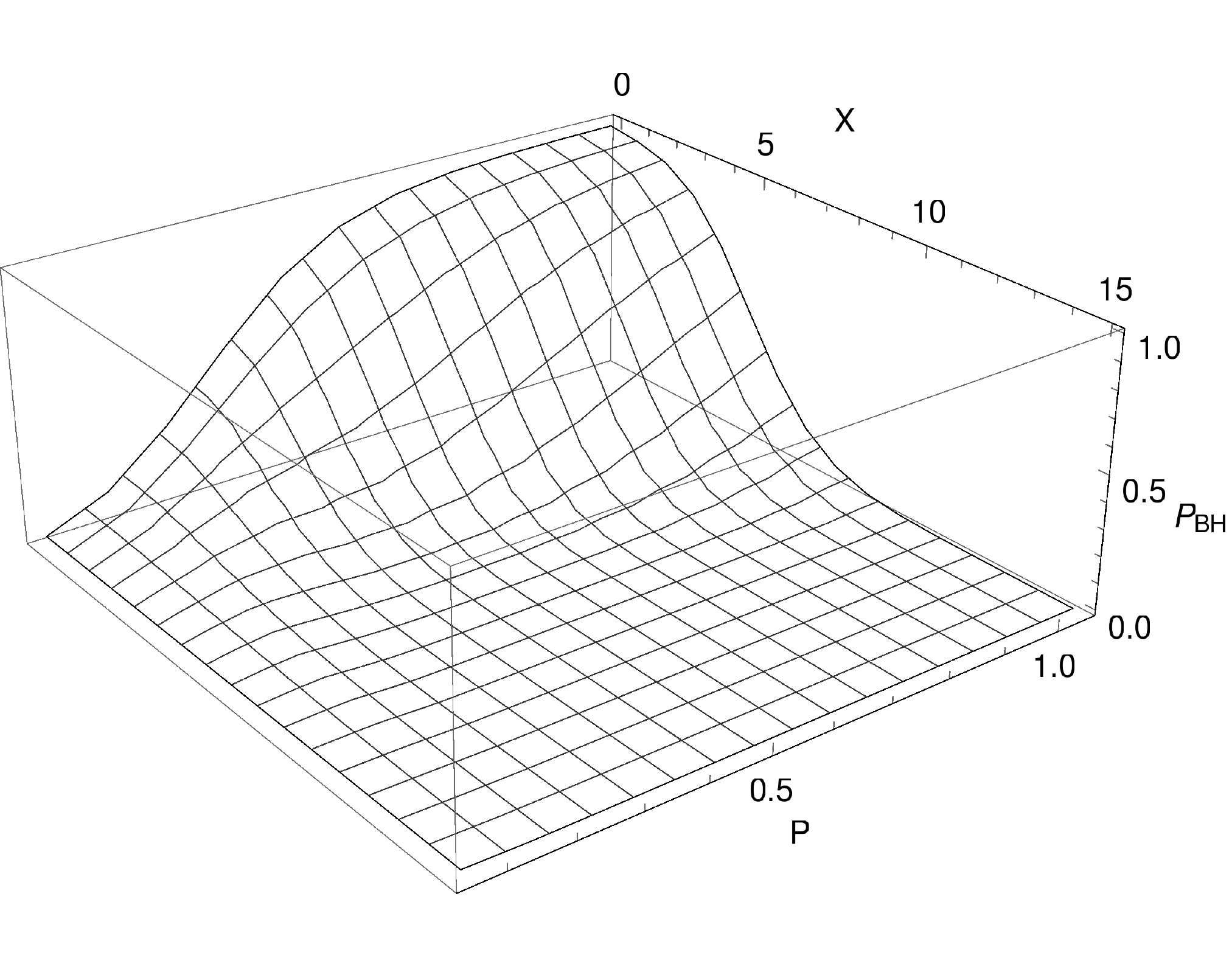}
\caption{Probability the two-particle system is a BH as a function of $X$ and
$P$ (in units of Planck length and mass respectively).}
\label{P_BH}
\end{figure*}
\par
The energy of the system is solely determined by the momentum, and in fact
we notice from Fig.~\ref{PBHx} that $\mathcal{P}_{\rm H}= |\psi_{\rm H}(\rh)|^2$ 
does not vary much with $X$, but is clearly affected by $P$ (see Fig.~\ref{PBHp}).
Moreover, the peak of the probability density $\mathcal{P}_{\rm H}$ is 
always located around $\rh\simeq 2\,\lp\,(2 P/\mpl)$.
\par
The final step is to compute the probability~\eqref{PBH} that the two-particle
system is a BH as a function of the distance $X$ of each particle from the
center-of-mass and the total energy $2P$ (see Fig.~\ref{P_BH}).
One may argue that a rough estimate of the time evolution is given by
considering this function along lines of constant $P$ and decreasing $X$.
In fact, it is easy to see that the probability increases to a
maximum for $X=0$, when the two particles overlap exactly.
Hence, there is a large probability that the collision forms a BH,
e.g.~$P_{\rm BH}(X,2P\gtrsim 2\mpl)\gtrsim 80\%$, when
\be
X
\lesssim
{2\,\lp}\left({2P}/{\mpl}\right)-{\lp}
=
\rh(2P)-\lp
\ .
\label{linearHoop}
\ee
The second term in the r.h.s.~can be viewed as a quantum correction
to the hoop formula~\eqref{hoop} for $E\simeq 2P\gtrsim 2\,\mpl$, and
becomes negligible for large (semi)classical BHs produced in collisions with
$2P\gg\mpl$.
Lowering $P$, we have that the region $P_{\rm BH}(X,2P\lesssim 2\mpl)\gtrsim 80\%$ 
corresponds to the momenta satisfying $2P\gtrsim \mpl\,(1+X^2/9\,\lp^2)$ 
and its boundary $P_{\rm BH}(X,2P\lesssim 2\mpl)\simeq 80\%$ 
can be approximated by
\be
2P-\mpl
\simeq
{\mpl\,X^2}/{9\,\lp^2}
\ ,
\label{quadHoop}
\ee
which crosses the axis $X=0$ for $2P\simeq \mpl$.
This curve represents a further correction to the hoop conjecture~\eqref{hoop},
and supports the conclusion that the mass of quantum BHs is bounded below
by about $\mpl$.
We should remark that, although these numerical values strongly depend 
on what probability $P_{\rm BH}$ is considered large enough,
the slope in Eq.~\eqref{linearHoop} agrees perfectly with Eq.~\eqref{hoop}.
One can thus conclude that, despite the great simplifications assumed in this
analysis, the HQM appears suitable to extend the hoop conjecture into the
quantum description of BH formation.
\subsection{Higher and lower dimensional models}
The idea that the number of dimensions of space-time is not exactly four
as we experience, was proposed in order to explain some puzzles of the
Standard Model, like the hierarchy problem, or for consistence with string theory.
Remarkably, in $D>3$ spatial dimensions, the fundamental gravitational mass
$m_D\ll\mpl$, and $\ell_D=\hbar/m_D\gg\lp$, where
\be
G_D
=
\frac{\ell_D^{D-2}}{m_D}
\ .
\label{GD}
\ee
This opens up the possibility of having much lighter BHs, possibly within the reach of
current high-energy experiments~\cite{giddings,dimopoulos}.
This happens both in the ADD~\cite{ArkaniHamed:1998rs,Antoniadis:1998ig} 
and the Randall-Sundrum~\cite{Randall:1999ee,Randall:1999vf} 
models (for a comprehensive, see Ref.~\cite{Maartens:2010ar}).
In Ref.~\cite{Casadio:2015jha} the HQM probability that BHs form in the ADD
scenario was computed, with some interesting consequences.
\par
It is also instructive to study theories with less than three spatial dimensions, since
the corresponding quantum theories are simpler and can be solved exactly~\cite{Mureika:2012na}.
In recent years, interest in such theories was also revived by the possibility that
the number of spatial dimensions effectively decreases when approaching $\lp$,
regardless of the model under consideration.
This effect is called ``spontaneous dimensional reduction''
and has been extended to various contexts, most of which with special focus 
on the energy dependence of the spectral dimension, including causal dynamical
triangulations~\cite{cdt1,Ambjorn,Ambjorn2} and non-commutative geometry inspired 
mechanisms~\cite{lmpn,Nicolini:2012fy,Carr:2015nqa,Mureika:2012fq}.
An alternative approach is built on the claim that the effective
dimensionality of space-time increases as the ambient energy
scale decreases~\cite{vd1,vd2,jrmds1,jrmds2,nads,Stojkovic:2014lha}.
\subsubsection{$(1+D)$-dimensional Schwarzschild metric}
In $D$ spatial dimensions, the generalised Schwarzschild metric is given by
\be
\d s^2
=
-\left(1-\frac{R_D}{r^{D-2}}\right) \, \d t^2
+\left(1-\frac{R_D}{r^{D-2}}\right)^{-1} \, \d r^2
+r^{D-1} \, \d\Omega_{D-1}
\ ,
\ee
where the classical horizon radius is
\be
R_{D}
=
\left(\frac{2 \, \gd \, m}{|D-2|}\right)^{\frac{1}{D-2}} 
=
\begin{cases} 
\strut\displaystyle\frac{1}{2 \,G_1\, m} & \quad \text{if }\ D=1
\\
&
\\
\strut\displaystyle\left(\frac{2 \,\gd\, m}{D-2}\right)^{\frac{1}{D-2}}
& \quad \text{if }\ D > 2 
\ .
\\
\end{cases}
\label{SchwD}
\ee
Note that $D=2$ is excluded because in that case there exists no
asymptotically flat BH, and we do not want to include a cosmological constant.
\par
The source of the gravitational field is still described by a Gaussian wave-function,
that is
\be
\psi_{\rm S}(r)
=
\frac{e^{-\frac{r^2}{2\,\ell^2}}}{(\ell\, \sqrt{\pi})^{D/2}}
\ ,
\label{GaussD}
\ee
whose momentum space counterpart is
\be
\tilde{\psi}_{\rm S}(p)
=
\frac{e^{-\frac{p^2}{2\,\Delta^2}}}{(\Delta\, \sqrt{\pi})^{D/2}}
\ ,
\label{momGauss}
\ee
where $\Delta=m_D \,\ell_D/\ell$ and, taking again Eq.~\eqref{LLc}, $\ell\sim m^{-1}$,
we recover $\Delta\simeq m$.
As in $D=3$, we assume the relativistic mass-shell relation in flat space~\eqref{mass-shell},
and, for $D>3$, one obtains the HWF
\be
\psi_{\rm H}
&=&
\left\{
\frac{D-2}{\ell_D^{\,D} \, \pi^{D/2}}
\left[\frac{(D-2)\,\ell}{2\,\ell_D}\right]^{\frac{D}{D-2}}
\, \frac{\Gamma\left(\frac{D}{2}\right)}
{\Gamma\left(\frac{D}{2D-4},1\right)}
\right\}^{1/2}
\nonumber
\\
&&
\times
\Theta(\rh-R_D)\, \exp\left\{ -
\frac{(D-2)^2}{8} \, \frac{\ell^2\,\rh^{2(D-2)}}{\ell_D^{2(D-1)}} \right\}
\ ,
\label{HWFD}
\ee
whose normalisation was fixed in the scalar product
\be
\pro{\psi_{\rm H}}{\phi_{\rm H}}
=
\Omega_{D-1}\int_0^\infty
\psi_{\rm H}^*(\rh)\,\phi_{\rm H}(\rh)\,\rh^{D-1}\,\d \rh
\ ,
\ee
where $\Omega_{D-1}$ is the volume of the $D-$sphere.
\par
For $D=1$, there is an important change of sign in the argument of the step function.
In fact, the generalisation~\eqref{CondTheta} of the hoop conjecture~\eqref{hoop}
is now satisfied when $0\le \rh\le R_1$ and the HWF reads
\be
\psi_{\rm H}
=
\sqrt{\frac{2/\ell}
{\Gamma\left(-\frac{1}{2},1\right)}}
\, \Theta(R_1-\rh)\, \exp\left\{ -
\frac{\ell^2}{8\,\rh^2} \right\}
\ ,
\label{HWF1}
\ee
which otherwise is the same as~\eqref{HWFD} with $D=1$.
\subsubsection{BH probability}
\begin{figure}[t]
\centering
\includegraphics[scale=0.43]{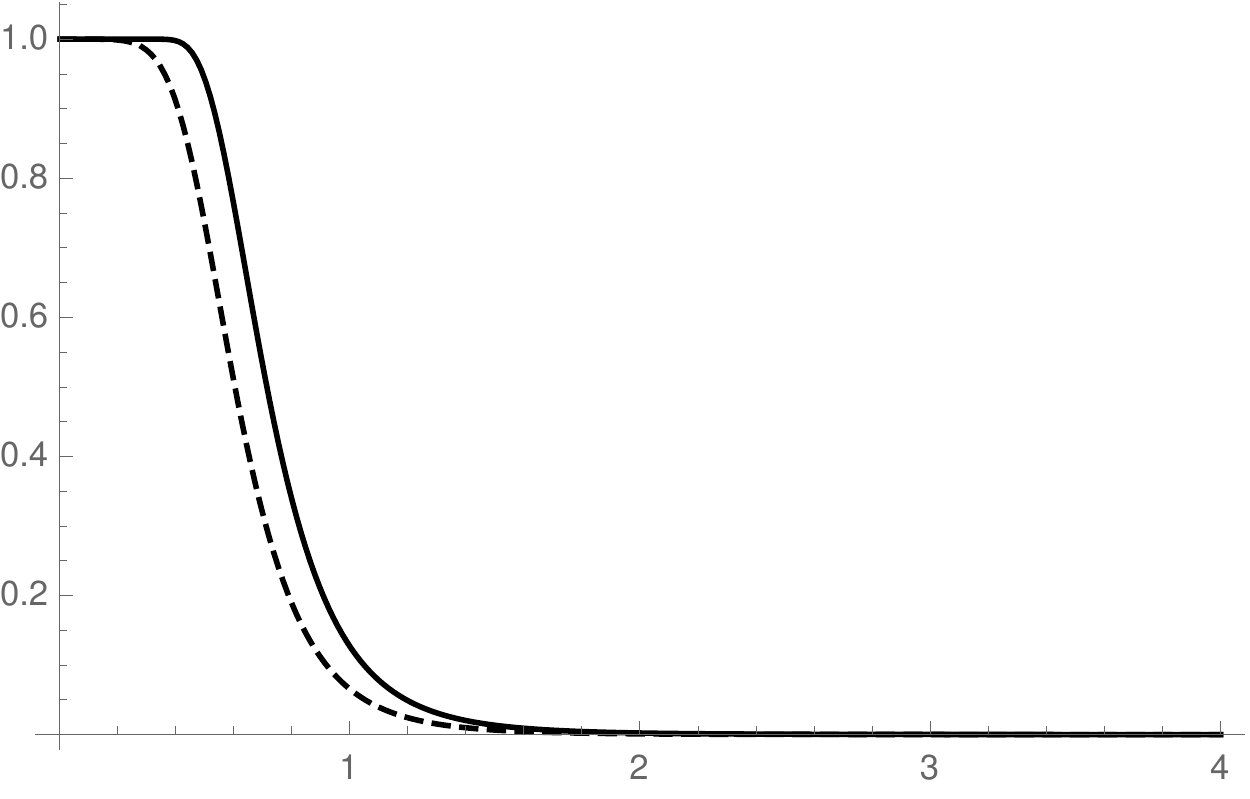}
$\ $
\includegraphics[scale=0.43]{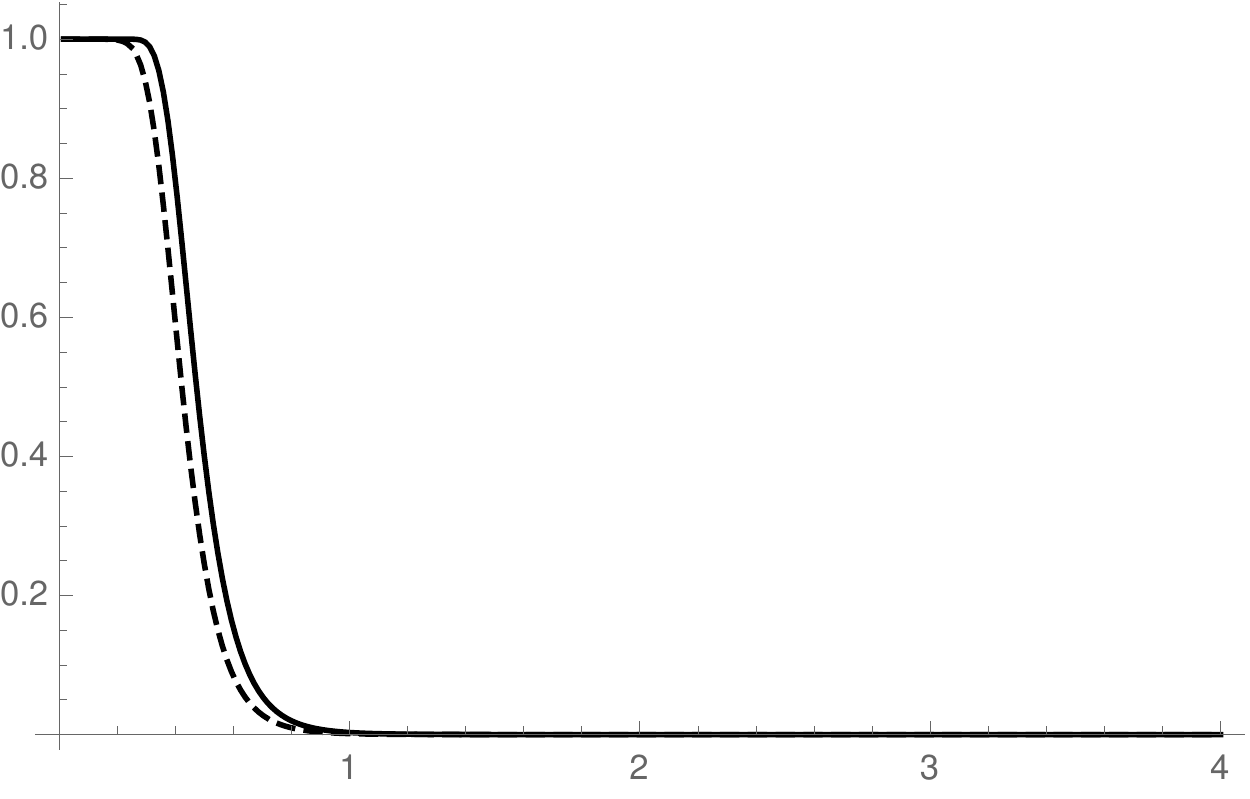}
{$\frac{\ell}{\ell_D}$}
\\
$D=5$
\hspace{5cm}
$D=9$
\caption{
Probability $P_{\rm BH}(\ell)$ of a particle to be a BH (straight line)
compared to its analytical approximation (dashed line), for $D=5$ and $9$.
\label{prob2}}
\end{figure}

It is straightforward to write down the probability for the particle to be inside a
$D$-dimensional ball of radius $\rh$,
\be
P_{\rm S}(r<\rh)
=
\Omega_{D-1}\int_0^{\rh} |\psi_{\rm S}(r)|^2 \, r^{D-1} \, \d r
\ ,
\ee
and the probability density that the gravitational radius equals $\rh$ is
\be
\mathcal{P}_{\rm H}(\rh)
=
\Omega_{D-1} \, r^{D-1} \, |\psi_{\rm H}(\rh)|^2
\ .
\ee
Omitting the details, one then finds
\be
\mathcal{P}_<
&=&
\frac{2}{\ell_D^{\,D}}\, \left[\frac{(D-2) \, \ell}{2 \,\ell_D}\right]^{\frac{D}{D-2}}
\frac{D-2}{\Gamma\left(\frac{D}{2\,D-4},1\right)\Gamma\left(\frac{D}{2}\right)}
\,\Theta(\rh-R_D)
\nonumber
\\
&&
\times\,
\gamma\left(\frac{D}{2},\frac{\rh^2}{\ell^2}\right) \,\exp\left\{ -
\frac{(D-2)^2}{4}\,\frac{\ell^2\,\rh^{2(D-2)}}{\ell_D^{2(D-1)}} \right\}
\, \rh^{D-1}
\label{P<D}
\ee
and the BH probability is
\be
P_{\rm BH}
&=&
\frac{2(D-2)}{\Gamma\left(\frac{D}{2D-4},1\right) \Gamma\left(\frac{D}{2}\right)}
\notag
\\
&&
\times
\int_1^\infty \gamma\left(\frac{D}{2},
\left[\frac{2}{D-2}
\,\left(\frac{\ell_D}{\ell}\right)^{D-1}\right]^{\frac{2}{D-2}}\,x_D^2\right)
e^{-x^{2(D-2)}_D} \, x_D^{D-1}\,\d x_D
\ ,
\label{PBHexplicit}
\ee
where we defined
$
x_D^{D-2}
=
(D-2)\,\ell\, \rh^{D-2}/2\,\ell_D^{D-1}
$.
Eq.~\eqref{PBHexplicit} depends, as usual, on the Gaussian width $\ell$, but
also on the number $D$ of spatial dimensions (with $D=3$ reproducing Eq.~\eqref{Pbh}).
Since the above integral cannot be performed analytically for a general $D$, 
in Fig.~\ref{prob2} we show the numerical dependence on $\ell$ of the above probability
for different spatial dimensions, and compare it with the approximation obtained by
taking the limit $R_D\to 0$.
\par
The most important fact here is that the probability $P_{\rm BH}=P_{\rm BH}(m,D)$
at a given $m$ decreases significantly for increasing $D$, and for large
values of $D$ a particle of mass $m\simeq m_D$ is most likely not a BH.
This result should have a strong impact on the number of BHs
produced in particle collisions.
In fact, one expects the effective production cross-section
$\sigma(E)\sim P_{\rm BH}(E)\,\sigma_{\rm BH}(E)$, where
$\sigma_{\rm BH}\sim 4\,\pi\,E^2$ is the usual expression following from
Eq.~\eqref{hoop}.
Since $P_{\rm BH}$ can be very small, $\sigma(E)\ll \sigma_{\rm BH}(E)$
for $D>4$, and much less BHs should be produced than estimated
previously~\cite{dimopoulos}.
\begin{figure}[t]
\centering
\raisebox{4.5cm}{ $P_{\rm BH}$}
\includegraphics[width=8cm,height=5cm]{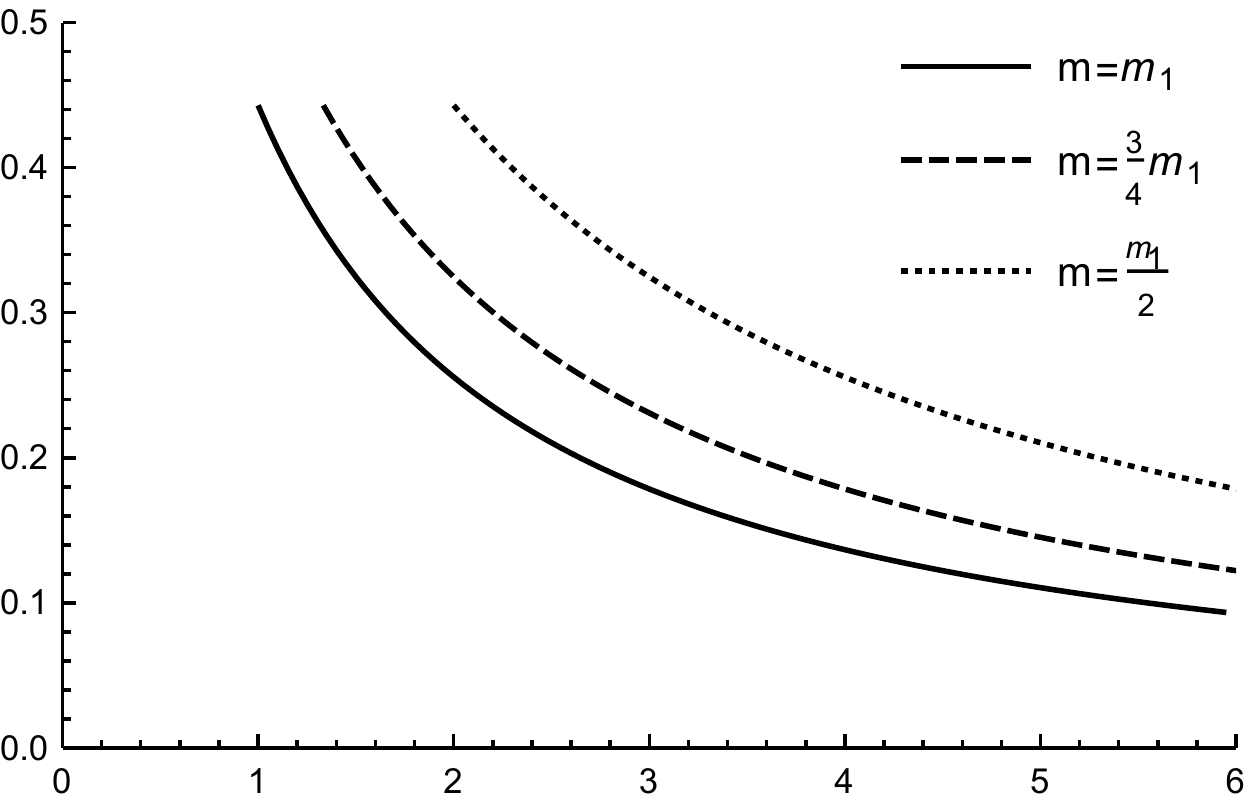}
{ $\frac{\ell}{\ell_1}$}
\caption{Probability $P_{\rm BH}(\ell,m)$ for a particle to be
a BH in $D=1$, for $m=m_1$ (solid line), $m=3\,m_1/4$ (dashed line)
and $m=m_1/2$ (dotted line).
\label{prob1D}}
\end{figure}
\par
For $D=1$ and $\ell=\lambda_m$, we can integrate the density
\be
\mathcal{P}_{\rm H}
=
\frac{2/\ell}
{\Gamma\left(-\frac{1}{2},1\right)}
\, \Theta(R_1-|\rh|)\,\mathrm{erf}\left(\frac{\rh}{\ell}\right)\, \exp\left\{ -
\frac{\ell^2}{4\,\rh^{2}}\right\}
\ ,
\ee
obtained from Eq.~\eqref{HWF1}, and find
\be
P_{\rm BH}
=
\frac{1}{\Gamma\left(-\frac{1}{2},1\right)}
\int_0^1
\mathrm{erf}\left(\frac{x_1}{2}\right)\,e^{-\frac{1}{x^2_1}} \, \d x_1
\simeq
0.44
\ ,
\ee
where $x_1=2\,\rh/\ell$, which can also be obtained 
from Eq.~\eqref{PBHexplicit} by setting $D=1$. 
This last equation reveals a striking difference between $D=1$ and 
higher-dimensional space-times.
The maximum probability that a BH may form is independent of the mass
of the source.
This result is supported by the fact that the one-dimensional gravitational
constant $G_1=\hbar$ and
\be
\expec{\hat r_{\rm H}}
\simeq 
R_1(m)
\simeq
\lambda_m
\ ,
\ee
for any possible mass, and hence no source can be treated in a classical way.
Moreover, for more general cases with $\ell>\lambda_m$,
particles with masses considerably lower than the mass scale $m_1$ still
have a relatively large probability to be BHs (see Fig.~\ref{prob1D})~\cite{Casadio:2015jha}.
Another important feature of the HWF in $D=1$ is that 
\be
\Delta\rh
\simeq
\ell
\simeq
\Delta p^{-1}
\ ,
\ee
that is the uncertainty in the horizon radius shows the same dependence on the
momentum uncertainty found in the Heisenberg relation.
This implies that we cannot obtain a GUP in $D=1$ by combining (linearly) the above
the two uncertainties, unlike in the three-dimensional case~\eqref{effGUP}.
In fact, all of these results agree with the notion that two-dimensional BHs are strictly 
quantum objects~\cite{Mureika:2012fq}.
\subsubsection{GUP from HWF in higher dimensions}
For $D>3$, we have
\be
\expec{\hat{r}}
=
\frac{2^{1-D}\sqrt{\pi}\, (D-1)!}{\Gamma\left(\frac{D}{2}\right)^2} \, \ell
\ee
and
\be
\expec{\hat{r}^2}
=
\frac{D}{2} \, \ell^2
\ .
\ee
Moreover, 
\be
\Delta p
=
\sqrt{A_D} \,m
=
\sqrt{A_D} \, m_D \, \frac{\ell_D}{\ell}
\ ,
\label{DeltaPD}
\ee
so that
\be
\frac{\Delta r}{\ell_D}
=
\sqrt{A_D} \, \frac{\ell}{\ell_D}
=
A_D\,\frac{m_D}{\Delta p}
\ ,
\label{DrQM}
\ee
where
\be
A_D
\equiv
\frac{D}{2}
-\left(\frac{2^{1-D}\sqrt{\pi}}{\Gamma\left(\frac{D}{2}\right)^2} \, (D-1)!\right)^2
\ .
\ee
\par 
From the HWF~\eqref{HWFD}, we likewise obtain the expectation values
\be
\expec{\hat{r}_{\rm H}}
=
\frac{\E_{\frac{D-5}{2D-4}}(1)}{\E_{\frac{D-4}{2D-4}}(1)} \, R_{D}
\ee
and
\be
\expec{\hat{r}_{\rm H}^2}
=
\frac{\E_{\frac{D-6}{2D-4}}(1)}{\E_{\frac{D-4}{2D-4}}(1)} \, R_{D}^2
\ ,
\ee
in terms of the exponential integral~\eqref{ExpInt}, so that
\be
\frac{\Delta \rh}{\ell_D}
=
C_D\,
\left(\frac{\ell_D}{\ell}\right)^{\frac{1}{D-2}}
=
B_D\,
\left(\frac{\Delta p}{m_D}\right)^{\frac{1}{D-2}}
\ ,
\label{deltarhD}
\ee
where $B_D=A_D^{-\frac{2}{D-2}}\,C_D$ and
\be
C_D
=
\sqrt{
\frac{\E_{\frac{D-6}{2D-4}}(1)}{\E_{\frac{D-4}{2D-4}}(1)} -
\left(\frac{\E_{\frac{D-5}{2D-4}}(1)}
{\E_{\frac{D-4}{2D-4}}(1)}\right)^2 } \, 
\left(\frac{2}{D-2}\right)^{\frac{1}{D-2}}
\ .
\ee
By combining the two uncertainties~\eqref{DrQM} and \eqref{deltarhD}
linearly, one finally finds
\be
\frac{\Delta r}{\ell_D}
=
A_D\, \frac{ m_D}{\Delta p}
+
\xi\,B_D
\left(\frac{\Delta p}{m_D}\right)^{\frac{1}{D-2}}
\ ,
\label{DrGUPD}
\ee
where, like before, the coefficient $\xi$ is a dimensionless parameter. 
\par
\begin{figure}[t!]
\centering
\raisebox{3.3cm}{$\frac{\Delta r}{\ell_D}$}
\includegraphics[scale=0.4]{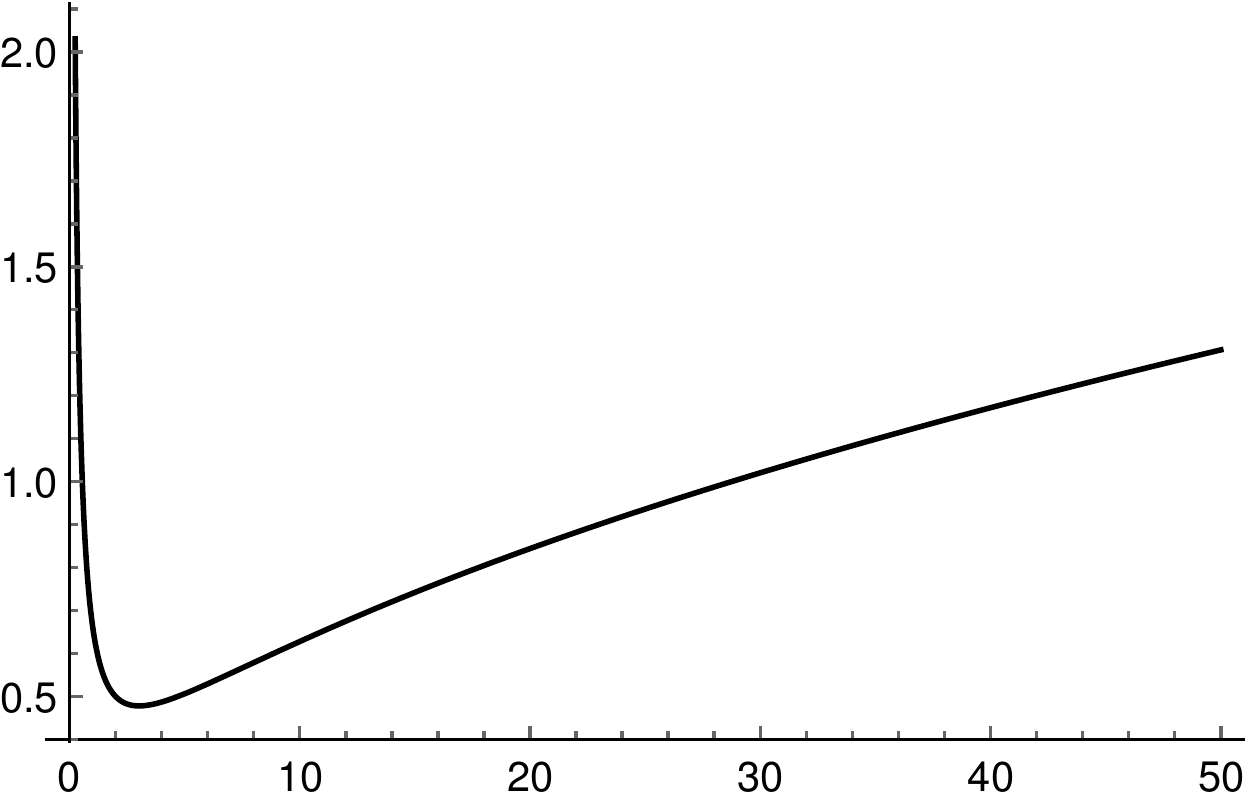}
$\quad$
\includegraphics[scale=0.4]{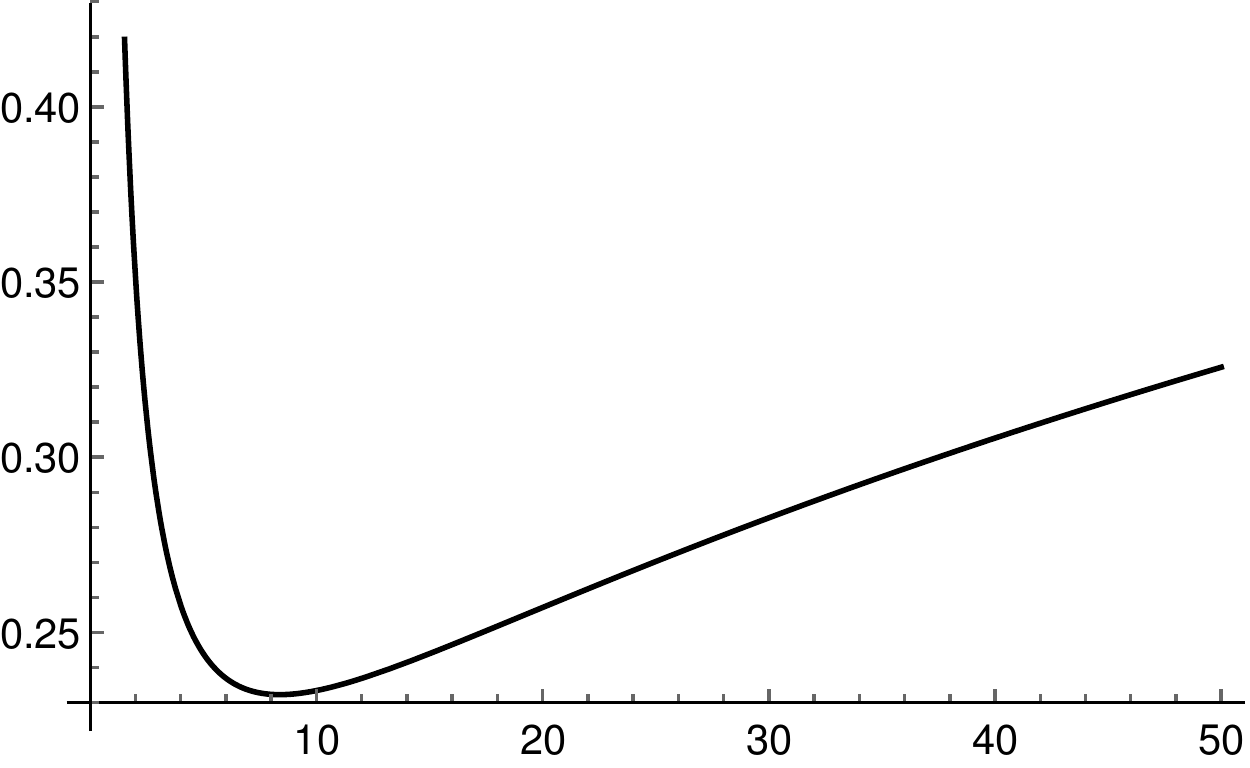}
$\frac{\Delta p}{m_D}$
\\
$D=4$
\hspace{4cm}
$D=5$
\caption{Uncertainty $\Delta r$ as function of $\Delta p$ for $D=4$ and $5$
 and $\xi=1$.
\label{delta-r}}
\end{figure}
\begin{figure}[h!]
\centering
\raisebox{3.3cm}{\tiny ${\frac{L_D}{\ell_D}}$}
\includegraphics[scale=0.4]{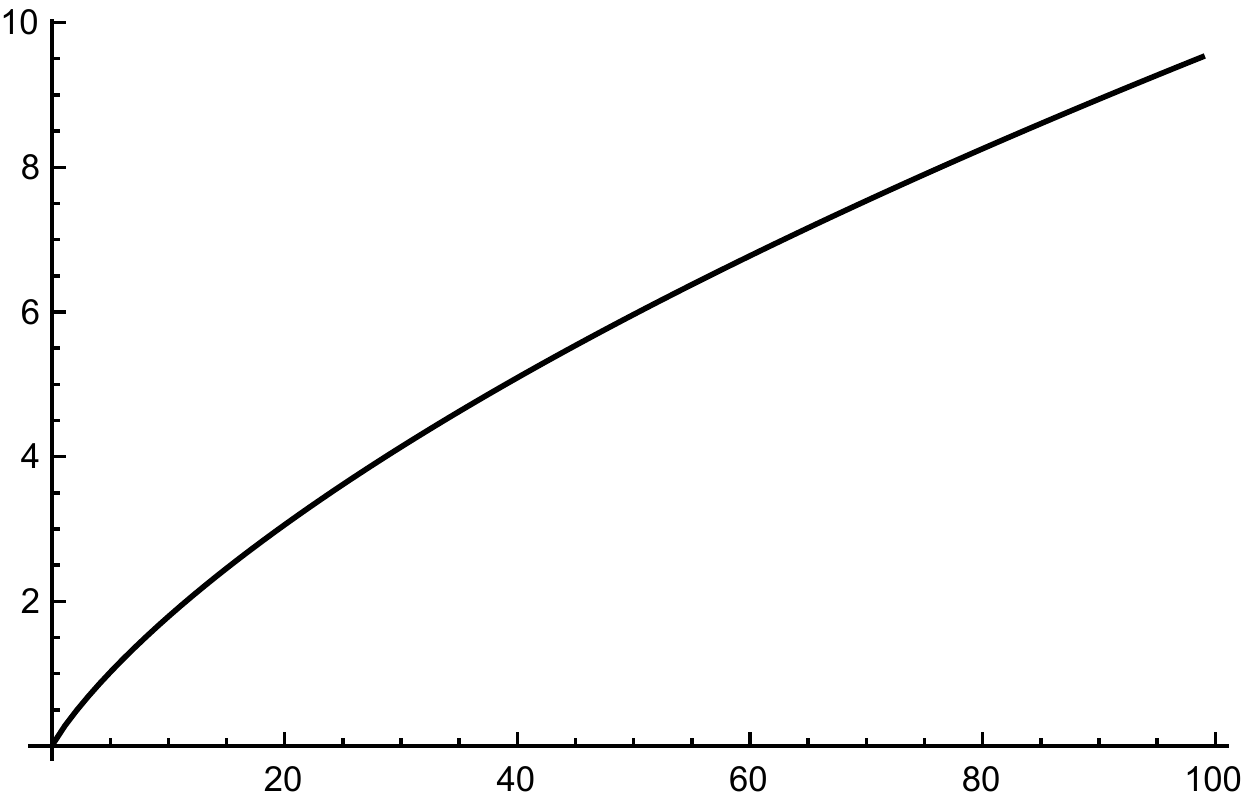}
$\quad$
\includegraphics[scale=0.4]{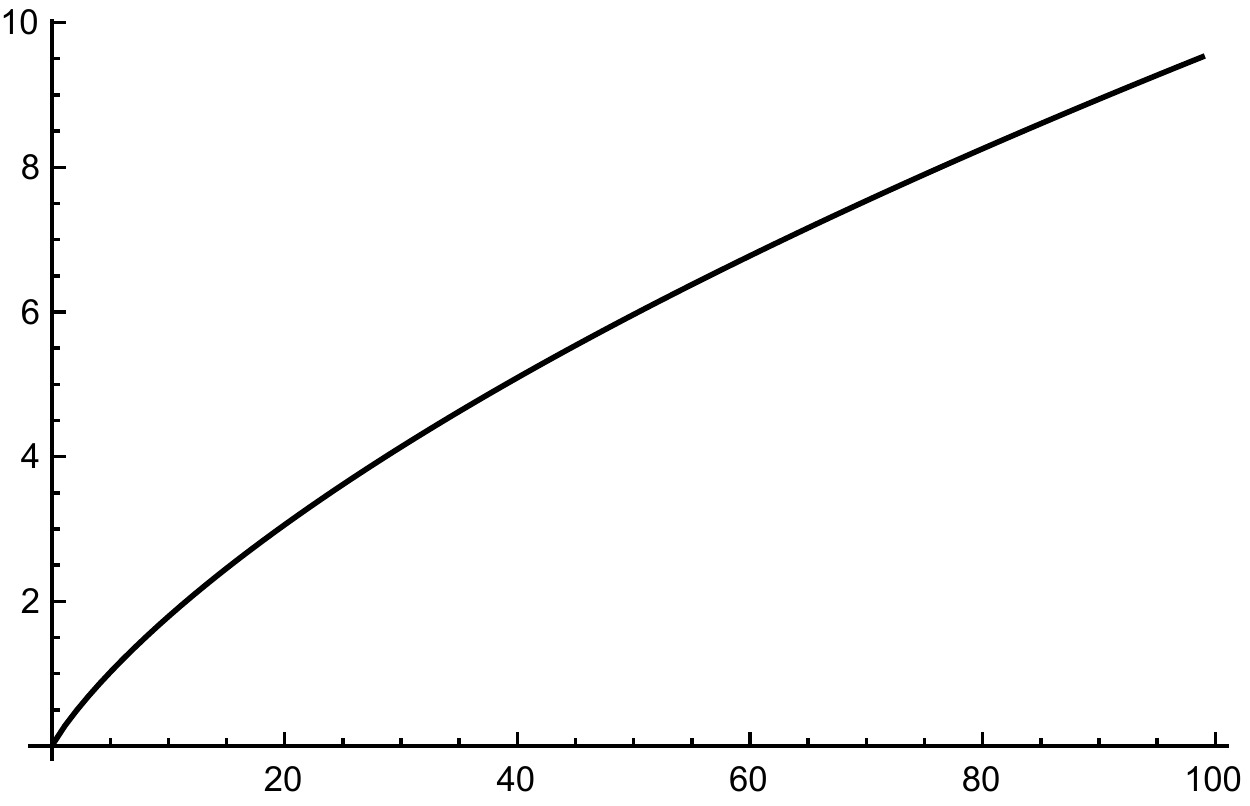}
$\xi$
\\
$D=4$
\hspace{4cm}
$D=5$
\caption{Minimum scale $L_D$ as function of the parameter $\xi$
for $D=4$ and $5$.
\label{scL}}
\end{figure}
\begin{figure}[h!]
\centering
\raisebox{3.3cm}{\tiny ${\frac{M_D}{m_D}}$}
\includegraphics[scale=0.4]{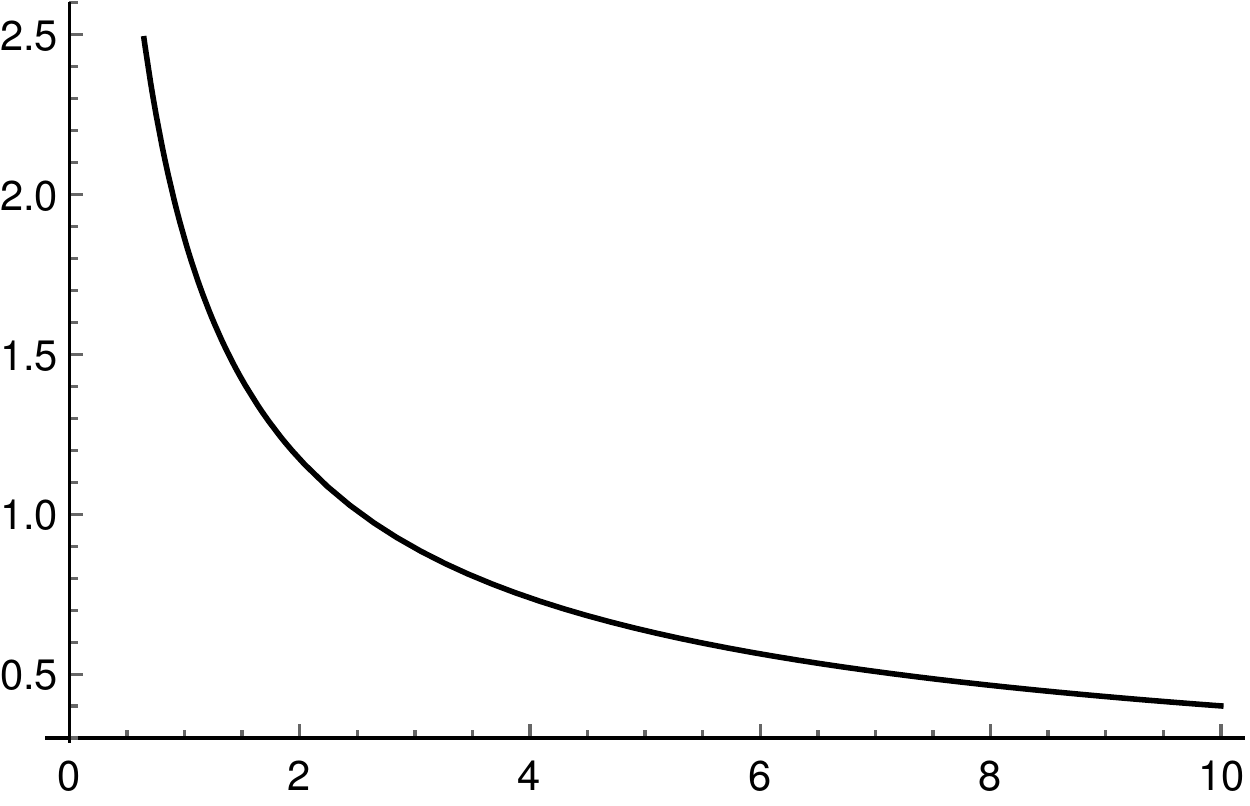}
$\quad$
\includegraphics[scale=0.4]{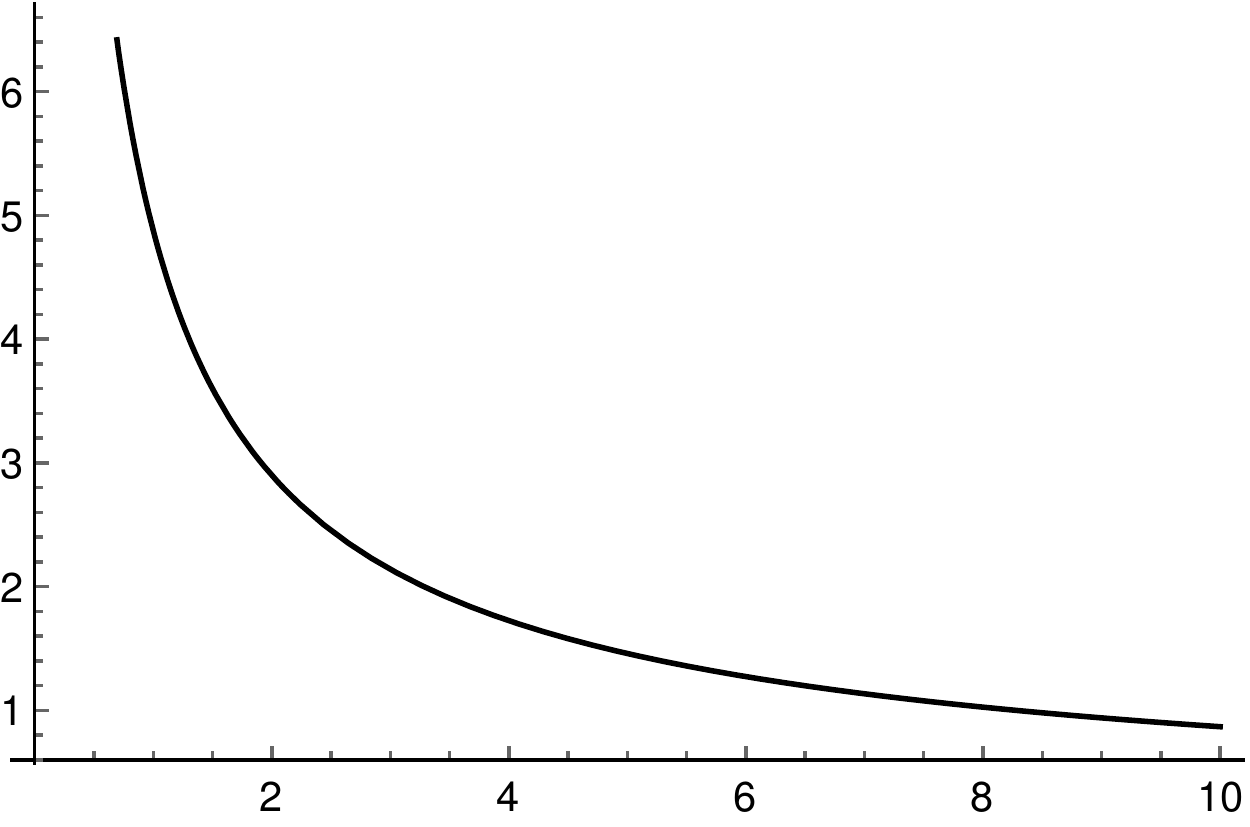}
$\xi$
\\
$D=4$
\hspace{4cm}
$D=5$
\caption{Minimum scale $M_D$ as function of the parameter $\xi$
for $D=4$ and $5$.
\label{scM}}
\end{figure}

Fig.~\ref{delta-r} shows the total uncertainty $\Delta r$ for different numbers of spatial
dimensions (and $\xi=1$).
It is clear that in higher dimensions, one obtains the same qualitative behaviour
as in $D=3$, with Eq.~\eqref{DrGUPD} being again minimised by a length $L_D$
corresponding to an energy scale $M_D$, which we plot in Figs.~\ref{scL} and~\ref{scM}
as functions of the parameter $\xi$.
From these plots we can infer that, for every value of $D$ considered here, the assumption
$M_D\simeq m_D$ makes large values of $\xi$ significant, whilst the opposite
happens if we set $L_D\simeq \ell_D$.
\section{Causal time evolution}
\label{Tevo}
So far, time evolution was not considered.
In the case of the two colliding wave-packets, one could sort of infer how the probability
for the system of particles to form a BH evolves by looking at the plot representing this
probability as a function of the distance between the two particles.
However, in this crude approximation, nothing would forbid the two particles from
crossing each other, and the probability $P_{\rm BH}$ to reach one and then
decrease.
How a non-negligible BH probability could affect the evolution of a quantum state
was addressed in Ref.~\cite{Casadio:2014twa} for the usual spherically symmetric
Gaussian wave-packet~\eqref{Gauss}.
In order to simplify the analysis, all Standard Model interactions are neglected and the
point of view is taken of an observer placed at a very large distance from this particle.
It seems therefore sensible to assume that, if the particle is not a BH ($P_{\rm BH}\ll 1$),
the time evolution is governed by the standard Schr\"odinger equation with the Hamiltonian
$H=E$ of the mass-shell Eq.~\eqref{mass-shell}.
If instead the system is a BH ($P_{\rm BH}\simeq 1$), no evolution should appear
to occur at all (Hawking evaporation is also neglected in this toy model).
The pictured considered in Ref.~\cite{Casadio:2014twa} is therefore of a BH as a
``frozen star''~\footnote{Historically, this name was commonly used for gravitationally
collapsed objects before the term BH was introduced~\cite{frozen}.}.
\par
When the wave-packet $\psi_{\rm S}$ does not fall into one of the above two
limiting conditions, the evolution for arbitrarily ``short'' time intervals $\delta t$
is taken to be described by means of the combination
\be
\psis(r,t+\delta t)
=
\left[
\mu_{\rm H}(t)\,\hat{\mathbb{I}}
+
\bar\mu_{\rm H}(t)\,e^{-\frac{i\,\delta t}{\mpl\lp}\,\hat H}
\right]
\psis(r,t)
\ ,
\label{dpsi}
\ee
where $\hat{\mathbb{I}}$ is the identity operator and the coefficients
\be
\mu_{\rm H}(t)\simeq P_{\rm BH}(t)
\simeq
1 - \bar\mu_{\rm H}(t)
\ ,
\ee
so that the two limiting behaviours are included by construction and
unitarity is preserved,
\be
1
=
\mu_{\rm H}^2+\bar\mu_{\rm H}^2
+2\,\bar\mu_{\rm H}\,\mu_{\rm H}\,\cos\left(\frac{\delta t}{\mpl\lp}\,\hat H\right)
\simeq
\left(\mu_{\rm H}+\bar\mu_{\rm H}\right)^2
\ ,
\label{muNre}
\ee
for $\delta t$ sufficiently short (see below about this very important point).
In this limit, Eq.~\eqref{dpsi} results in the effective Schr\"odinger equation
\be
i\,\mpl\lp\,\frac{\delta \psis(r,t)}{\delta t}
\simeq
\left[1-P_{\rm BH}(t)\right]
\hat H\,
\psis(r,t)
\ ,
\label{schro}
\ee
which reproduces the standard quantum mechanical evolution in the limit
$P_{\rm BH}\to 0$.
Since the (now time-dependent) probability $P_{\rm BH}=P_{\rm BH}(t)$ is determined
by the entire wave-function $\psi_{\rm S}=\psi_{\rm S}(r,t)$ and its associated HWF,
the apparently trivial correction it introduces is instead non-local, and cannot be 
reproduced by means of  a local interaction term of the form $H_{\rm int}=H_{\rm int}(r,t)$.
This insight makes it evident that it will be generally very hard to solve
Eq.~\eqref{schro}  for a finite time interval.
\par
By employing the spectral decomposition at fixed time $t$,
\be
\psis(r,t)
=
\sum_E
C_E(t)\,j_0(E,r)
\ ,
\label{CEt}
\ee
where $j_0$ is a spherical Bessel function of the first kind, Eq.~\eqref{dpsi} can
be written as
\be
i\,\mpl\lp\,\delta C_E(t)
\simeq
\left[1-P_{\rm BH}(t)\right]
E\,
C_E(t)\,
{\delta t}
\ .
\label{CETeq}
\ee
One can now determine $\delta C_E(t)$ provided $\psis(t)$ is known, and reconstruct
both $\psis$ and $\psih$ at the time $t+\delta t$, in order to proceed to the next time step. 
%
%
%
%
%
%
\par
\begin{figure}[t]
\centering
\raisebox{4cm}{${\mathcal P}_{\rm S}$}
\includegraphics[width=8cm,height=5cm]{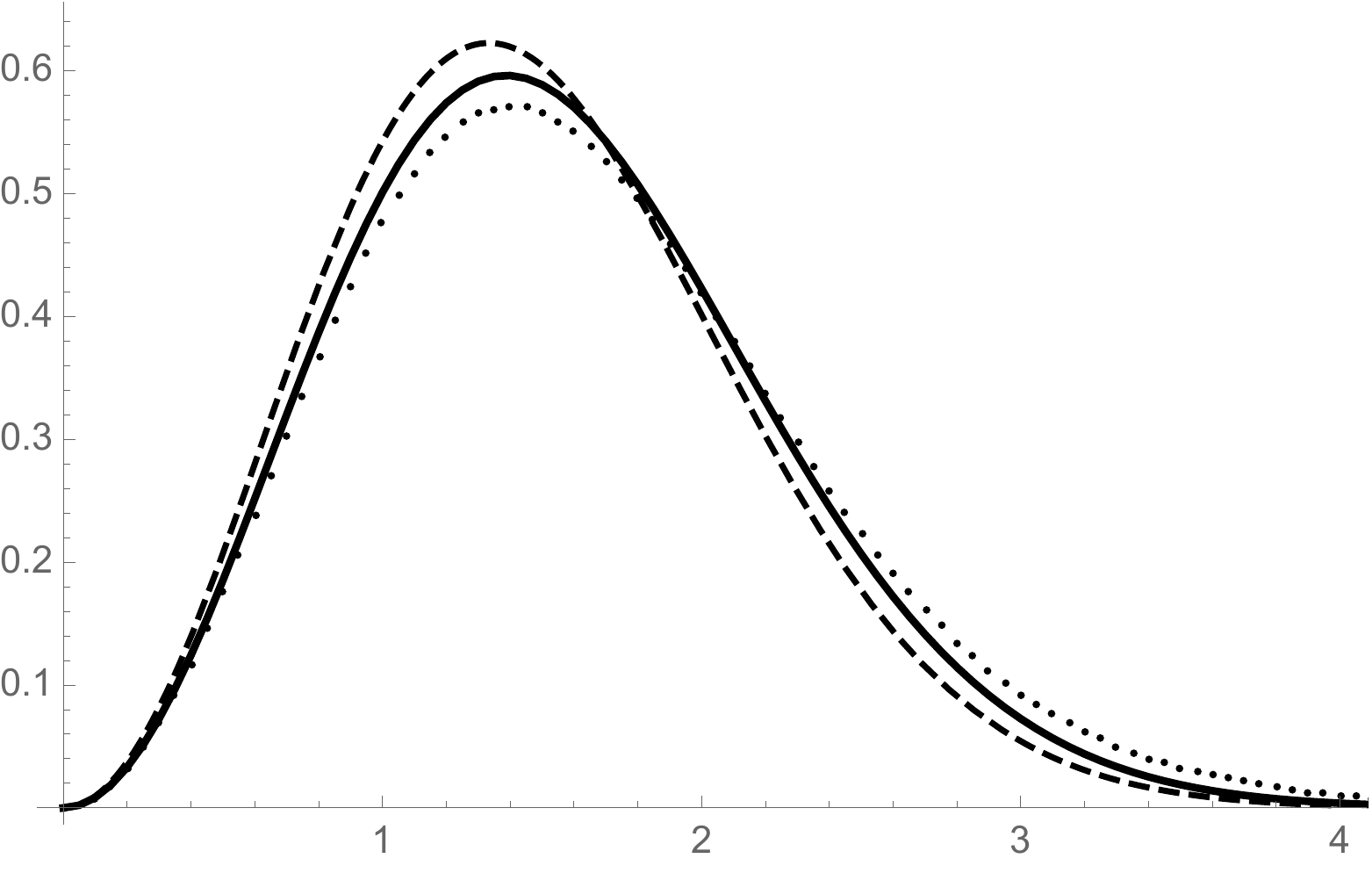}
\raisebox{0cm}{$\frac{r}{\lp}$}
\caption{Time-evolution of the probability density for the initial Gaussian
packet~\eqref{Gauss} with $m=3\,\mpl/4$ and $\ell=\lambda_m=4\,\lp/3$ (dashed line)
according to standard quantum mechanics (dotted line) compared to its causal
evolution~\eqref{schro} (solid line)
for $\delta t=\lp$.
\label{T1_2}}
\end{figure}

If $\psis(r,t=0)$ is the Gaussian wave-function~\eqref{Gauss}, the corresponding
$P_{\rm BH}(t=0)=P_{\rm BH}(\ell,m)$ discussed in Section~\ref{sec:neutral}, 
and this state will likely be a BH only if $m\gtrsim\mpl$ and $\ell\lesssim\lp$.
In particular, by setting $E\simeq\mpl$, we expect the evolution
equation~\eqref{CETeq} holds for
\be
\delta t\lesssim \lp\frac{\mpl}{E}\simeq \lp
\ ,
\label{dtp}
\ee
and even shorter intervals for modes with energy $E\gg\mpl$,
which is a form of the natural duality $(E>\mpl)\Leftrightarrow (\delta t<\lp)$.
One can now solve Eq.~\eqref{CETeq} with a time step satisfying~\eqref{dtp},
and subsequently obtain the wave-function $\psis(r,t=\delta t)$ by inverting the
decomposition~\eqref{CEt}.
Fig.~\ref{T1_2} shows the probability density 
${\mathcal P}_{\rm S}=4\,\pi\,r^2\,|\psis(r,t)|^2$ at $t=0$ and $t=\delta t=\lp$ for 
$m=3\,\mpl/4$ and $\ell=\lambda_m=4\,\lp/3$.
One can make a comparison with the density arising from the standard free evolution
during the same interval of time $\delta t=\lp$. 
In this case, the initial state is characterised by the minimum gravitational radius
$\Rh=1.5\,\lp$ given in Eq.~\eqref{CondTheta}, the expectation value of the energy
$\expec{E}\simeq 1.15\,\mpl$, the Schwarzschild radius
$\expec{\hat r_{\rm H}}\simeq 2.3\,\lp$, and initial probability $P_{\rm BH}\simeq 0.8$.
One immediately notices that the modified evolution makes the packet more confined
than the usual quantum mechanical one.
However, since the packet will keep on spreading, it is reasonable to guess that
$P_{\rm BH}(t+\delta t)<P_{\rm BH}(t)$, and the effect of the horizon 
will mitigate over time.
%
%
%
%
%
\begin{figure}[t!]
\centering
\raisebox{4cm}{${\mathcal P}_{\rm S}$}
\includegraphics[width=7cm,height=4.5cm]{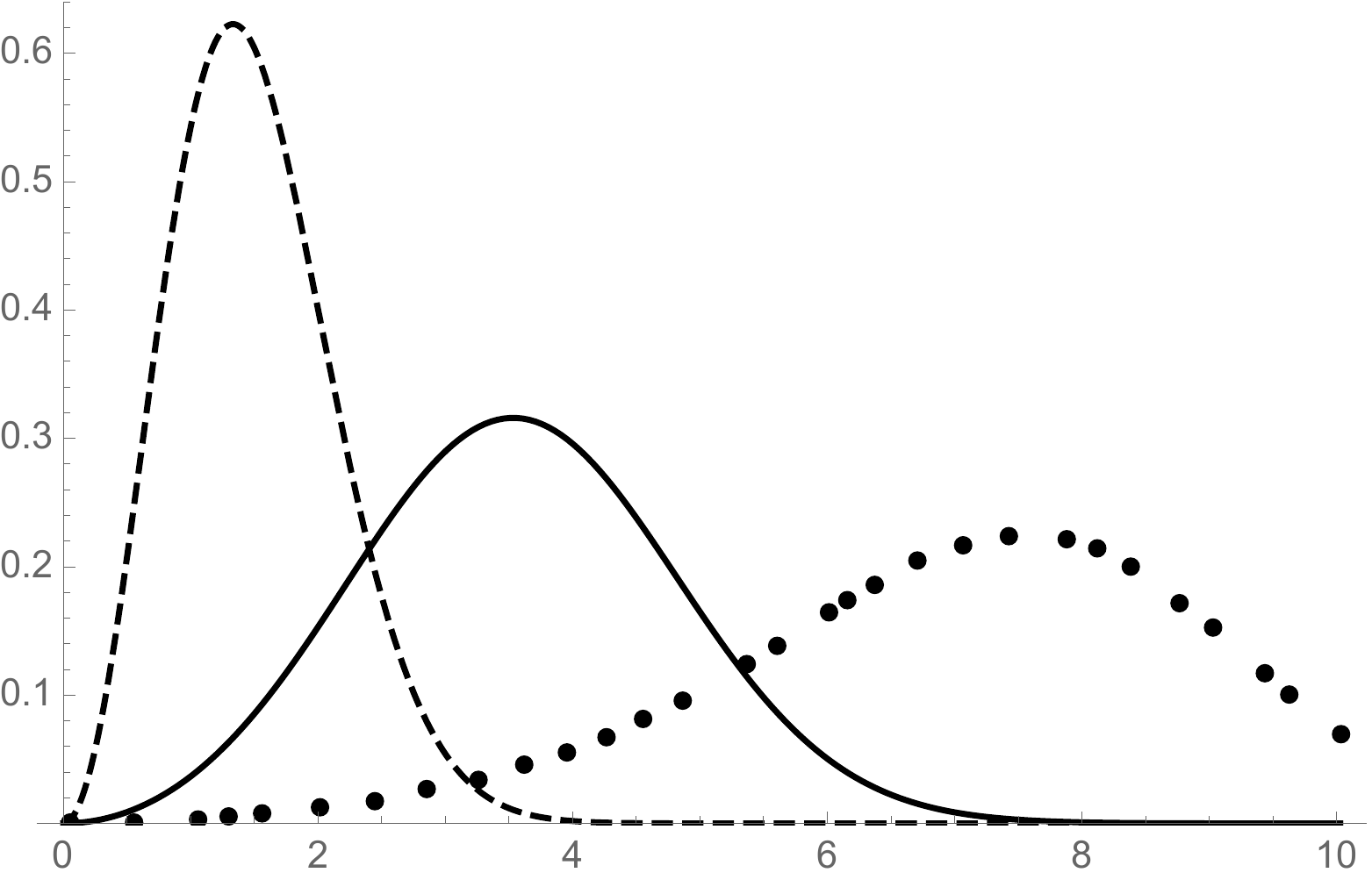}
{$\frac{r}{\lp}$}
\\
\raisebox{4cm}{${\mathcal P}_{\rm H}$}
\includegraphics[width=7cm,height=4.5cm]{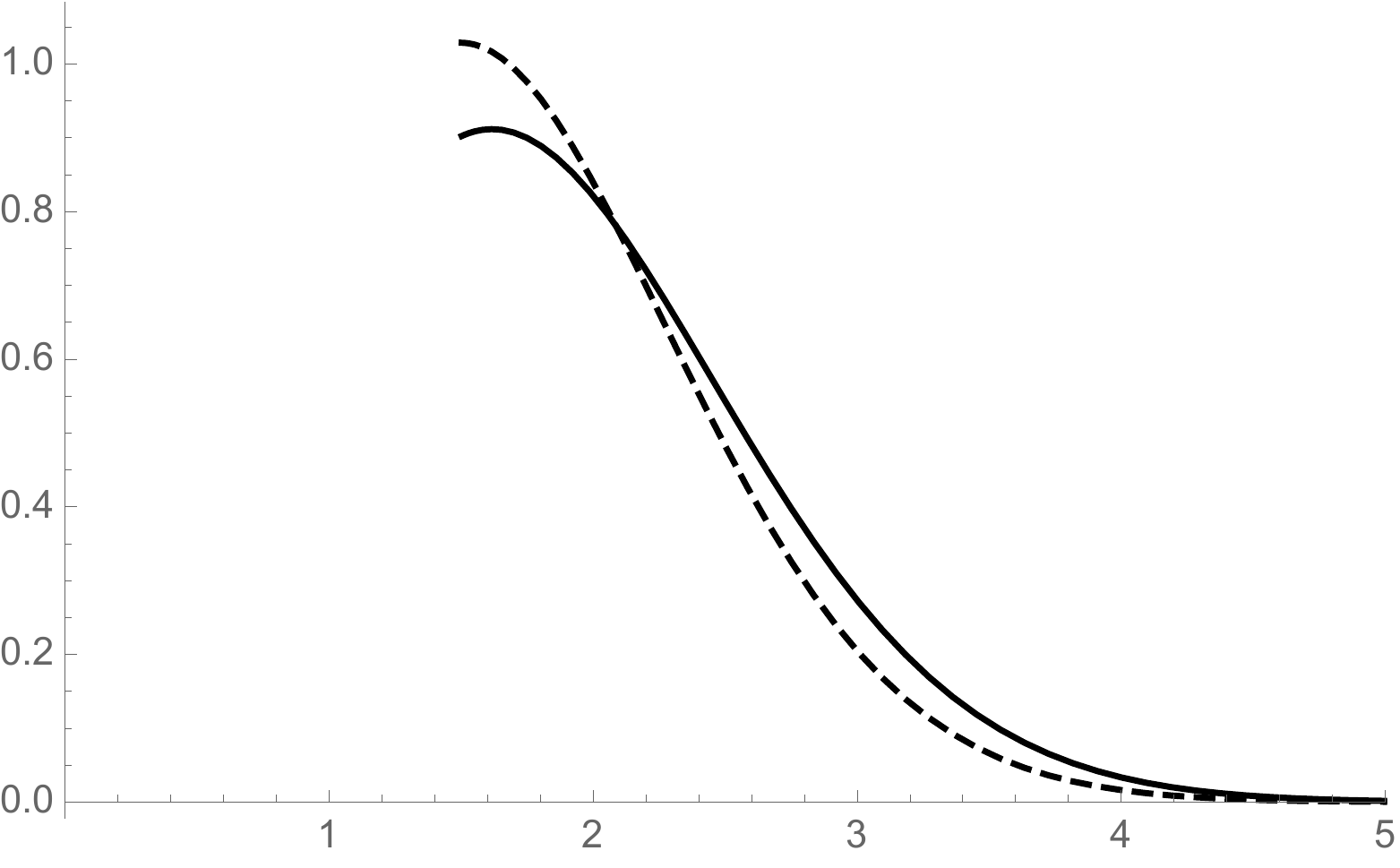}
{$\frac{\rh}{\lp}$}
\caption{Upper panel: probability density from the final wave-packet 
$\psis(r,10\,\lp)$ with $\ell=4\,\lp/3$ from the modified evolution~\eqref{schro}
(solid line) compared to the freely evolved packet (dotted line) and initial packet 
$\psis(r,0)$ (dashed line). 
Bottom panel: horizon probability density for the Gaussian particle in the
upper panel at $t=0$ (dotted line) and $t=10\,\lp$ (solid line).
Note that $\psih(\rh <\Rh,t)=0$, for $\Rh\equiv 1.5\,\lp$.
\label{Ppsit}}
\end{figure}
\begin{figure}[t]
\centering
\raisebox{4cm}{$P_{\rm BH}$}
\includegraphics[width=8cm,height=4.5cm]{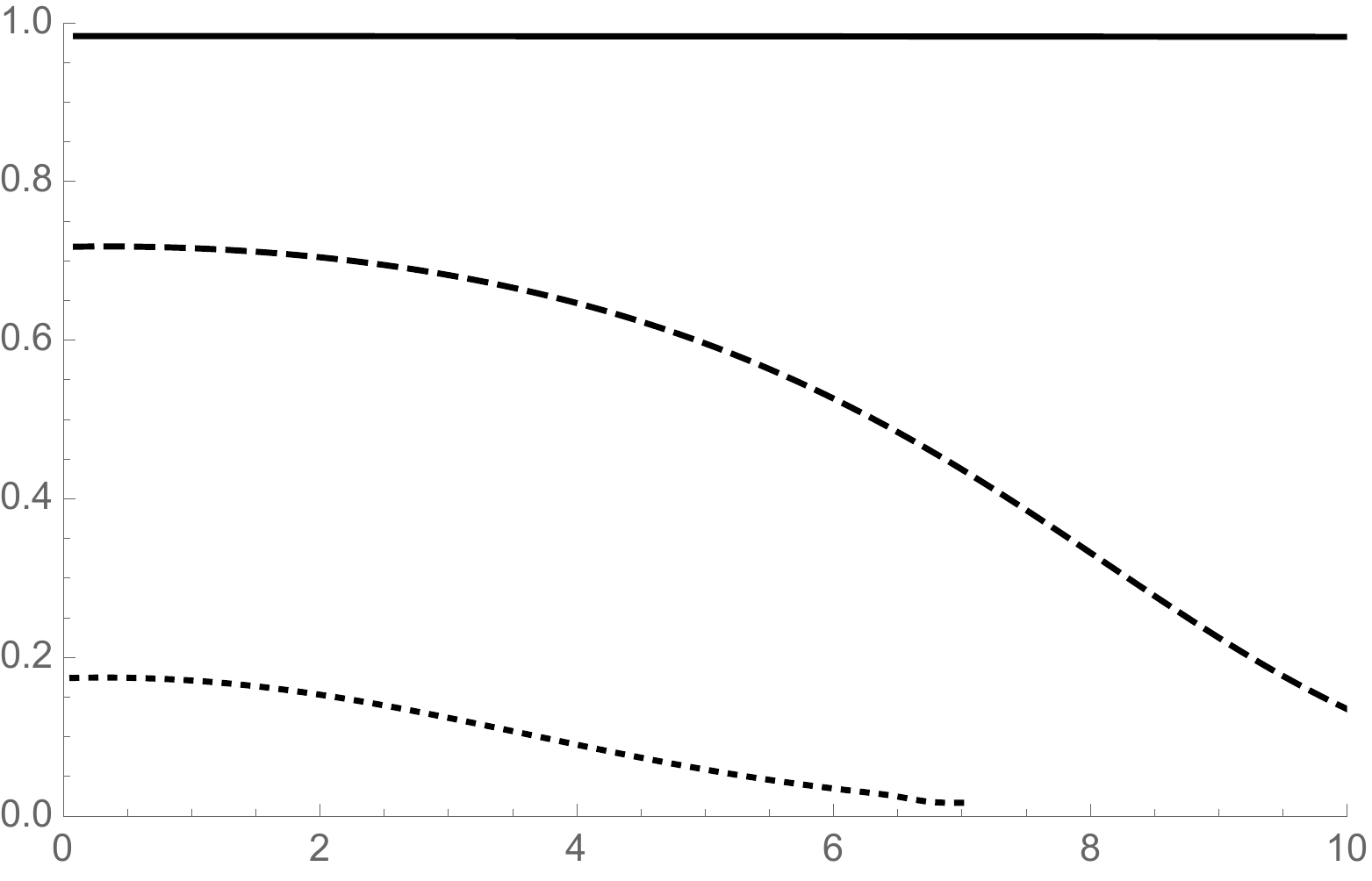}
{$\frac{t}{\lp}$}
\caption{Time-evolution of the probability $P_{\rm BH}$ for the Gaussian
wave-function~\eqref{Gauss} for $\ell=\lp$ (solid line),
$\ell=4\,\lp/3$ (dashed line) and $\ell=2\lp$ (dotted line).
\label{Pbht}}
\end{figure}
\par
Longer time evolutions can be obtained by discretising the time as
$t=n\,\delta t$, where $n$ is a positive integer and the time step $\delta t$
is bounded by~\eqref{dtp} for all relevant energies $E$ in the spectrum~\eqref{CEt}.
In Ref.~\cite{Casadio:2014twa} a numerical approach was employed in order to keep all
these features under control.
Fig.~\ref{Ppsit} shows the probability densities ${\mathcal P}_{\rm S}$ and 
${\mathcal P}_{\rm H}$ for $m=3\,\mpl/4$ and $\ell=\lambda_m=4\,\lp/3$,
at the time $t=10\,\delta t=10\,\lp$.
The broadening of ${\mathcal P}_{\rm S}$ is clearly slower than in
the standard quantum evolution, but still leads to a decreasing BH probability
density.
The time evolution of the BH probability is displayed in Fig.~\ref{Pbht} 
for $\lambda_m=\ell=\lp$, $4\,\lp/3$ and $2\,\lp$.
As usual, whenever the Gaussian width exceeds the Planck length, $\ell>\lp$, 
the BH probability tends to vanish very fast.
A possible interpretation of this result is that the initial quantum BH decays and 
its own Hawking radiation is simulated by the widening of the wave-function~\cite{Casadio:2014twa}.
\section{Conclusions}
\label{Conclusions}
Since Schwarzschild solved the field equations of General Relativity and BHs
entered the scene of contemporary physics, it was clear that they would have played a big
role in the correspondence between large gravitational structures and the geometry of
the space-time. Unfortunately, while giving very accurate corrections to Newtonian
gravity, General Relativity fails at punching through the realm of quantum physics,
which is renowned for giving a more reliable description of (microscopic) reality than the
one given in classical terms.
It seems therefore a prominent necessity to find a way which allows us to quantise the
gravitational interaction (and perhaps the geometry) as our understanding
of nature improves.
This review introduces the reader to the investigation of the quantum properties 
of the geometrical structures of space-time by means of the HWF, 
a tool that endorses the gravitational radius with properties expected of a quantum
mechanical observable.
\par
This HQM is best elucidated by modelling a spherically symmetric massive particle
with a Gaussian wave-function, which appears to be a viable description for
sources around the Planck scale, that is potential quantum BHs.
In fact, one finds a neutral particle is most likely inside its own horizon (i.e.~the BH
probability $P_{\rm BH}\simeq 1$) when its width $\ell$ reaches into the quantum
gravitational scale, $\ell\sim\lp$, or equivalently, the mass $m\sim\mpl$.
Moreover, the characteristic uncertainty in the horizon radius combined with the one
in the size of the quantum source, results in having a GUP and a minimum measurable
length (and corrections to the Hawking decay rate).
This procedure is also applicable to space-times with more than one horizon, 
like the Reissner-Nordstr\"om metric.
When the specific charge $\alpha<1$, one finds that for a considerably broad
interval of masses only the outer horizon has a large probability to form, while the
probability for the inner horizon to exist is negligible.
This result is counter-intuitive in the classical description and it is
a phenomenological prediction resulting directly from considering the quantum
nature of the causal structure of space-time.
The formalism also allows one to dive into the over-charged regime, where
one classically expects to have a naked singularity. 
It is possible to continue, albeit in a non-unique way, the HWF past the extremal
$\alpha=1$ case, but the specific charge is still limited by an upper value above which 
the basic properties of quantum systems, such as unitarity (which in turn
follows from the normalisability of the HWF) cannot be preserved.
\par
Forming BHs via particle collisions is a fascinating and straightforward implication
of Thorne's hoop conjecture.
In the strong approximation of a one-dimensional space (case in which the impact parameter
is zero), the probability for a trapping surface to appear as a result of the collision between 
two gaussian wave-packets was computed, lending support to a quantum version of the
hoop conjecture.
The main correction with respect to its classical version is that a minimum BH mass of the
order of the Planck scale is again confirmed, thus pushing the BH production by particle
collisions way beyond our experimental capabilities.
Of course, the picture could drastically change in scenarios with extra spatial dimensions,
where the fundamental gravitational mass that replaces the Planck scale could be within
our reach.
The HQM leads to significant corrections to the production cross-sections in the ADD
models, with a possible larger and larger suppression in higher and higher dimensions.
Particular emphasis was also given to lower-dimensional space-times, based on the recent
claim that quantum BHs could be effectively one-dimensional objects.
The HQM further supports the view that BHs in $(1+1)$ dimension cannot indeed be
classical.
\par
Most of the review has dealt with static configurations, and even the case of particle
collisions was treated in this perspective.
However, one should not forget the proper (classical) meaning of a horizon is to trap
matter inside of it, and one can hardly overlook how this property must affect the
time-evolution of the system deeply.
A possible time-dependent HQM is governed by a modified Schr\"odinger equation, 
in which the probability for the particle to lie inside its own horizon affects 
the evolution in such a way that a state with probability $P_{\rm BH}=1$ does no
longer evolve in time.  
As expected, even on qualitative grounds, this modified quantum dynamics 
slows down the spread of a Gaussian packet.
\par
The above cases show some of the uses of the HQM and open up many
perspectives for future works.
First of all, one could apply the HQM to simple models of spherically symmetric
gravitational collapse and estimate the chance that it actually leads to the formation
of a BH.
After extending the formalism to electrically charged wave-packets, the next natural
thing to do is then to investigate rotating sources.
This extensions is important for quantum BHS because elementary particles can have
non-vanishing spin, and because the impact parameter in particle collisions is generally
not zero, so that the resulting BH is expected to have angular momentum in most physical
cases.
\subsection*{Acknowledgments}
It is a pleasure to thank X.~Calmet, R.T.~Cavalcanti, J.~Mureika, A.~Orlandi,
F.~Scardigli, D.~Stojkovic for fruitful collaboration,
and A.~Davidson, G.~Dvali, A.~Giusti, C.~Gomez,
B.~Harms, F.~Kuhnel, P.~Nicolini, N.~Wintergerst for stimulating discussions.
R.~C.~and A.~G.~are partly supported by the INFN grant FLAG.
%
\appendix
\section{Useful integrals}
\label{Integrals}
In this review, we made use of integrals of the form
\be
{\rm I}_3
=
\int_1^\infty \gamma\left(\frac{3}{2}, A \, x^2 \right) \, e^{-x^2} \, x^2 \, \d x
\ee
where $A$ is a positive real parameter.
From 
\be
x\,e^{-x^2}=-\frac{1}{2}\,\frac{\d }{\d x}e^{-x^2}
\ee
and
\be
\frac{\d}{\d y} \, \gamma(s,y)
=
y^{s-1} \, e^{-y}
\ ,
\ee
upon integrating by parts, one obtains
\be
{\rm I}_3
&=&
\int_1^\infty \gamma\left(\frac{3}{2}, A^2 \, x^2 \right) \, e^{-x^2} \, x^2 \, \d x
\notag
\\
&=&
\frac{1}{2\,e}\,\gamma\left(\frac{3}{2}, A^2 \right)+\frac{1}{2}
\int_1^\infty e^{-x^2} \, \frac{\d}{\d x}
\left[x\,\gamma\left(\frac{3}{2}, A^2 \, x^2 \right)\right] \, \d x
\notag
\\
&=&
\frac{1}{2\,e}\,\gamma\left(\frac{3}{2}, A^2 \right)+A^3
\int_1^\infty e^{-\left(1+A^2\right)x^2} \, x^3\, \d x +\frac{1}{2}
\int_1^\infty e^{-x^2} \, \gamma\left(\frac{3}{2}, A^2 \, x^2 \right) \, \d x
\notag
\\
&=&
\frac{1}{2\,e}\,\gamma\left(\frac{3}{2}, A^2 \right)
+\frac{A^3\,\Gamma\left(2,1+A^2\right)}{2\left(1+A^2\right)^2}
+\frac{1}{2} \int_1^\infty e^{-x^2} \, \gamma\left(\frac{3}{2}, A^2 \, x^2 \right) \, \d x
\ ,
\ee
From the property
\be
\gamma\left(\frac{3}{2},A^2\,x^2\right)
&=&
\frac{1}{2}\gamma\left(\frac{1}{2},A^2\,x^2\right)-A\,x\,e^{-A^2\,x^2}
\notag
\\
&=&
\frac{\sqrt{\pi}}{2}\,\erf(Ax)-A\,x\,e^{-A^2\,x^2}
\ ,
\ee
the integral
\be
\int_1^\infty e^{-x^2} \, \gamma\left(\frac{3}{2}, A^2 \, x^2 \right) \, \d x
&=&
\frac{\sqrt{\pi}}{2}\int_1^\infty \erf(Ax) \, e^{-x^2} \, \d x
-A\int_1^\infty e^{-(1+A^2)x^2} \, \d x
\notag
\\
&=&
-\frac{A\,e^{-(1+A^2)}}{2(1+A^2)} +\frac{\pi}{4}[1-\erf(1)\erf(A)]
\notag
\\
\quad &&
-\pi\, T\left(\sqrt{2}\,A,\frac{1}{A}\right)
\ ,
\ee
where $\erf(x)$ is an error function and $T(a,b)$ is the Owen's $T$ distribution
defined as
\be
T(a,b)
=
\frac{1}{2\pi}\int_0^a \frac{e^{-\frac{1}{2}\,b^2\,(1+x^2)}}{1+x^2}\,\d x
\ .
\label{Towen}
\ee 
Finally, putting everything together yields
\be
{\rm I}_3
&=&
\frac{1}{2\,e}\, \gamma\left(\frac{3}{2},A^2\right)+\frac{\pi}{8}[1-\erf(1)\erf(A)]
+\frac{A^3\,\Gamma\left(2,1+A^2\right)}{2\left(1+A^2\right)^2}
\notag
\\
&&
-\frac{A\,e^{-(1+A^2)}}{4(1+A^2)}-\frac{\pi}{2}\, T\left(\sqrt{2}\,A,\frac{1}{A}\right)
\notag
\\
&=&
\frac{\pi}{8}\erfc(A)+\frac{\sqrt{\pi}}{4}\,\Gamma\left(\frac{3}{2},1\right)\,\erf(A)
-\frac{A}{2}e^{-(1+A^2)}
+\frac{A^3\,(2+A^2)\,e^{-(1+A^2)}}{2\left(1+A^2\right)^2}
\notag
\\
&&
-\frac{A\,e^{-(1+A^2)}}{4(1+A^2)}-\frac{\pi}{2}\, T\left(\sqrt{2}\,A,\frac{1}{A}\right)
\notag
\\
&=&
\frac{\pi}{8}\erfc(A)+\frac{\sqrt{\pi}}{4}\,\Gamma\left(\frac{3}{2},1\right)\,\erf(A)
-\frac{A(3+A^2)}{4(1+A^2)^2}e^{-(1+A^2)}
\notag
\\
&&
-\frac{\pi}{2}\, T\left(\sqrt{2}\,A,\frac{1}{A}\right)
\label{I3}
\ .
\ee
%
%
%
%
%
%

%
%
\end{document}